\documentclass[11pt]{article}

   \usepackage[margin=1in]{geometry}
  
\usepackage{algorithm, algpseudocode}

\usepackage[utf8]{inputenc}
\usepackage[pdftex]{graphicx}
\usepackage{afterpage}

\usepackage{amsmath}
\usepackage{breqn}
\usepackage{bm}
\usepackage{subcaption}
\usepackage{comment}
\usepackage{hyperref}
\usepackage{setspace}
\usepackage{threeparttable}
\usepackage[round]{natbib}

\usepackage[title]{appendix}
\usepackage{caption}
\usepackage{amsmath}
\usepackage{booktabs}
\usepackage{lscape}
\renewcommand{\thefootnote}{\arabic{footnote}}
\usepackage{adjustbox}
\usepackage{rotating}
\usepackage{tikz}

\usepackage{float}
\usepackage{footmisc}
\usepackage{amsmath,amssymb}
\usepackage{amsthm}
\usepackage{enumerate}

\usepackage[inline]{trackchanges}

\newtheorem{mythm}{Theorem} 
\newtheorem{myprop}[mythm]{Proposition}
\newtheorem{mylemma}{Lemma}
\newtheorem{mydef}{{D}e{f}inition}

\newtheorem{myassump}{Assumption}
\newtheorem{mycor}{Corollary}

\usepackage{chngcntr}
\usepackage{apptools}
\AtAppendix{\counterwithin{mylemma}{section}}

\usepackage{bbold}

\newcommand{\ex}[1]{\mathbb{E}\left[#1\right]}
\newcommand{\lrb}[1]{\bigg[#1\bigg]}
\newcommand{\expe}{\mathbb{E}}

\newcommand{\prob}{\mathbb{P}}

\DeclareMathOperator{\sgn}{sgn}

  \title{The Adoption of Blockchain-Based Decentralized Exchanges
}
 \date{\today}

 \author{
 \large Agostino Capponi\thanks{\scriptsize Columbia University, Department of Industrial Engineering and Operations Research, Email: \href{mailto:ac3827@columbia.edu}{\texttt{ac3827@columbia.edu}}}, Ruizhe Jia\thanks{\scriptsize  Columbia University, Department of Industrial Engineering and Operations Research, Email: \href{rj2536@columbia.edu}{\texttt{rj2536@columbia.edu}}.}
}
\begin{document}
\vspace{-1cm}
\maketitle
\vspace{-1.2cm}

\begin{abstract}

We investigate the market microstructure of Automated Market Makers (AMMs), the most prominent type of blockchain-based decentralized exchanges. We show that the order execution mechanism yields token value loss for liquidity providers if token exchange rates are volatile. AMMs are adopted only if their token pairs are of high personal use for investors, or the token price movements of the pair are highly correlated. 
A pricing curve with higher curvature reduces the arbitrage problem but also investors’ surplus. Pooling multiple tokens exacerbates the arbitrage problem. We provide statistical support for our main model implications using transaction-level data of AMMs.

\end{abstract}



	\newpage

\section{Introduction}
Since the emergence of Bitcoin in 2008, practitioners and academics have argued that financial innovations such as 
tokenization of assets and decentralized ledgers, along with the backbone blockchain technology, will disrupt traditional financial services (see, e.g.,  \cite{cryptofinance}, \cite{Yermack}, \cite{blockchaindisruption},  \cite{blockchainsettlement}, \cite{tokenizationcoporatefinance}, \cite{ICO1}). However, despite thousands of crypto tokens have been created and the total capitalization of cryptocurrencies has exceeded 1.7 trillions as of early 2021, no blockchain-based financial service providers has yet truly challenged traditional financial intermediaries. 
Ironically, most transactions of crypto tokens still rely on unregulated centralized intermediaries that expose investors to the risk of thefts and exit scams (see, e.g.,  \cite{GANDAL201886}).

In the mid of 2020, a new type of blockchain application called decentralized finance, and commonly referred to as {D}e{F}i, has emerged (see \cite{introtodf} for an overview). {D}e{F}i utilizes open-source smart contracts on blockchains to provide financial services which typically rely on centralized financial intermediaries{.} {One of the most prominent DeFi innovations are decentralized exchanges which run on the Ethereum blockchain. Most of these exchanges utilize an Automated Market Maker (AMM) smart contract, which makes the market according to a deterministic algorithm, instead of utilizing order books and relying on central intermediaries.} The largest of these decentralized exchanges, Uniswap, has become the fourth-largest cryptocurrency exchange by daily trading volume, just a few months after its launch (\cite{bloomberg}). 

The AMMs built on blockchains are revolutionary in many ways. First, the settlement of transactions is instantaneous, after they are confirmed and included on the blockchain. Prior to settlement, traders still retain full control of their tokens. This eliminates counterparty risk for the users. Second, users {of} AMMs do not need to be paired to complete a transaction. Rather, they gain immediate access to available liquidity by interacting with the smart contract. Thirdly, different from traditional centralized exchanges where liquidity providers are typically professional market makers, any token holder can become a liquidity provider by depositing their tokens and earning fees from trading activities.

The mechanism underlying decentralized exchanges is fundamentally different from that of traditional exchanges. Hence, existing market microstructure literature has little to say about promises and pitfalls of this new form of exchanges. Many important questions remain unanswered: Can AMM provide sufficient incentives for provision of liquidity? Will there be any transaction breakdown where the liquidity reserve of the AMM is drained? What kind of tokens are most suitable for AMMs?  
How does the AMM structure affect trading activities and economic incentives of market participants?  

In this paper, we develop a game theoretical model to answer the questions above, and provide empirical support to our main model implications. 
We show that, even without any information asymmetry, liquidity providers face arbitrage problems if the token exchange rate fluctuates. This stands in contrast with the adverse selection problem that market makers face in a centralized limit-order-book exchange (see, for instance, \cite{GLOSTEN198571}). While market makers in a centralized limit-order-book exchange can alleviate adverse selection through bid-ask spreads, liquidity providers in the AMM can neither charge a bid-ask spread nor front-run because of the order execution structure of blockchain.



We analyze the incentives of liquidity providers and characterize the subgame perfect equilibrium of the game. 
At equilibrium, if the exchange rate of tokens is {too} volatile, liquidity providers do not {deposit their tokens in the AMM}, which leads to a ``{\it liquidity freeze}". In such case, the expected fee revenue from providing liquidity is lower than the expected arbitrage loss plus the opportunity cost from holding the tokens to deposit in the AMM.  Moreover, a ``liquidity freeze'' is less likely to occur if investors extract large private benefits from using their tokens on the corresponding platforms, or if the price movement of two tokens are highly correlated. In the former case, investors would have strong incentives to trade, hence resulting in large trading volumes and high fee revenues for liquidity providers; in the latter case, the prices of two tokens are more likely to co-move, thus leading to less arbitrage opportunities. Last, but not least, we argue that a ``liquidity freeze" is more likely to occur for tokens with low expected returns, 
as liquidity providers incur a large opportunity cost for holding those tokens in their portfolios.

Our equilibrium results also help correct the widely spread misconception that the token value loss due to arbitrage is {\it impermanent} and vanishes when the exchange rate reverts to its initial value. We argue that the long-established benchmark used to measure the ``impermanent loss" is unable to fully account for the opportunity cost of depositing in the AMM. 
{We show that, contrary to the popular belief, liquidity providers' incentives to deposit in the AMM may become stronger if the expected ``impermanent loss'' increases and  the probability of exchange rate reversion decreases.} 

We show that if the AMM is largely adopted, both the expectation and variance of transaction fees charged by the blockchain underlying the exchange increase. This imposes negative externalities on other decentralized applications operating on the same blockchain. Higher and more volatile transaction fees may prevent consumers from using these applications or cause high execution delays. Hence, our results indicate that not only does the blockchain affect the {D}e{F}i applications built on it, but the {D}e{F}i applications also affect the underlying blockchain.


We use our model to analyze the design of an AMM. There currently exist hundreds of AMMs, which mainly differ in terms of the pricing curve used and the number of managed tokens. (See, for instance, \cite{xu2021sok}). 
First, we show that the curvature of the pricing function determines the severity of the arbitrage problem and the fee revenue generated by trading activities. Pricing functions with larger curvature reduce the arbitrage problem, but also decrease  investors'  surplus. {We construct an optimal pricing function, which maximizes welfare and deposit efficiency, and minimizes the occurrence of a ``liquidity freeze''.} Second, we explore how pooling more than two different tokens in the AMM affects {the occurrence and profitability of} { exploitable arbitrage opportunities.} Our analysis warns against a typical fallacy---pooling more tokens in the AMM reduce the token value loss due to diversification effects. We show that pooling multiple tokens not only increases the {probability that an arbitrage occurs} but also allows the arbitrageur to extract a larger portion of deposited tokens.




We provide empirical support to the main testable  implications of our model using transaction-level data from {Uniswap V2  and Sushiswap AMMs}. We identify the deposit, withdrawal, and swap orders from the raw transaction history of 80 AMMs {offering the most actively traded token pairs}, in the period from Dec 22, 2020 to June 20, 2021. 
Consistently with our theoretical predictions, we find that token exchange rate volatility has a negative effect on deposit flow rate, while trading volume has a positive effect. Both effects are 
statistically and economically significant. An increase in weekly log spot rate volatility by 0.04 (around one-standard-deviation) decreases the weekly deposit flow rate by 0.06 (around 25\% of the standard deviation). Moreover, an increase in trading volume by  one standard deviation increases the weekly deposit flow rate by around 35\% of the standard deviation. 

We exploit the segmentation of AMMs by dividing them into two groups: ``stable pairs" and ``non-stable pairs". ``Stable pairs" consist of two stable coins which are each pegged to one US dollar, and thus these pairs have very low token exchange rate volatility compared with ``unstable pairs". {Consistent with our theoretical predictions, we find that gas fees for transactions of  ``stable pairs" are 8\% lower than those of ``non-stable pairs", and that the weekly volatility of gas fees for ``stable pairs" is about 40\% lower than for ``non-stable pairs". } This implies that the size of (negative) externalities imposed by ``stable pairs" on other platforms using the Ethereum blockchain is smaller. 


The rest of paper is organized as follows. Section \ref{sec: institutional details} gives institutional details of crypto exchanges and describes the AMM. Section \ref{sec:model setup} develops the game theoretical model. Section \ref{sec:analysis} solves for the subgame perfect equilibrium of the model, and analyzes its economic implications. Section \ref{sec:ext} discusses  the design of AMMs. Section \ref{sec:empirical} tests statistically the main model implications.  We conclude in Section \ref{sec conclu}. 

\paragraph{Literature Review.}

Our paper contributes to the so-far scarce, but rapidly growing, literature on blockchain-based decentralized finance.  Relevant contributions include \cite{angeris2021analysis}, who show that the AMM can track the market price closely under no-arbitrage conditions; \cite{bartoletti2021theory} who abstract away from the economic mechanisms behind AMMs, and prove a set of structural properties; \cite{frontrunning} who provide empirical evidence for the existence of arbitrage at AMM. We refer to \cite{introtodf} for an excellent and comprehensive survey of {D}e{F}i applications. {In their survey, they highlight the potential token value loss faced by liquidity providers, often referred to as ``impermanent loss".} We show that the widely accepted measure of ``impermanent loss" can be misleading and lead to sub-optimal liquidity management strategy of liquidity providers.

To the best of our knowledge, our paper is the first to explore theoretically and empirically the market microstructure of AMMs and their design. It is worth mentioning two recent complimentary studies to ours. \cite{anotherAMM} discusses front-running arbitrage, a different form of arbitrage that arises in blockchain-based crypto exchanges, and \cite{anotherAMM2} {compare liquidity provision at a centralized exchange with liquidity provision at an AMM which utilizes a constant product function}. 

Our results also add to existing literature on crypto trading. \cite{Tether}, \cite{washtrade1}, and \cite{li_shin_wang_2018} analyze the trading activities and price manipulations at centralized crypto exchanges. We contribute to this strand of literature by analyzing the economic incentives behind trading and provision of liquidity at decentralized exchanges, whose trading volume has been growing steadily. 


Our paper is broadly related to the the stream of literature studying blockchain technologies. Some of these studies investigate miners' incentives and how blockchain protocols achieve decentralized consensus. Noticeable contributions in this direction include \cite{blockchaineconomics},  \cite{Folk}, \cite{axiomatic}, \cite{PoS},  \cite{economicllimit}, \cite{POSEVOLUATION}, \cite{hinzen_john_saleh_2019}, and \cite{miningpool}. \cite{hinzen_john_saleh_2019} argue that the Bitcoin's protocol, despite ensuring decentralization, may result in limited adoption of Bitcoin as a payment system. Our results show that while gas fees incentivize miners and guarantee decentralization on the Ethereum blockchain, they also make the arbitrage problem on {D}e{F}i exchanges unavoidable and consequently reduces adopting.  

A related branch of literature analyzes blockchain in the context of crypto transactions and pricing. Contribution in this direction include  \cite{transaction1},  \cite{weixiong},  \cite{miningtotrade}, \cite{pagnotta2018equilibrium}, \cite{SIMPLEBTC}, \cite{athey2016bitcoin}, and  \cite{irresberger_john_mueller_saleh_2020}. Unlike these studies which view blockchain as the technology underlying payment systems, we investigate the pros and cons of blockchain as an infrastructure for decentralized exchanges. In their work, \cite{irresberger_john_mueller_saleh_2020} highlight the functionality of public blockchain as an infrastructure supporting {D}e{F}i applications.  


\section{Crypto Exchanges} \label{sec: institutional details}
In this section, we provide institutional details of crypto exchanges and discuss the mechanics of AMMs.

\subsection{Institutional Details}
Tokenization has gained increasing popularity since the introduction of bitcoin in 2008 by \cite{bitcoin}. As of January 2021, there are over 4,000 crypto tokens created, distributed, and circulated (\cite{bagshaw_2020}). 
With the increasing adoption of cryptocurrencies, many exchanges have been created specifically for trading of crypto tokens.  Those exchanges usually fall into two categories: centralized exchanges and decentralized exchanges (often called DEX).

A centralized cryptocurrency exchange is a trusted intermediary which  monitors and facilitates crypto trades as well as securely stores tokens and fiat currencies. Similar to the equity market, centralized exchanges are often in the form of limit order books, and many of them also provide leverage and derivative trading. However, different from the equity market, most of the centralized cryptocurrency exchanges are unregulated and lack of proper insurance for the assets stored. This presents concerns for safety, trustworthiness, and potential manipulation.\footnote{
The study of \cite{GANDAL201886} highlights price manipulation behavior on crypto exchanges including Mt.Gox which, at its peak, was responsible for more than 70\% of bitcoin trading. In early 2014, Mt.Gox suddenly closed its platform and filed for bankruptcy claiming that the platform's wallet was hacked and a great amount of assets were stolen. Other centralized exchanges subject to thefts and exit scams include Binance, BitKRX, BitMarket, PonziCoin, and so on.}

Because of the concerns presented by centralized crypto exchanges, decentralized exchanges are becoming alternative platforms for the purchase and sale of crypto tokens. As of August 2020, DEX account for more than 5\% of the total crypto trading, and their market shares have been increasing steadily (\cite{mcsweeney_2020}). Different from centralized exchanges, DEX are blockchain-based smart contracts which can operate without a trusted central authority. Most of them  utilize an AMM smart contract that tracks a constant product function. {The two largest DeFi exchanges, Uniswap and Sushiswap,} respectively make up for about 50\% and 20\% of the trading volume at decentralized exchanges (\cite{smith_das_2021}).

Orders are typically executed on the scale of milliseconds in a centralized exchange (\cite{sedgwick_2018}). Unlike centralized exchanges which facilitate trades and manage all orders using their own infrastructure, decentralized exchanges take in, manage, and execute orders through a blockchain-based platform (typically Ethereum). Instead of sending the order directly to the exchange, users submit their orders\footnote{Users maintain full custody of their tokens and can overwrite their orders before they are confirmed. As a result, users are not exposed to counterparty risk, that is, the risk that the exchange defaults on its obligations to deliver tokens after the transaction is confirmed. Upon confirmation, the delivery of tokens is instantaneous.} to the blockchain network and attach a transaction fee (the gas price and gas limit in the Ethereum network). The miner that mines the very next block will prioritize orders based on the attached gas price, from highest to lowest, and append them to the blockchain. Since each block has a maximum size, the number of orders a miner can include in a single block is limited. Hence, an order with a too low transaction fee may need to wait a few minutes before being confirmed and executed. 


\subsection{AMM with Constant Product Function}
We provide a brief description of the most common AMM {smart contract}, which deploys a constant product function. We refer to \cite{adams_2020} for a more extensive overview.

An AMM does not rely on a limit-order book for transactions. Rather, it develops a new market structure called liquidity pool. A liquidity pool allows for a direct exchange of two crypto tokens, say A and B tokens, instead of first selling A tokens for fiat currency and then purchasing B tokens using proceeds from the asset sale. Each liquidity pool {typically} manages a pair of tokens.\footnote{It is also possible to pool multiple tokens together, despite pools with a single pair are the most common. It is worth emphasizing that the trading mechanism of pools with more than two tokens is almost identical to that of pools with a single pair of tokens.} {We remark that any pair of tokens can form a liquidity pool. Hence, liquidity pools can support trading activities for tokens not yet listed on centralized exchanges.}

The liquidity pool works by incentivizing owners to deposit their tokens into the smart contract. Assume a liquidity pool manages the exchange of two tokens, A and B, where each A token is worth $p_A$ and each $B$ token is worth $p_B$. Anyone who owns both A and B tokens can choose to be a liquidity provider by depositing an equivalent value of each underlying token in the AMM and in return, receiving pool tokens which prove his share of the AMM. For example, if the current reserve in the liquidity pool contains $10$ 
tokens A and $5$ tokens B, and the current value is 1 {dollar} for one A token and 2 {dollars} for one B token, then the liquidity provider must deposit A tokens and B tokens in the ratio 2:1. After depositing $10$ tokens A and $5$ tokens B, the liquidity provider can claim pool tokens that account for half of the total tokens in the current liquidity reserve.  The provider can exit the liquidity pool by trading in his pool tokens, and receiving his share of the liquidity reserve in the  AMM. For instance, if the liquidity reserve contains 25 tokens A and 9 tokens B when the liquidity provider exits and no one deposits after her, then she receives 12.5 tokens A and 4.5 tokens B. 

{Suppose a new investor arrives at the AMM and wants to exchange A for B tokens. To complete such a trade, this investor does not need to search for a counterparty who is willing to exchange B for A tokens. Rather, she directly interacts with the AMM by submitting a swap order through which she deposits an amount $\Delta_A$ of A tokens and withdraws an amount $\Delta_B$ of B tokens.} The quantities $\Delta_A$ and $\Delta_B$ satisfy $(20+\Delta_A) (10 - \Delta_B) = 20*10= 200$. {That is, multiplying the amount of both tokens in the AMM must yield a constant.} Formally, assume the initial liquidity reserve in the AMM contains $y_A$ A tokens and $y_B$ B tokens. Then the trade needs to satisfy $(y_A+\Delta_A) (y_B - \Delta_B) = y_Ay_B$. In addition to the amount $\Delta_A$ of tokens A exchanged, the investor must pay an additional amount $f\Delta_A$ of tokens A as trading fee. Most of this fee is then added to the liquidity pool.\footnote{In practice, a small portion of the trading fees may be collected by the underlying platform. This can be incorporated in the model by multiplying the trading fee collected by the liquidity provider with a constant term. Because such an additional feature would not qualitatively change our results, we opted for leaving it out.} Hence, the trading fee increases both the total liquidity reserve of the AMM and the AMM share of liquidity providers. Liquidity providers are incentivized to deposit because they earn the trading fee. For Uniswap V2 and Sushiswap, this trading fee is currently set to $0.3\%$ {of the tokens that investors trade in}. The ratio between the amount of two tokens in the liquidity pool equals the spot exchange rate when the trading size is infinitesimally small.\footnote{It can be easily verified that if $\Delta_A \rightarrow 0$, then $\frac{\Delta_A}{\Delta_B} \rightarrow \frac{y_A}{y_B}$.}  

It is worth noticing that the relationship $(y_A+\Delta_A) (y_B - \Delta_B) = y_Ay_B$ can, in principle, be replaced by $F (y_A+\Delta_A , y_B - \Delta_B) = F (y_A, y_B)$, where $F(x,y)$ is an arbitrary pricing curve referred to as the pricing function. AMMs differ in terms of the chosen pricing function. The constant production function is just a special case, where $F(x,y) =xy.$ We refer to \cite{xu2021sok} for an overview of pricing functions used by different AMMs. 

\section{Baseline AMM Model} \label{sec:model setup}
Our baseline model consists of $2$ periods indexed by $t$, $t =
 1, 2.$ Each period has 3 sub-periods. There are three kinds of agents: liquidity providers, an arbitrageur, and investors. The discount factor of each agent is equal to $1$.

\subsection{The Pricing Function of the AMM}
The agents have access to two different tokens, referred to as A and B tokens. The tokens can be used directly on the corresponding platforms A and B. Alternatively, they can be exchanged for a single  consumption good used as a numeraire at prices which are public information for all agents. We denote the amount of consumption good that a single A and B token can exchange for at the end of sub-period $s$ of period $t$,  as $p_A^{(t,s)}$ and $p_B^{(t,s)}$ respectively.  {We will  refer  to $p_A^{(t,s)}$ and $p_B^{(t,s)}$ as the token prices of A and B tokens respectively interchangeably throughout the paper.} We will refer to the ratio of token prices, $\frac{p_A^{(t,s)}}{p_B^{(t,s)}}$, as the fair value exchange rate.  

There is a smart contract built on a blockchain that functions as the AMM for A and B tokens. The smart contract utilizes a  twice continuously differentiable pricing function $F(x,y): \mathbb{R}^2 \rightarrow \mathbb{R} $ to decide the exchange rate for any trade. If the AMM contains an amount $y_A$ of A tokens and $y_B$ of B tokens, any trade that exchanges $\Delta_A$ A tokens for $\Delta_B$ B tokens needs to satisfy the relation $F(y_A+\Delta_A , y_B - \Delta_B) = F (y_A, y_B), 0 \leq  \Delta_B \leq y_B $. An additional amount $f\Delta_A$ of A tokens needs to be added to the AMM as a trading fee. 

\begin{myassump}\label{curve}
The function $F(x,y): \mathbb{R}^2 \rightarrow \mathbb{R} $  satisfies the following properties:
\begin{enumerate}
    \item $F_x >0, F_y > 0$.
    \item $F_{xx}<0, F_{yy}<0, F_{xy}>0$.
    \item $ \forall c \geq 0, c^{l}F(x,y) = F(cx,cy)$ for some $l>0$.
    \item $\lim_{x \rightarrow 0}\frac{F_x}{F_y} = \infty, \lim_{x \rightarrow \infty}\frac{F_x}{F_y} = 0$, $\lim_{y \rightarrow 0}\frac{F_x}{F_y} = 0, \lim_{y \rightarrow \infty}\frac{F_x}{F_y} = \infty$.
\end{enumerate}
\end{myassump}


The first assumption ensures that a positive amount of A tokens can be exchanged for a positive amount  of B tokens from the AMM.  The second assumption {guarantees} that the 
curve $F(x,y) = C$, where $C$ is a constant, is convex. This implies that if the demand of A tokens goes up, the exchange rate used to convert from B to A tokens correspondingly increases. Symmetrically, if the demand for B tokens increases, a higher amount of A tokens is required to exchange for a single B token. The third assumption states that the function $F$ is homogeneous of degree $l$, and ensures that the exchange rate at the AMM  does not change significantly if the amount of deposited tokens {goes up}. {The last condition ensures that the AMM supports trading for any token exchange rate in the interval $[0, \infty)$.} {These four properties can be verified to hold for the majority of existing AMMs}. (see, for instance, \cite{xu2021sok}).

\subsection{Liquidity Providers}

 There are $n>1$ liquidity providers, indexed by $\mathcal{N}= \{1,2,...,n\}$, and each endowed with {a} positive amount of consumption good at $t=0$. We use $e_i^{(0)}$ to denote the initial endowment of liquidity provider 
 $i\in \mathcal{N}$, where $e_i^{(0)}>0$. The aggregate initial endowment of liquidity providers is 
  $\sum_{i=1}^n e_i^{(0)} = e^{(0)}$. 
 
 At sub-period 1 of each period $t, t =1,2$, liquidity providers choose their portfolios. Specifically, they decide whether to exchange their consumption good for A and B tokens, and how much to exchange for. In addition, they decide whether to deposit their tokens in the AMM and how much to deposit. The liquidity providers maximize the amount of consumption good that they can exchange for at the end of period $ 2$. We also impose the following tie-breaking rule:
\begin{myassump}\label{asstiebreak}
The liquidity providers do not deposit their tokens if they are indifferent between whether or not to deposit. 
 \end{myassump}  We denote the amount of A and B tokens deposited in the AMM at the beginning of period $t$, respectively 
 by $y_A^{(t,1)}$ and $y_B^{(t,1)}$. Liquidity provider $i$ deposits $w_i^{(t)}y_A^{(t,1)}$ and $w_i^{(t)}y_B^{(t,1)}$ amount of A and B tokens, respectively, where $\sum_1^n w_i^{(t)} = 1, w_i^{(t)} \geq 0$. The AMM requires tokens to be deposited at the current fair value exchange rate\footnote{As an example, an AMM which utilizes a constant product function requires the deposited tokens to have equal value, i.e., $y_A^{(2)} p_A^{(2)} = y_B^{(2)} p_B^{(2)}$.}, i.e., $ \frac{ F_x}{F_y}\bigg\rvert_{(x,y) = (y_A^{(t,1)}, y_B^{(t,1)})} = \frac{p_A^{(t,1)}}{p_B^{(t,1)}}$.

\subsection{Investors' Arrival and Token Price Shocks}
In sub-period 2 of each period $t, t =1,2$, after liquidity providers decide on their token holdings and deposit in the AMM, one of the following three mutually exclusive and collectively exhaustive events occurs: ``the arrival of an investor'', ``the arrival of a token price shock that hits A token or B token'', and “neither a shock hits, nor an investor arrives". 

\paragraph{Investors' Arrival.}
With probability $\kappa_I$ , an investor arrives to the AMM. An investor is characterized by an intrinsic type, that is “type A” or “type B”. A “type A” investor extracts a private benefit of $(1+\alpha)p_A^{(t,1)}$ from using one A token on its corresponding platform. A “type A” investor does not use platform B, so she only receives $p_B^{(t,1)}$ for each B token. Symmetrically, a “type B” investor receives $(1+\alpha)p_B^{(t,1)}$ for each B token and $p_A^{(t,1)}$ for each A token.\footnote{For instance, the platform that issues token B20 is Metapurse. On this platform, one can use token B20 to claim ownership of NFT collectibles. The Ethereum network is a platform, where one can use ETH tokens as a cryptocurrency to exchange for goods and other tokens, or to run applications. If an investor prefers to gain exposure to NFT collectibles rather than fiat currencies, then she prefers token B20 over stable coins pegged to USD, such as USDT, USDC. If an investor needs to run applications on Ethereum networks, then she can extract a private benefit from token ETH. If an investor prefers low volatility, high liquidity and wants to exchange  her tokens for fiat currencies, then she can extract a private benefit from holding stable coins.}
The investor arriving to the AMM is of “type A” or of “type B”  with equal probability. The investor chooses the traded quantity to maximize her {total surplus from the transaction}. 

\paragraph{Token Price Shock.} In each period $t$, the prices of A and B tokens may be hit by exogenous shocks. With probability $\theta$, the price changes of A and B tokens are driven by a common shock $\zeta_{com}$: 
\begin{equation}
    \zeta_{com} \thicksim Bern{(\kappa_{com})}, p_{i}^{(t,2)} = (1+\beta \zeta_{com} )p_i^{(t,1)}, i = A,B.
\end{equation}
That is, with probability $0 < \kappa_{com} < 1$, the common shock causes a price increase of $\beta$ for both A and B tokens. 

With probability $1-\theta$, the price changes of A and B tokens are caused by independent, idiosyncratic shocks $\zeta_{A},\zeta_{B}$:
\begin{equation}
    \begin{split}
         & \zeta_{A} \thicksim Bern{(\kappa_1)},  \zeta_{B} \thicksim Bern{(\kappa_2)}, \zeta_{A} \perp \zeta_{B}, \kappa_1 > \kappa_2,\\
  & p_{i}^{(t,2)} = (1+\beta \zeta_{i} )p_i^{(t,1)}, i = A,B.
    \end{split}
\end{equation}
That is, with probability $0 < \kappa_1 < 1$, the idiosyncratic shock $\zeta_{A} $ realizes and increase the price of A tokens by $\beta$; with probability  $0 < \kappa_2 < 1$, the idiosyncratic shock $\zeta_{B}$ realizes, and increases the price of B tokens by $\beta$. Notice that $\kappa_1 > \kappa_2$. This means that the expected return of holding B tokens is smaller than the corresponding return of holding A tokens, and thus holding B tokens presents an opportunity cost.

\subsection{Arbitrage and Token Withdrawal}
The arrival of an investor or the occurrence of a token price shock at sub-period 2 presents an opportunity for the arbitrageur. At sub-period 3 of each period $t$, the arbitrageur submits an arbitrage order and the liquidity provider submits a withdrawal order.

\paragraph{Arbitrage Opportunity} After the investor arrives and trades, the ratio between the amount of A and B tokens deviates from $\frac{y_A^{(t,1)}}{y_B^{(t,1)}}$. Since the spot exchange rate is solely decided by the ratio between the amount of tokens, this deviation may present arbitrage opportunities. 

Similarly, upon realization of the token price shock, the fair  exchange rate of two tokens may change. However, the spot exchange rate in the AMM remains unchanged, and this creates an  arbitrage opportunity. The arbitrageur is incentivized to trade in the token not hit by the shock for the token which becomes more valuable after the shock. This, in turn, yields a loss for the liquidity providers, who then have strong incentives to withdraw their tokens ahead of the arbitrageur. 

\paragraph{Arbitrageur.} The arbitrageur does not use tokens on either platform and only exchange tokens for consumption good. It takes advantage of the price deviation at the  AMM and trades to maximize its profit from arbitrage $p_B^{(t,2)}\Delta q_B^{(t,3)}+ p_A^{(t,2)}\Delta q_A^{(t,3)}$, where $\Delta q_A^{(t,3)}$ and $\Delta q_B^{(t,3)}$ are respectively the {change in the} amount of A and B tokens held by the arbitrageur.  This trading opportunity can be exploited by the arbitrageur only if its order is confirmed by the blockchain before others. The arbitrageur adds a transaction fee $g_{arb}^{(t,3)}$ to its order. We will also refer to such fee as ``gas fee"\footnote{For Ethereum, the total gas fee is equal to the gas price multiplied by the gas amount needed to execute the transaction. The gas price is defined as the amount of ETH paid per unit of gas used, and the gas {amount} measures the computational resources needed to execute a transaction on Ethereum. The unit of gas price is Gwei, that is, $10^{-9}$ ETH token. Transactions with higher gas prices are confirmed first, because miners prioritize them to maximize their fee revenue. Since transactions executed on the same AMM use similar gas amounts, the transactions with higher gas prices have higher total gas fees. In our model, we assume that the agents directly submit a gas fee instead of a gas price.} interchangeably throughout the paper, following its institutional counterpart. The orders will be included on the underlying blockchain and executed in decreasing order of of gas fees, and any tie will be broken uniformly at random. We assume that gas fees attached to unconfirmed orders are observable by everyone.\footnote{This assumption is consistent with current practices. All submitted, pending orders to the AMMs are stored at the mempool which is publicly accessible.} Following \cite{GLOSTEN198571}, we assume that the arbitrageur earns zero profit from each trade {net of the paid gas fee}.

\begin{myassump} \label{assarb}
 The arbitrageur attaches a gas fee equal to the highest possible profit from an arbitrage order.
\end{myassump}

Such an outcome can be attained in a competitive environment with many arbitrageurs. Suppose an arbitrageur submits an arbitrage order and attaches a gas fee smaller than the profit earned from the arbitrage. Another arbitrageur can undercut the first order by submitting the exact same order and attaching a slightly higher gas fee. Only if an arbitrageur submits an {optimal arbitrage} order (i.e., one which maximizes its trading profits) and bids a gas fee equal to its profit, then other arbitrageurs are unable to undercut the submitted order. 

\begin{figure}[t]
    \centering
    \includegraphics[trim={1cm 2cm 10cm 1cm},clip,width=\linewidth]{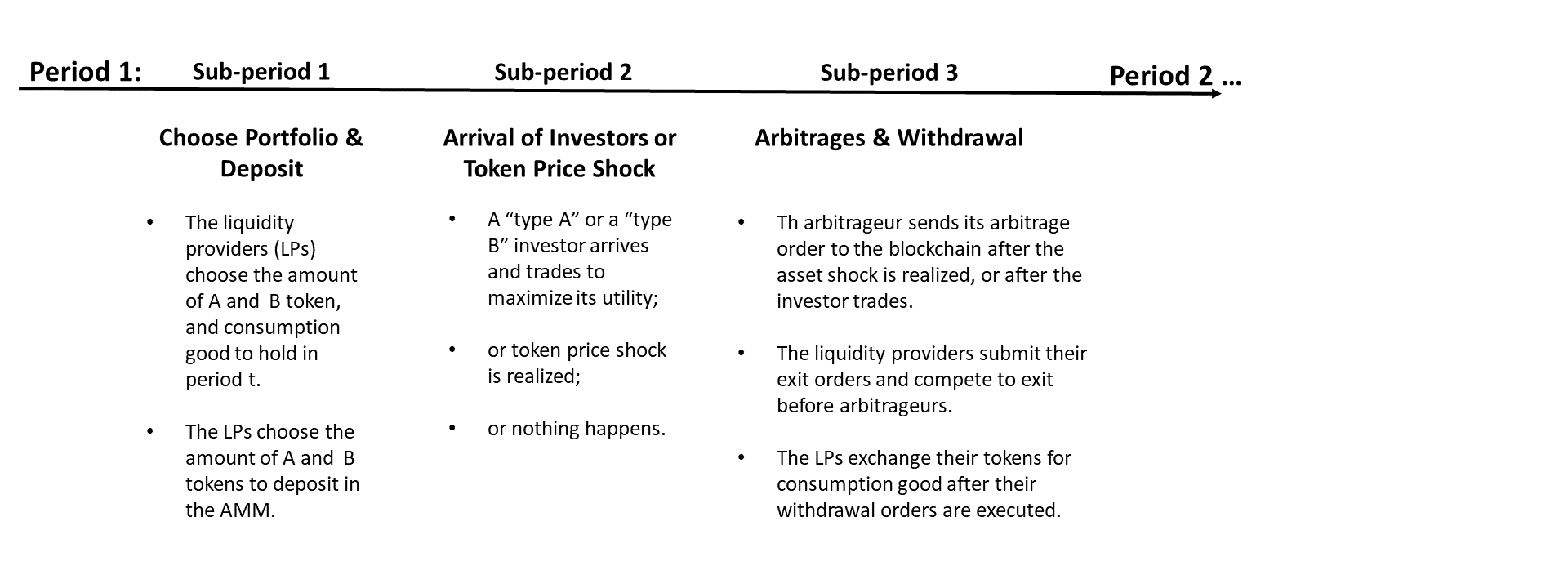}
    \caption{Timeline of The Model.}
    \label{fig:timeline}
\end{figure}

\paragraph{Token Withdrawal. } In each period $t$, at sub-period 3, the liquidity provider $i$ withdraws his tokens from the AMM by submitting an order and attaching a non-negative gas fee {$g_{(lp,i)}^{(t)} $} to it.  When the withdraw order is executed, the liquidity provider $i$ pays the attached gas fee and receives an amount $y_A' w_i^{(t)}$ of A tokens and an amount $y_B' w_i^{(t)}$ of B tokens, where $y_A', y_B'$ are the total reserves in the AMM before the first withdrawal. We recall here that $w_i^{(t)}$ is the share of reserves that liquidity provider $i$ deposited at the beginning of period $t$. Upon receiving their tokens, liquidity providers exchange them for the consumption good, and  period $t$ ends. We denote the amount of consumption good that liquidity provider $i$ owns at the end of period $t$ by $e_i^{(t)}$. 
We visualize {the timeline of the model} in Figure \ref{fig:timeline}. 

\subsection{States, Actions, Strategy Profiles, and Equilibrium}
In this section, we formally define the states, action space, and strategy profiles and payoffs of liquidity providers, investors, and arbitragers.

\paragraph{States.} We denote by $\Omega$ the set of possible states. We denote the initial state of the game as $\omega^{(0)}$. The state $\omega^{(t,s)} = \{P^{(t,s)}, H^{(t,s)}\} $ at sub-period $s$ of period $t$ consists of two components:
\begin{enumerate}
    \item $P^{(t,s)} = \{(p_A^{(1,1)},p_B^{(1,1)}), ..., (p_A^{(t,s)},p_B^{(t,s)}))\} $, where $p_A^{(t,s)},p_B^{(t,s)}$ are the price of  A and B tokens at the end of sub-period $s$ of period $t$. $P^{(t,s)}$ is the history of token prices until the end of sub-period $s$ of period $t$.
    \item $H^{(t,s)} = \{ h^{(1,1)},...,h^{(t,s)}\} $, where 
    $h^{(t,s)} = \{(w_i^{(t)}y_A^{(t,s)}, w_i^{(t)}y_B^{(t,s)},c_i^{(t,s)},x_{A_i}^{(t,s)},x_{B_i}^{(t,s)})\}_{i\in \mathcal{N}}$ collect the portfolios of liquidity providers at the end of sub-period $s$ of period $t$. $w_i^{(t)}y_A^{(t,s)}$ and $w_i^{(t)}y_B^{(t,s)}$ are, respectively, the amounts of A and B token in the AMM that belong to liquidity provider $i$; $c_i^{(t,s)}$ is the amount of consumption good he holds;  $(x_{A_i}^{(t,s)},x_{B_i}^{(t,s)})$ are the amounts of A and B token that liquidity provider $i$ holds (apart from {his} deposit in the AMM). 
\end{enumerate}


\paragraph{Action Space.} Liquidity provider $i$ chooses the portfolio at sub-period 1 of each period $t$, $(w_i^{(t)}y_A^{(t,1)}, w_i^{(t)}y_B^{(t,1)},c_i^{(t,1)},x_{A_i}^{(t,1)},x_{B_i}^{(t,1)})$, and the gas fee $g_{(lp,i)}^{(t)}$ attached to the withdrawal order at sub-period 3 of each period $t$. Liquidity provider $i$ finances his portfolio at $t$ using his endowment $e^{(t-1)}_i$ at the end of period $t-1$, i.e. he is subject to the following budget constraint:

$$p_A^{(t,1)}(w_i^{(t)}y_A^{(t,1)}+x_{A_i}^{(t,1)})+ p_B^{(t,1)}(w_i^{(t)}y_B^{(t,1)}+x_{B_i}^{(t,1)}) +c_i^{(t,1)} = e^{(t-1)}_i.$$

When the investors arrive at the AMM, they choose the trading quantities, $(\Delta Q_A^{(t,2)}, \Delta Q_B^{(t,2)})$. At sub-period 3, the arbitrageur chooses the arbitrage order, $(\Delta q_A^{(t,3)}, \Delta q_B^{(t,3)})$.

\paragraph{Strategy.} A (pure) strategy of the liquidity provider $i$, $\sigma_{(lp,i)}$ consists of four mappings: the first is from the initial state $\omega^{(0)} $ to a  portfolio $h^{(1,1)}$ {at sub-period 1 of period 1},  the second is from $\omega^{(1,3)} \in \Omega$ to a portfolio $h^{(2,1)}$ at sub-period 1 of period 2, and the third and fourth are from a  state $\omega^{(t,2)} \in \Omega$ to a gas fee $g_{(lp,i)}^{(t)}$, for $t=1,2, $ respectively. We denote the strategy profile of the $n$ liquidity providers by $\sigma_{lp} = \{\sigma_{(lp,i)}\}_{i \in \mathcal{N}}$.

The strategies of an arriving  ``type A'' and ``type B'' investor are denoted, respectively, by $\sigma_{(inv,A)}$ and $\sigma_{(inv,B)}$. They are mappings from a state $\omega^{(t,1)} \in \Omega$ to $\Delta Q_A^{(t,2)}$ and $ \Delta Q_B^{(t,2)}$, i.e., the  amount of A and B tokens  traded, respectively, by the ``type A'' and ''type B`` investors. We denote  the strategy profile of the investors by $\sigma_{inv} = \{\sigma_{(inv,A)},\sigma_{(inv,B)}\}$.

A (pure) strategy of the arbitrageur  $\sigma_{arb}$ is a mapping from  a state $\omega^{(t,2)} \in \Omega$ to the  amount of A and B tokens resulting from its  arbitrage, respectively $\Delta q_A^{(t,3)}$ and $ \Delta q_B^{(t,3)}$. 


\paragraph{Payoffs. } The {expected} payoff of liquidity provider $i$ is $\ex{e_i^{(2)}}$, that is, the total amount of consumption good that liquidity provider $i$ {is expected to possess} at the end of period $2$. 
The {expected} payoff of the arbitrageur is $\sum_{t=0}^2 \ex{p_B^{(t,2)}\Delta q_B^{(t,3)}+ p_A^{(t,2)}\Delta q_A^{(t,3)} - g_{arb}^{(t,3)}}$, that is, the expected total profit from the arbitrage order net of the gas fee paid.  

The  payoff of an investor arriving at period $t$ is defined by her surplus from the transaction, that is, $(1+\alpha)p_A^{(t,1)} \Delta Q_A^{(t,2)} + p_B^{(t,1)} \Delta Q_B^{(t,2)}$ for “type A” investors, and $(1+\alpha)p_B^{(t,1)}\Delta Q_B^{(t,2)} + p_A^{(t,1)}\Delta Q_A^{(t,2)}$ for “type B” investors.



\paragraph{Equilibrium.} The states, the strategy profile, and the payoffs above  define our dynamic game. Our equilibrium concept is that of a subgame perfect equilibrium. 
\section{{Equilibrium Trading and Liquidity Provision}} \label{sec:analysis}
 We analyze the game theoretical model developed in the previous section. In Section~\ref{sec:adverse}, we show that the liquidity providers face an arbitrage problem after a token price shock occurs.   
 In Section \ref{sec: trader}, we study the trading strategy of investors. 
 In Section \ref{sec equilibrium}, we characterize the subgame perfect equilibrium of the game, and study under which conditions liquidity providers do not find it incentive compatible to deposit tokens. 
 
 

\subsection{{The Arbitrage Problem}} \label{sec:adverse}

{Upon the occurrence of a token price shock in sub-period 2, liquidity providers and the arbitrageur submit their respective orders to the blockchain.} An arbitrageur decides the amount of A or B tokens traded in as well as the gas fee attached to the order. A liquidity provider decides the gas fee attached to his exit order only. We show that a liquidity provider will be subject to an arbitrage problem and token value loss, if token prices do not co-move and the price change is large enough. 

Recall that the amount of A and B tokens in the AMM at sub-period 2 of period $t$, is denoted respectively, by $y_A^{(t,2)}$ and $y_B^{(t,2)}$. {When the common shock occurs and the price of the two tokens co-move,} the fair value exchange rate remains unchanged: $\frac{p_A^{(t,2)}}{p_B^{(t,2)}}=\frac{p_A^{(t,1)}}{p_B^{(t,1)}}$. {However, if a shock hits either token A or token B only}, then the fair value exchange rate $\frac{p_A^{(t,2)}}{p_B^{(t,2)}}$ deviates from the spot exchange rate $\frac{p_A^{(t,1)}}{p_B^{(t,1)}}$. Without loss of generality, we assume the shock hits B tokens\footnote{The case where the price shock yields an appreciation of A tokens can be handled symmetrically.}, and the price rises from $p_B^{(t,1)}$ to $p_B^{(t,2)} = (1+\beta)p_B^{(t,1)}$. To profit from this deviation, the arbitrageur submits an order to the AMM and exchanges an amount $- \Delta q_A^{(t,3)}$ of A tokens for $\Delta q_B^{(t,3)}$ B tokens. The arbitrageur aims for the optimal arbitrage, i.e., chooses the buy order which solves the following optimization problem:
\begin{equation}\label{eq arbitrageroptimization}
\begin{aligned}
\max_{\Delta q_A^{(t,3)}, \Delta q_B^{(t,3)}} \quad &  p_A^{(t,2)}(1+f) \Delta q_A^{(t,3)} + p_B^{(t,2)}\Delta q_B^{(t,3)}\\
\textrm{s.t.} \quad & F(y_A^{(t,2)},y_B^{(t,2)}) = F(y_A^{(t,2)}-\Delta q_A^{(t,3)} , y_B^{(t,2)}-\Delta q_B^{(t,3)}) \\
  &\Delta q_A^{(t,3)} \leq 0, y_B^{(t,2)} \geq \Delta q_B^{(t,3)} \geq 0,    \\
\end{aligned}
\end{equation}
 where $p_A^{(t,2)}(1+f) \Delta q_A^{(t,3)} $ is the trading cost, that is, the value of A tokens traded in plus the trading fee paid to the liquidity providers, and $p_B^{(t,2)}\Delta q_B^{(t,3)} = p_B^{(t,1)}(1+\beta)\Delta q_B^{(t,3)}$ is the value of B tokens received by the arbitrageur from the order. Solving for the optimal arbitrage yields the following result:

\begin{mylemma}\label{adverseselction order1}
   If a price shock hits only one token and the price change of that token exceeds the paid fee, i.e., $\beta > f $, then {the} arbitrageur earns a positive profit  $\pi(y_A^{(t,2)},y_B^{(t,2)},p_B^{(t,2)}, p_A^{(t,2)})>0$ from the optimal arbitrage trade. Moreover, such profit  and the {unique} optimal trading size $|\Delta q_A^{(t,3)*}|$ for A tokens and $|\Delta q_B^{(t,3)*}|$ for B tokens are increasing in $\beta$ and decreasing in $f$.
\end{mylemma}

When the prices of two tokens do not co-move, if the realized token price change is sufficiently large, it is profitable for the arbitrageur to exchange the token not hit by the positive price shock for the other more valuable token. The larger {the realized price change}, the higher the payoff attained from the arbitrage, and the larger the token value loss for liquidity providers. Moreover, the higher the trading fee charged by the  AMM, the higher the trading cost for the arbitrageur, which in turn reduces the arbitrage profit and the order size. 



The profit of the arbitrageur equals the loss in token value incurred by the liquidity providers. Formally, liquidity provider $i$ incurs a token value loss  $w_i'\pi(y_A^{(t,2)},y_B^{(t,2)},p_B^{(t,2)}, p_A^{(t,2)})$ if an optimal arbitrage order is executed. Hence, this liquidity provider has an incentive to submit a withdrawal order to the AMM. To avoid the loss, the withdrawal order must be executed and included in the underlying blockchain before the order submitted by the arbitrageur. This means that the liquidity provider must pay a gas fee higher than the one attached to the arbitrage order. {By Assumption} \ref{assarb}, the gas fee paid by the arbitrageur for its order matches exactly its gain $\pi(y_A^{(t,2)},y_B^{(t,2)},p_B^{(t,2)}, p_A^{(t,2)})$ from the optimal arbitrage. Hence, the arbitrageur makes a zero net profit. Observe that it is never profitable for liquidity provider $i$ to pay a gas fee higher than $w_i' \pi(y_A^{(t,2)},y_B^{(t,2)},p_B^{(t,2)}, p_A^{(t,2)})$, because he would otherwise incur a cost higher than the loss from arbitrage. Hence, for liquidity providers, the problem is not avoidable by exiting the contract before the arbitrage is exploited.\footnote{One can easily derive the same result by introducing multiple arbitrageurs and modelling competition for  first execution as a first-price auction with any tie-breaking rule. For any arbitrageur, the first execution has value $\pi(y_A^{(t,2)},y_B^{(t,2)},p_B^{(t,2)}, p_A^{(t,2)})$, and for liquidity provider $i$, the first execution has a value $w_i' \pi(y_A^{(t,2)},y_B^{(t,2)},p_B^{(t,2)}, p_A^{(t,2)})$. At  equilibrium, an arbitrageur wins the auction, and the gas fee that the winning arbitrageur pays for the first execution is exactly $\pi(y_A^{(t,2)},y_B^{(t,2)},p_B^{(t,2)}, p_A^{(t,2)})$. In the absence of competition, the liquidity providers can exit with zero additional gas fee after the first execution. } 
This also implies that liquidity providers end up submitting their exit orders with zero gas fee attached, because there is no benefit for bidding up. 
The next proposition formalizes the above discussion.

\begin{myprop}\label{adverse selection does happen3}
  If $\beta > f$ and the token price shock hits only one token in sub-period 2 of period $t, t=1,2$, then an optimal arbitrage order is the first order executed in sub-period 3. The arbitrage yields a loss $w_i' \pi(y_A^{(t,2)},y_B^{(t,2)},p_B^{(t,2)}, p_A^{(t,2)})$ for liquidity provider $i$, and the gas fee $g_{arb}^{(t,3)}$ attached to the arbitrage order is $\pi(y_A^{(t,2)},y_B^{(t,2)},p_B^{(t,2)}, p_A^{(t,2)})$. If $\beta \leq f$, or the prices of two tokens co-move, then the arbitrageur does not trade. 
\end{myprop}

It is worth noticing that the arbitrage  problem in the AMM is fundamentally different from the adverse selection problem arising in typical open limit-order book markets. In a limit-order book market, studied for instance in \cite{GLOSTEN198571} and \cite{glosten1994}, market markers can be adversely selected by other investors who have private information about future realization of asset returns. However, in the AMM, the arbitrage exists even if there is complete information, due to the order execution mechanism of decentralized exchanges built on blockchains. Moreover, market makers in traditional open limit-order book markets can offset the adverse selection problem by placing a bid-ask spread, or they can even front-run the orders when they are able to predict the direction of order flow. In contrast, the AMM does not charge a bid-ask spread, and the liquidity providers are also unable to front-run the arbitrage order because the execution priority is decided by the gas fee attached to the order. As a result, the liquidity providers participating in the AMM are subject to token value loss, and must be compensated with enough trading fees. 

A question which often puzzles liquidity providers is as follows: does the token value loss still exist if the token exchange rate reverts back to its initial level in subsequent periods? Some commentators have argued that if the token exchange rate is hit by a shock in the opposite direction, another arbitrage will occur and bring the ratio of deposits back to the initial ratio. Hence, there would be no token value loss from arbitrage.  This is the reason why token value loss from arbitrage is often referred to as ``impermanent loss" by practitioners (see also \cite{introtodf}).

\begin{mydef}
 The ``impermanent loss" is defined as:
\begin{equation}\label{impermanent loss}
 IL\left(\frac{p_A^{(0)}}{p_B^{(0)}}, \frac{p_A^{(2,3)}}{p_B^{(2,3)}} \right) := 1- \frac{p_A^{(2,3)}x_2 + p_B^{(2,3)}y_2}{p_A^{(2,3)}x_1 + p_B^{(2,3)}y_1},
   \end{equation}
where $x_1,y_1,x_2,y_2>0$ are specified by the following constraints: $$F(x_1,y_1)=F(x_2,y_2), \frac{F_x(x_1,y_1)}{F_y(x_1,y_1)} = \frac{p_A^{(0)}}{p_B^{(0)}}, \frac{F_x(x_2,y_2)}{F_y(x_2,y_2)} = \frac{p_A^{(2,3)}}{p_B^{(2,3)}}.$$
\end{mydef}
In the definition above, $x_1, y_1$ are the amount of A and B tokens, respectively deposited in the AMM if the initial fair value exchange rate is $\frac{p_A^{(0)}}{p_B^{(0)}}$. If the fair value exchange rate changes to $\frac{p_A^{(2,3)}}{p_B^{(2,3)}}$ at the end of the investment horizon, then after the arbitrage is exploited, the amount of A and B tokens in the AMM will be $x_2, y_2$ respectively (assuming zero fee). Moreover, the constraint imposed by the pricing curve, $F(x_1,y_1)=F(x_2,y_2)$, needs to be satisfied. Thus, if $x_1, y_1$ are the amount of deposited tokens, $p_A^{(2,3)}x_2 + p_B^{(2,3)}y_2$ is the total value of deposited tokens  after the price change. If the tokens are not deposited, $p_A^{(2,3)}x_1 + p_B^{(2,3)}y_1$ is the total value of tokens. The expression in \eqref{impermanent loss} aims at capturing the magnitude of token value loss from depositing  relative to not depositing. The ``impermanent loss'' is indeed zero if the token price reverts, i.e., $\frac{p_A^{(0)}}{p_B^{(0)}} = \frac{p_A^{(2,3)}}{p_B^{(2,3)}}.$  All this leads to the seemingly logical liquidity management strategy to minimize impermanent loss---whenever a token value occurs, ignore token price movements in short term, continue depositing in the AMM,  and wait for the reversion of exchange rate to occur. (see, for instance, \cite{davis_2021}). However, as we show in Section \ref{sec equilibrium}, the above claim is fallacious.


\subsection{Investors' Trading} \label{sec: trader}
Each investor arrives to the exchange and decides the amount of A and B tokens to trade. Moreover, investors' trades may leave an exploitable arbitrage opportunity for the  arbitrageur. 



Assume a ``type A'' investor arrives at sub-period 2 of period $t$. The case of a ``type B'' investor arriving first follows from symmetry arguments. Since a ``type A'' investor has personal use for A tokens, she uses B tokens to exchange for A tokens. Formally, when the ``type A'' investor decides the desired amount $\Delta Q_A^{(t,2)} \geq 0$ of A tokens from the trade, she maximizes her total surplus from the transaction subject to the constraints imposed by the pricing function of the AMM: 
\begin{equation} \label{eq: investorincentiveequation}
\begin{aligned}
\max_{\Delta Q_A^{(t,2)}, \Delta Q_B^{(t,2)}} \quad & (1+\alpha)p_A^{(t,1)} \Delta Q_A^{(t,2)} +(1+f) p_B^{(t,1)} \Delta Q_B^{(t,2)}\\
\textrm{s.t.} \quad & F(y_A^{(t,1)},y_B^{(t,1)}) = F(y_A^{(t,1)}-\Delta  Q_A^{(t,2)} , y_B^{(t,1)}- \Delta Q_B^{(t,2)}) \\
  &  y_A^{(t,1)} \geq  \Delta Q_A^{(t,2)} \geq 0,  \Delta Q_B^{(t,2)} \leq 0,  \\
\end{aligned}
\end{equation}
where $(1+\alpha)p_A^{(t,1)} \Delta Q_A^{(t,2)}$ is {the total private benefit of} the ``type A'' investor after trading, and $-(1+f) p_B^{(t,1)} \Delta Q_B^{(t,2)}$ is the total value of B tokens paid by the investor.  We use $s_A(y_A^{(t,1)},y_B^{(t,1)},p_B^{(t,1)}, p_A^{(t,1)})$ and $s_B(y_A^{(t,1)},y_B^{(t,1)},p_B^{(t,1)}, p_A^{(t,1)})$, to denote the maximum surplus of {an arriving ``type A'' investor, respectively  ``type B'' investor, in period $t$.}

\begin{mylemma}\label{traderoptimization}
 If $f< \alpha$, then the arriving {``type i'' }investor, $i\in\{A,B\}$, trades and earns a positive surplus from the transaction, i.e., $s_i (y_A^{(t,1)},y_B^{(t,1)},p_B^{(t,1)}, p_A^{(t,1)})>0$. Moreover, the maximum surplus and the optimal trading quantities of an arriving investor, $|Q_A^{(t,2)*}|, |Q_B^{(t,2)*}|$, are strictly positive, increasing in $\alpha$, and decreasing in $f$. If instead $f \geq \alpha$, then the arriving investor does not trade. 
\end{mylemma}


Trading generates fees for the liquidity providers, which are then compensated for the arbitrage problem they face. 
After the investor arrives and trades, the amounts of A  and B tokens in the AMM  become $y_A^{(t,2)}$ and $y_B^{(t,2)}$, respectively. The trade by the investor alters the ratio  $\frac{y_A^{(1)}}{y_B^{(1)}}$ of A and B tokens in the AMM, which may lead to an arbitrage opportunity. {Again, assume a ``type A'' investor arrives at $t=1$. After the investor trades, the spot rate at which A tokens are exchanged for B tokens is higher than the fair value exchange rate: $$\frac{ F_x}{F_y}\bigg\rvert_{(x,y) = (y_A^{(t,2)},y_B^{(t,2)})}  > \frac{ F_x}{F_y}\bigg\rvert_{(x,y) = (y_A^{(t,1)},y_B^{(t,1)})} = \frac{p_A^{(t,1)}}{p_B^{(t,1)}} = \frac{p_A^{(t,2)}}{p_B^{(t,2)}}.$$ The arbitrageur then exchanges A  for B tokens and chooses the exchange order which solves the  optimization problem \eqref{eq arbitrageroptimization}}.

\subsection{The Adoption of the AMM } \label{sec equilibrium}

In this section, we first characterize the subgame perfect equilibrium of the game. {We then provide the conditions under which a ``liquidity freeze'' occurs at equilibrium.}


\begin{myprop} \label{propostion unique equilibrium}
  For any $\alpha,  \theta, \beta, f, \kappa_I, \kappa_{com}, \kappa_1,  \kappa_2$, there exists a  subgame perfect equilibrium $ (\sigma_{lp}^*,\sigma_{inv}^*, \sigma_{arb}^* ).$  \footnote{Multiplicity of equilibria only arises if the liquidity providers  are indifferent between investing A tokens and B tokens in period 1, and depositing tokens in the AMM in period 1 yields lower expected payoff than investing in either A token or B token. This means that the deposit amounts of A and B tokens in the AMM are the same for all possible equilibria.  Moreover, for all the possible equilibria, the expected payoff of all the agents are the same.  }
\end{myprop}



We then examine the conditions under which at the equilibrium, the liquidity providers are incentive compatible to deposit their tokens into the decentralized exchange. If none of the liquidity providers deposits their tokens at period $t$, no trading activities occur at that period. We call this market breakdown a ``liquidity freeze". The following proposition characterizes the condition under which a ``liquidity freeze" occurs.

\begin{myprop}\label{freezeequilibrium5}
     A ``liquidity freeze'' occurs surely in period 1 and  2 if and only if  $\beta \geq \overline{\beta_{frz}}$, where $ \overline{\beta_{frz}} \in [0, +\infty]$.     Moreover,  the threshold  $\beta_{frz}$ is increasing in $\alpha, \kappa_I, \theta, \kappa_2 $, and decreasing in $ \kappa_1$. 
     

\end{myprop}

The above proposition states that when the token exchange rate is sufficiently volatile, the arbitrage problem is severe and a ``liquidity freeze'' occurs. 
The comparative static results are intuitive. First, when tokens become more attractive for investors ($\alpha$ increases) and the arrival rate of investors goes up ($\kappa_I$ increases), the expected trading volume increases and thus liquidity providers collect a higher trading fee. Hence, a ``liquidity freeze'' is less likely to occur. Second, when the tokens are more likely to be hit by a common shock ($\theta$ increases) and co-move, arbitrage opportunities are less likely, i.e., the arbitrage problem faced by liquidity providers is less severe.  Third, when the token exchange rate is more volatile (magnitude of the price shock $\beta$ and arrival rate of the shock $\kappa_1$ both increase), the arbitrage becomes more costly for liquidity providers, and thus a ``liquidity freeze'' is more likely. Fourth, when $\kappa_2$ increases, the difference in expected return of the two tokens $(1+\beta)(\kappa_1 - \kappa_2)$ decreases, and thus the opportunity cost of holding both tokens and providing liquidity decreases. In this way, the incentive of liquidity providers to adopt the AMM becomes stronger.

The above result suggests that the AMM may be more suitable for adoption of pairs whose token prices are highly correlated and not volatile, such as a pair of stable coins.  Moreover, adoption is higher for tokens which provide investors with high personal use value, such as BTC and ETH. Because those tokens attract investors who in turn generate high trading volumes, liquidity providers can earn large trading fees and be compensated for the arbitrage problem they face. Last, but not least, adoption is very unlikely for pairs of tokens whose expected returns are low, such as the majority of Altcoins\footnote{Altcoins is the term used to refer to all alternative cryptocurrencies that were launched after the massive success achieved by Bitcoin.} which have no value and  discernible purpose. This is because owning and providing liquidity for this kind of tokens is very risky and presents large opportunity costs. 

\begin{mycor} \label{cor1}
Suppose that the fair value exchange rate changes from $\frac{p_A^{(0)}}{p_B^{(0)}}$ to  $\frac{p_A^{(0)}}{(1+\beta)p_B^{(0)}}$ in period 1. Then, the probability that the fair value exchange rate reverts to $\frac{p_A^{(0)}}{p_B^{(0)}}$ in period 2  decreases in $\kappa_2$, and the expected ``impermanent loss'' increases in $\kappa_2$. The expected marginal fee revenue from deposits does not change in $\kappa_2$. However, in period 2, liquidity providers deposit in the AMM at equilibrium  if and only if $\beta < \beta_{frz}^{(2)} $, and the threshold $\beta_{frz}^{(2)} $ increases in $ \kappa_2$.
\end{mycor}
{It is a very common misconception that the token value loss from arbitrage is "impermanent" and will decrease to zero if  the exchange rate reverts back to the initial level. This leads to the widely spread liquidity management strategy--- after arbitrage loss occurs, if the token exchange rate is likely to revert to its initial level, then it is optimal for the liquidity providers to keep depositing and wait for exchange rate reversion. However, the above corollary shows that this claim is not sound, and the strategy is sub-optimal.  As $\kappa_2$ decreases,  even though the probability of token exchange rate reversion increases, the expected ``impermanent loss'' decreases, and the  expected marginal fee revenue from deposits is unaffected, liquidity providers' incentive to deposit becomes weaker. This is because  if the token exchange rate reverts in period 2, then the liquidity providers who deposit will only suffer from another token value loss due to arbitrage orders which trade B tokens for A tokens.  No matter whether token exchange reverse or not, each realized arbitrage occurred still yields a permanent loss that cannot be offset by the previous arbitrages. In this way, when the reversion probability is high, instead of depositing and waiting for reversion, the optimal action of liquidity providers in period 2 should be holding A tokens in the portfolio  and not providing liquidity at the AMM.  }

 The benchmark upon which the {widely accepted} measure of ``impermanent loss" (formally defined in \eqref{impermanent loss}) is based,  is  misleading: it compares the return of depositing tokens in the AMM with the return of holding the same tokens in a portfolio out of the AMM for the whole time. Maintaining such a fixed portfolio is by no means the optimal strategy ex-ante or the best alternative of depositing tokens for the whole time, and this is why the measure of ``impermanent loss" fails to fully account for the opportunity cost of depositing tokens at the AMM. 
 The optimal liquidity management strategy at equilibrium requires the liquidity providers to account for opportunity costs, and decide their portfolio based on the expected return  calculation for the following periods. Using this strategy as a benchmark, liquidity providers can quantify the  opportunity cost of depositing in the AMM  for the entire investment horizon. 

\begin{myprop} \label{feesgoup6}
  The expectation  and variance of the gas fee in period $t$, $ \expe_{(t,1)} [g_{arb}^{(t,3)}]$\footnote{ $\expe_{(t,s)}[ X ]$ denotes the {expectation of the random variable $X$ conditional on the information available at the end of sub-period $s$ of period $t$.}} and $Var_{(t,1)} [g_{arb}^{(t,3)}]$, are both increasing in the amount of token $y_A^{(t,1)}$ and $y_B^{(t,1)}$ deposited by liquidity providers. 
\end{myprop}

The above proposition captures an important, yet undesirable, consequence of AMM adoption. Intuitively, if  a large amount of tokens is deposited at the AMM, we expect more profitable arbitrage opportunities and thus gas fee surges due to arbitrageur's bidding. Hence, as the AMM becomes more popular and widely adopted, the instability of its underlying infrastructure---blockchain--may increase. A surge of gas fees implies that other decentralized applications built on the same blockchain may lose their costumers due to high transaction fees, or significantly delay the execution of orders by their costumers. Additionally, if those fees become less predictable due to larger variance, the usage of the blockchain decreases.  This means that the large adoption of the AMM may impose negative externalities on other decentralized applications  using  the same blockchain infrastructure.  

\section{The Design of AMMs} \label{sec:ext}
{Since Uniswap first introduced AMMs with constant product function,} numerous AMMs have been developed. {They differ in terms of the pricing functions they utilize, the number of different tokens handled, and the charged transaction fees.} 

In this section, we study how such choices affect the equilibrium outcome. In Section \ref{subsec: curvature}, we show that the curvature of the curve $F(x,y) =C$ determines the severity of the arbitrage problem and the {\it deposit efficiency}, i.e., the expected trading volume per unit token deposited. We also solve for the optimal curvature of the pricing function, {i.e., the curvature at which a ``liquidity freeze'' is least likely to occur and aggregate welfare maximized.} In  Section \ref{subsec: more asset}, we argue why pooling more than two tokens in the AMM does not reduce the arbitrage problem. 
 

 
 \subsection{The Curvature of Pricing Curve} \label{subsec: curvature}

 We begin by recalling that at sub-period $s$ of period $t$, any trade satisfies the relation: 
 \begin{equation}\label{general curve}
     F (y_A^{(t,s-1)}+\Delta_A , y_B^{(t,s-1)} + \Delta_B) = F (y_A^{(t,s-1)} , y_B^{(t,s-1)}),   \Delta_B  \geq -y_B^{(t,s-1)}, \Delta_A  \geq -y_A^{(t,s-1)},
 \end{equation}
 where  $\Delta_A, \Delta_B$ are, respectively, the amounts of A and B tokens added to (or withdrawn from, if the sign is negative) the AMM. The slope of the curve, $-\frac{\frac{\partial F}{\partial \Delta_A}}{\frac{\partial F}{\partial \Delta_B}}$, is the negative of marginal exchange rate, and the curvature of the curve at each point captures the rate of change of the marginal exchange rate. 
 
 
 For example, if the pricing function $F$ is linear, $ F_0^{(t)} (
x,y) =p_A^{(t,1)} x +p_B^{(t,1)}  y$, then the slope of the curve is $-\frac{p_A^{(t,1)}}{p_B^{(t,1)}}$ and its curvature is 0, i.e., the marginal exchange rate is fixed at the fair rate at sub-period 1 and equal to $\frac{p_A^{(t,1)}}{p_B^{(t,1)}}$. Another example is the product function used by Uniswap V2 and Sushiswap, i.e., $F_1 (x,y) = x y$. The slope of the curve is $-\frac{y}{x}$, which means that the marginal exchange rate depends on the {deposited tokens} at the AMM. Moreover, the curvature of the exchange rate curve is positive.  This implies that the marginal exchange rate from  A to  B tokens decreases in the amount of A tokens traded in, $\Delta_A $, which leads to the so called ``slippage''.\footnote{``Slippage'' occurs when the  rate $\frac{|\Delta_B|}{|\Delta_A|}$ {at which A tokens are exchanged for B tokens} (respectively $\frac{|\Delta_A|}{|\Delta_B|}$ if B tokens are exchanged for A tokens), is worse than the spot exchange rate  $-\frac{\frac{\partial F}{\partial \Delta_A}}{\frac{\partial F}{\partial \Delta_B}}$ (respectively $-\frac{\frac{\partial F}{\partial \Delta_B}}{\frac{\partial F}{\partial \Delta_A}}$ if B tokens are exchanged for A tokens).}  The size of the slippage, which is determined by the curvature of the pricing curve, affects the equilibrium outcomes. To see this, consider the following family of pricing functions:
$$F^{(t)}_k (x,y) = (1-k) \; A \; F_0^{(t)} (
x,y) + k \; F_1 (
x,y),$$
where $ k \in [0,1]$, and $A = \left(\frac{y_A^{(t,1)} y_B^{(t,1)}}{p_A^{(t,1)} p_B^{(t,1)}}\right)^{1/2}$ is a scaling coefficient. The curvature of the pricing curve $F^{(t)}_k (x,y) = C$ is increasing in $k$. If $k=0$, the pricing curve is a straight line with zero curvature; if $k=1$, the pricing curve is the constant product function, and it has the largest curvature among all $F^{(t)}_k$'s, $k\in [0,1]$.

\begin{mylemma} \label{curvature lemma1}
Suppose $y_A^{(t,1)},y_B^{(t,1)} >0$. The following claims hold:
  \begin{enumerate}
         \item  the expected arbitrage loss ratio in period t, that is, the expected loss from arbitrage divided by the total value of tokens in the AMM, $\expe_{(t,1)}\bigg[\frac{\pi(y_A^{(t,2)},y_B^{(t,2)},p_B^{(t,2)}, p_A^{(t,2)})}{p_A^{(t,1)}y_A^{(t,1)}+p_B^{(t,1)}y_B^{(t,1)}}\bigg] $, decreases in $k$;
      \item the  investors' surplus ratio  in period t, that is, {the sum of ``type A" investor and  ``type B" investor's}  maximum surplus  divided by the  total  value of tokens in the AMM, given by $\frac{\sum_{i=A,B}s_i(y_A^{(t,1)},y_B^{(t,1)},p_B^{(t,1)}, p_A^{(t,1)})}{p_A^{(t,1)}y_A^{(t,1)}+p_B^{(t,1)}y_B^{(t,1)}} $, is decreasing in $k$. 
  \end{enumerate}
 
\end{mylemma}

When $k$ increases, the curvature of the pricing curve also increases, and the exchange rate adjusts more quickly to the increased exchange amount. This yields a higher slippage for  trades of the arbitrageur and of investors. Hence, the amount of a token the arbitrageur or investors exchange for the other token increases. Because trading costs increase, the arbitrageur extracts a lower profit from the arbitrage opportunity, thus the expected token value loss of liquidity providers decreases. This explains why employing a pricing curve with a larger curvature reduces the severity of the arbitrage problem. 

A higher curvature does not benefit investors, who see their total trading surplus decrease as a result of higher trading costs. In a few cases, the fee revenue of liquidity providers would decrease if the curvature increases. 

\begin{myprop} \label{curvature propostion}
Suppose that $\alpha > \beta.$ Then there exists a critical threshold $k^* \in (0,1)$ such that 
 \begin{enumerate}
     \item The expected payoff of liquidity providers at equilibrium is increasing in $k$, for $k \in [0,k^*]$, and decreasing in $k$, for $k \in [k^*,1]$. 
     \item ``Liquidity freeze'' is least likely to occur if $k=k^*$. That is, for any state $\omega^{(t-1,3)}$, {$t=1,2$}, if there is a ``liquidity freeze'' at period $t$ when $k=k^*$, then there is a ``liquidity freeze'' at period $t$ for any other $k \in [0,1]$.
 \end{enumerate}
 
\end{myprop}
When the curvature of the pricing curve is {too} small, that is, $k<k^*$, the exchange rate adjusts very slowly to the exchange amount, and the slippage of trading is very small. In this case, an arriving investor only needs to deposit a very small amount of A (respectively B) tokens to take out all the B (respectively A) tokens in the AMM. As a result, the fee revenue generated by investors' trades is very small. Moreover, because the slippage has a small size, the arbitrage problem is severe. Hence, increasing the curvature of the pricing function leads to a higher fee revenue and a smaller arbitrage loss for investors, which in turn increases their payoffs. Conversely, when the curvature of the pricing function is {too} high, that is, $k > k^*$, then so is the slippage of trading. Despite making the arbitrage problem less severe, investors become more reluctant to trade, and the fee revenue of liquidity providers is reduced. Hence, decreasing the curvature of the pricing function increases the payoffs of  liquidity providers. If $k=k^*$, the pricing curve achieves a balance between these two economic forces: generating a higher fee revenue and increasing the severity of arbitrage problem. As a result, liquidity providers have the strongest incentives to deposit their tokens, and the occurrence of a ``liquidity freeze'' is minimized.


\begin{myprop} \label{proposition welfare}
  Suppose that $\alpha > \beta.$ The equilibrium deposit efficiency in period $t, t= 1,2, $  measured by the expected investors' trading volume divided by the  total value of tokens deposited in period $t$, $\ex{\frac{p_A^{(t,1)} |\Delta Q_A^{(t,2)*}| +p_B^{(t,1)} |\Delta Q_B^{(t,2)*}|}{p_A^{(t,1)}y_A^{(t,1)}+p_B^{(t,1)}y_B^{(t,1)}}}$\footnote{We define the deposit efficiency to be zero when $p_A^{(t,1)}y_A^{(t,1)}+p_B^{(t,1)}y_B^{(t,1)}=0$.}, is maximized if $k=k^*$. Moreover, the socially optimal pricing curve, i.e., under which aggregate welfare measured by the sum of agents' expected payoffs  is maximized, is attained at $k^*$.
\end{myprop}


The arbitrage loss and transaction fees are merely transfers of wealth between agents in our model, whereas the gas fee paid by the arbitrageur to miners of underlying blockchain {is a deadweight} loss. Hence, the highest social welfare is attained when the aggregated investors' maximum surplus and liquidity providers' payoffs are maximized. Both liquidity providers' payoffs and deposit efficiency are maximized at $k=k^*$. {Hence, deposits of liquidity providers can support the highest number of trades for investors.} Additionally, the amount deposited in the AMM grows the fastest if liquidity providers' payoffs are maximized. This increases the expected aggregate trading volume and leads to higher social welfare.


\subsection{{Does Pooling More Tokens Reduce the Arbitrage Problem?}} \label{subsec: more asset}
In this section, we analyze how pooling more than two types of tokens in the AMM affects the token value loss of liquidity providers from arbitrage. Many practitioners have the seemingly convincing intuition that pooling more tokens may alleviate the arbitrage problem due to diversification effects. Few AMMs, including Balancer, have been designed based on this intuition. Interestingly, we find that if an AMM manages more than two types of tokens, the arbitrage problem becomes worse. 

Assume agents have access to three different tokens: A, B, and C token. There are two AMMs, one handles A and B tokens only, while the other handles all three tokens. Both AMMs utilize a constant product function, that is, $F_{AB}(x,y) =xy$ for the first AMM, and $F_{ABC}(x,y,z)=xyz$ for the second AMM. We denote the price of a single C token at  sub-period $s$ of period $t$ by $p_C^{(t,s)}$.

{As in the baseline model,} in each period $t$ the prices of tokens A, B, and C may move due to exogenous shocks. With probability $\theta$, the price movements of A, B, and C tokens are driven by a common shock $\zeta_{com}$:

\begin{equation}
    \zeta_{com} \thicksim Bern{(\kappa)}, p_{i}^{(t,2)} = (1+\beta \zeta_{com} )p_i^{(t,1)}, i = A,B, C.
\end{equation}

With probability $1-\theta$, they are determined by  independent, idiosyncratic shocks  $\zeta_{A},\zeta_{B},\zeta_{C}$, respectively:
\begin{equation}
    \begin{split}
         & \zeta_{i} \thicksim Bern{(\kappa)},  \zeta_{A} \perp \zeta_{B}, \perp \zeta_{B} \perp \zeta_{C}, \perp \zeta_{A} \perp \zeta_{C},  \\
  & p_{i}^{(t,2)} = (1+\beta \zeta_{i} )p_i^{(t,1)}, i = A,B, C.
    \end{split}
\end{equation}

Apart from this adjustment, the model remains the same as the baseline model. The following proposition compares the arbitrage loss of the two AMMs in equilibrium:

\begin{myprop} \label{more than one token proposition}
  Suppose $y_A^{(t,1)},y_B^{(t,1)} > 0$ for the AMM pooling two tokens and $y_A^{(t,1)},y_B^{(t,1)}, y_C^{(t,1)} > 0$ for the AMM pooling three tokens. At period t, the expected arbitrage loss ratio of the AMM pooling A and B tokens, $\expe_{(t,1)}\bigg[\frac{\pi_{AB}(y_A^{(t,2)},y_B^{(t,2)},p_B^{(t,2)}, p_A^{(t,2)})}{p_A^{(t,1)}y_A^{(t,1)}+p_B^{(t,1)}y_B^{(t,1)}}\bigg] $, is  smaller than the expected arbitrage loss ratio of the AMM which pools A, B, and C tokens, $\expe_{(t,1)}\bigg[\frac{\pi_{ABC}(y_A^{(t,2)},y_B^{(t,2)},y_C^{(t,2)},p_C^{(t,2)}, p_B^{(t,2)}, p_A^{(t,2)})}{p_A^{(t,1)}y_A^{(t,1)}+p_B^{(t,1)}y_B^{(t,1)}+p_C^{(t,1)}y_C^{(t,1)}}\bigg] $. 
\end{myprop}

The above result shows that the arbitrageur can extract more profit from the AMM pooling three tokens, leading to a higher loss for the liquidity providers. The reason is twofold. First, an arbitrage opportunity becomes more likely as the number of tokens in the AMM increases. Consequently, the probability of a token value loss for liquidity providers increases. Second, for each realized arbitrage opportunity, the arbitrageur can extract a larger portion of the shocked token from the AMM deposit. Intuitively, if the price of token A increases due to a shock in period $t$, the arbitrageur can only use token B to exchange for token A in the AMM pooling two tokens. The exchanged amount will be such that the marginal benefit of trading is equal to the marginal trading cost. However, in the AMM which pools three tokens, the arbitrageur can use both token B and C to exchange for the appreciated token A, and it will stop trading only when the marginal benefit of trading both B and C tokens for A tokens equals to the marginal trading costs. 

The logic described above carries through for AMMs which pool $M>3$ tokens, $M \in \mathbb{N}$. The main takeaway is that the arbitrage problem cannot be mitigated by pooling more tokens in the AMM. 

\section{Empirical Analysis}\label{sec:empirical}
In this section, {we provide empirical support to the main testable implications of our model.} We state the implications in Section \ref{subsec: listof results}. We describe our dataset in Section \ref{subsection: dataset}. We define the variables used in our empirical validation in Section~\ref{subsection:variable of interests}. We discuss the results from the regression analysis in  Section \ref{subsection:empirical results}.

\subsection{Testable Implications} \label{subsec: listof results}
Our model generates the following implications:

\begin{enumerate}[(1)]
    \item An increase in token exchange rate volatility decreases the amount of tokens deposited at the AMM. As shown in  Proposition \ref{freezeequilibrium5}, if the volatility\footnote{The standard deviation of a change of the log token exchange rate over one-period is $\log{(1+\beta)}\sqrt{(1-\theta)(\kappa_1+\kappa_2-2\kappa_1\kappa_2-(1-\theta)(\kappa_1-\kappa_2)^2)}$, which clearly increases in the parameter $\beta$.} of the token exchange rate increases, the arbitrage problem becomes more severe, and the liquidity providers have weaker incentives to deposit their tokens. We test this implication by examining how the amount of deposits in AMMs changes with the volatility of the token exchange rate.
    
    \item A higher trading volume results in a higher amount of tokens deposited at the AMM. As shown in  Proposition \ref{freezeequilibrium5}, if the token pairs attract more investors, both volume and trading fees increase, which gives liquidity providers stronger incentives to deposit. We test this implication by relating the change of deposits in AMMs with trading volumes.

    \item Average levels and volatility of gas fees attached to pairs with larger exchange rate volatility are higher. Proposition \ref{feesgoup6} implies that, as the volatility of the token exchange rate increases, arbitrage opportunities become more profitable. This in turn yields a higher expectation and variance of the gas fee. To test this implication, we group all token pairs into two categories, ``stable pairs" and ``unstable pairs", and examine whether transactions in ``stable pairs" are associated with lower average levels and volatility of gas fees. ``Stable pairs'' consist of two stable coins pegged to one US dollar, and have lower price volatility relative to ``unstable pairs''. 
\end{enumerate}

\subsection{Data}\label{subsection: dataset}
The dataset contains histories of all trades, deposits, and  withdrawals for a sample of 80 AMMs with actively traded pairs. Among the 80 AMMs, 40 of them are from Uniswap V2, and the rest are from Sushiswap. 7 pairs consist of only stable coins pegged to one US dollar, and they are denoted as ``stable pairs".

For each AMM, the transaction level data include the time stamp, the address of the investor, the gas price attached to the transaction, as well as the name and amount of tokens that the investor trades in or takes out of the AMM. If an investor trades in (takes out) both tokens in a transaction, then we identify the transaction as a deposit (withdrawal); if instead, the investor trades in one token and takes out the other token, we identify the transaction as a swap. We use the history to calculate and track the total liquidity reserve of both tokens {in each AMM}, and calculate the spot rate of the exchange from the liquidity reserves. 

This study covers the 25-week period\footnote{We choose this interval to have enough cross-sectional and time-series observations. A longer time period reduces the amount of AMMs available, as most of the AMMs are set up at the end of 2020.  } from Dec 22, 2020 to June 20, 2021. The number of AMMs initiated by Dec 22, 2020 is only 30. Among them, 16 are from Sushiswap, and the rest are from Uniswap; in particular, 3 pairs are ``stable pairs", and they all belong to Uniswap.

\subsection{Definitions of Variables}\label{subsection:variable of interests}
We next describe the main variables in our analysis. 

\paragraph{Token Exchange Rate Volatility.} We measure the volatility of the token exchange rate for AMM $j$ in week $t$ using the standard deviation of the log spot rate between two tokens deposited in the AMM $j$, in week $t$. This measurement is invariant with respect to the choice of base token\footnote{In the foreign exchange market, the first element of a currency pair is denoted as the base currency, and the second one is referred to as the quote currency. We follow the same convention for an AMM, and refer to the first token in a pair as the base token, and to the second token as the quote token.} of the pair and to scalar multiplication\footnote{Some tokens are more valuable than others, with Bitcoin being a prominent example. A normalized measure makes the volatility of the token exchange rate comparable across pairs} of token values. 

\paragraph{Deposit Inflow and Outflow.} Denote the pair of tokens in AMM $j$ as $A_j$ token and $B_j$ token.  We measure the change of deposits for AMM $j$ in week $t$ as follows:
\begin{equation}
    Depositflow_{jt} = \sgn(Deposit A_{jt}) \times \left(\frac{Deposit A_{jt}}{TokenA_{jt}} \times \frac{Deposit B_{jt}}{TokenB_{jt}}\right)^{1/2}
\end{equation}
where $Deposit A_{jt}$, $Deposit B_{jt}$ are the total $A_j$ tokens and $B_j$ tokens deposited (if positive), or withdrawn (if negative) by liquidity providers of AMM $j$, with deposit or withdrawal order during week $t$. $TokenA_{jt}$ and $TokenB_{jt}$ are, respectively, the total liquidity reserves of $A_j$ and $B_j$ tokens in AMM $j$ at the beginning of week $t$. $\sgn(Deposit A_{jt})$ is positive if the net deposit is larger than the net withdrawal, and negative otherwise. 

\paragraph{Trading Volume.} We measure the total trading volume in AMM $j$ during week $t$ as follows:
\begin{equation}
    Volume_{jt} =  \left(\frac{Trade A_{jt}}{TokenA_{jt}} \times \frac{Trade B_{jt}}{TokenB_{jt}}\right)^{1/2},
\end{equation}
where $Trade A_{jt}$, $trade B_{jt}$ are, respectively, the total {amount of} $A_j$ and $B_j$ tokens traded by investors with swap orders at  AMM $j$  in week $t$, and $TokenA_{jt}, TokenB_{jt}$ are, respectively, the total liquidity reserves of $A_j$ and $B_j$ tokens in AMM $j$ at the beginning of week $t$. The above measure captures the total trading volume by investors relative to the total reserve in AMM $j$ during week $t$.

\paragraph{Gas Price Volatility.} We measure the gas price volatility in AMM $j$ during week $t$ using the standard deviation of the gas price {attached to all transactions} executed on AMM $j$ in week $t$. All AMMs we consider are built on the same blockchain, i.e., Ethereum, and thus {levels and volatility of gas prices} are comparable across pairs.

Table \ref{table: Comparative statics} presents summary statistics of the data. Most of the variables have large in-sample variations. Since ``stable pairs" are pairs of stable coins pegged to one dollar, the log spot exchange rates are very close to 0, and their token exchange rate volatility is much lower than the volatility of ``unstable pairs".

\begin{table}[tb!] 
\caption{Summary statistics of the data set. It covers the 25-week period from Dec 22, 2020 to June 20, 2021. Weekly-level variables are included in Panel A, and transaction-level data are in Panel B. }\label{table: Comparative statics}
\begin{center}
 \begin{adjustbox}{width={1.05\textwidth},totalheight={1.2\textheight},keepaspectratio}
\renewcommand{\arraystretch}{1.4}
\begin{tabular*}{1.2\textwidth}{l@{\extracolsep{\fill}}rrrrrr}
\\  \hline   & N & Mean     & SD       & 10th  & 50th & 90th \\ \hline  \multicolumn{7}{c}{{Panel A: Weekly-level Data}} \\ \midrule Log Rate   Volatility, All     & 750 & 0.049   & 0.039   & 0.004  & 0.042   & 0.093   \\ 
Log Rate Volatility, Stable    & 75  & 0.004   & 0.002   & 0.002  & 0.003   & 0.005   \\
Log Rate Volatility, Unstable  & 675 & 0.054   & 0.038   & 0.015  & 0.045   & 0.094   \\ \\
Token Inflow Rate, All         & 750 & 0.013   & 0.273   & -0.164 & -0.001  & 0.160   \\
Token Inflow Rate, Stable      & 75  & 0.063   & 0.369   & -0.179 & -0.002  & 0.230   \\
Token Inflow Rate, Unstable    & 675 & 0.008   & 0.260   & -0.163 & -0.001  & 0.141   \\ \\
Trading Volume , All           & 750 & 1.987   & 1.980   & 0.200  & 1.365   & 4.672   \\
Trading Volume , Stable        & 75  & 1.033   & 1.018   & 0.173  & 0.762   & 2.032   \\
Trading Volume , Unstable      & 675 & 2.093   & 2.032   & 0.231  & 1.465   & 4.794   \\ \\
Gas Price Volatility, All      & 750 & 183.045 & 240.660 & 29.993 & 124.363 & 364.445 \\
Gas Price Volatility, Stable   & 75  & 84.888  & 67.732  & 17.458 & 75.391  & 157.683 \\
Gas Price Volatility, Unstable & 675 & 193.951 & 250.307 & 31.922 & 135.929 & 383.220 \\
 
\hline  \multicolumn{7}{c}{{Panel B: Transaction-level Data}} \\ \midrule
 Gas Price (Gwei),   All                      & 4,161,096 & 136.884 & 340.908 & 34 & 106.000 & 235.000 \\
Gas Price (Gwei), Stable                     & 223,318  & 117.471 & 113.586 & 40.000 & 99.000  & 202.000 \\
Gas Price (Gwei), Nonstable                  & 3,937,778 & 137.985 & 349.363 & 33.000 & 106.275 & 237.865 \\ \\
Absolute Value of Log Spot Rate, All         & 4,161,096 & 5.792   & 2.463   & 1.698  & 7.142   & 7.847   \\
Absolute Value of Log Spot Rate, Stable      & 223,318  & 0.003   & 0.003   & 0.000  & 0.002   & 0.005   \\
Absolute Value of Log Spot Rate,   Nonstable & 3,937,778 & 6.120   & 2.098   & 3.003  & 7.199   & 7.865   \\
\hline
\end{tabular*}\end{adjustbox} 
\end{center}
\end{table}
\subsection{Empirical Tests and Results} \label{subsection:empirical results}
In this section, we examine the three testable implications listed in Section \ref{subsec: listof results}. 

\subsubsection{Exchange Rate Volatility, Trading Volume, and Deposit}
We estimate the following panel regressions to measure the impact of token exchange rate volatility and trading volume on deposit flow rates:

\begin{equation}\label{regression1}
    Depositflow_{jt} = \gamma_j + \gamma_t + \rho_1 Volatility_{jt} + \epsilon_{jt}
\end{equation}
\begin{equation}\label{regression2}
    Depositflow_{jt} = \gamma_j + \gamma_t + \delta_1 Volume_{jt} + \epsilon_{jt}
\end{equation}
\begin{equation}\label{regression3}
    Depositflow_{jt} = \gamma_j + \gamma_t + \rho_2 Volatility_{jt} + \delta_2 Volume_{jt} + \epsilon_{jt},
\end{equation}
where $j$ indexes AMMs, $t$ indexes time, $Depositflow_{jt}$ is the deposit flow rate (inflow if positive and outflow if negative), $\gamma_j, \gamma_t$ are respectively the AMM and time fixed effects, $Volatility_{jt}$ is the volatility of the token exchange rate of AMM $j$  in week $t$, $Volume_{jt}$ is the trading volume at AMM $j$ in week $t$, and $\epsilon_{jt}$ is an error term. We cluster our standard errors at the AMM level. The coefficients $\rho_1, \rho_2$ quantify the sensitivity of deposit flow on token volatility, and the coefficients $\delta_1, \delta_2$ give the sensitivity of deposit flow on trading volume. 

Table \ref{regression: inflow} shows a negative, statistically significant relationship between the token exchange rate  volatility and the deposit flow rate, which is consistent with our theoretical prediction that $\rho_1, \rho_2 < 0$. 
After controlling for trading volume, a one-standard-deviation increase in weekly spot rate volatility (which is equal to 0.04) decreases the deposit flow  rate by 25\% standard deviations of that variable. 
Columns (b) and (c) show that there exists a positive, statistically significant relationship between trading volume and deposit flow  rate, confirming our theoretical prediction that $\delta_1, \delta_2>0$. After controlling for exchange rate volatility, a one-standard-deviation increase in trading volume (which is equal to 1.98) increases the deposit flow rate by 35\% standard deviations of that variable. 
 In summary, Table \ref{regression: inflow} confirms our model predictions from Section \ref{subsec: listof results} that the {amount of deposited tokens decreases with the exchange rate volatility, and increases with the trading volume.} {Additionally, the regression estimates indicate that these effects are economically significant.}

\begin{table}[t!] 
\caption{
 Results from regressing weekly deposit flow rates of AMMs on token exchange rate volatility and trading volumes. The data set covers 30 AMMs for a 25-week period from Dec 22, 2020 to June 20, 2021.  
 The dependent variable is the weekly deposit flow rate of AMMs. The independent variables are the weekly spot exchange rate volatility and the weekly trading volume of each AMM. Week fixed effects and AMM fixed effects are included for all regressions. Standard errors are clustered at the AMM level. Asterisks denote significance levels (***=1\%, **=5\%, *=10\%).}\label{regression: inflow}
\vspace{1cm}
\centering
\begin{tabular*}{\textwidth}{@{\extracolsep{\fill}} c c c c}
\\[-1.8ex]\hline
\hline \\[-1.8ex]
& \multicolumn{3}{c}{\textit{Dependent variable: Deposit Inflow Rate}} \
\cr \cline{2-4}
\\[-1.8ex] & (a) & (b) & (c) \\
\hline \\ [-1.8ex]
 Intercept & 0.023$^{}$ & $-0.103$ & $-0.018^{}$ \\
  & (0.077) & (0.073) & (0.070) \\
Exchange Rate Volatility &$-0.394^{**}$ & & $-1.451^{***}$ \\
  & (0.182) & & (0.405) \\
 Trading Volume & & $0.039^{***}$ & $0.052^{***}$ \\
  & & (0.015) & (0.020) \\ \hline \\[-1.8ex]
Week fixed effects? & yes & yes & yes \\
 AMM fixed effects? & yes & yes & yes \\
 Observations & 750 & 750 & 750 \\
 $R^2$ & 0.11 & 0.14 & 0.17 \\
\hline
\hline \\[-1.8ex]
\textit{Note:} & \multicolumn{3}{r}{$^{*}$p$<$0.1; $^{**}$p$<$0.05; $^{***}$p$<$0.01} \\
\end{tabular*}

\end{table}

\subsubsection{Token Exchange Rate Volatility and Gas Price}

We examine the impact of token exchange rate volatility on {levels and volatility of gas fees}. Specifically, we estimate the following two {linear} models: 

\begin{equation} \label{regressiongas1}
    GasVolatility_{jt} = \gamma_t+\gamma_{Uniswap}+ \kappa_1 \mathbb{1}_{StablePair} + \epsilon_{jt}
\end{equation}
\begin{equation} \label{regressiongas2}
    Gas_{js} = \gamma_t+\gamma_{Uniswap}+ \kappa_2 \mathbb{1}_{StablePair} + \epsilon_{js},
\end{equation}
where $j$ indexes AMMs, $t$ indexes time, and $s$ indexes transactions. {We run regressions at weekly} frequency\footnote{We choose weekly, instead of daily frequency, to have enough observations to estimate the volatility of gas price.} for the model in equation (\ref{regressiongas1}) and at daily frequency for the model in equation (\ref{regressiongas2}). $GasVolatility_{jt}$ is the volatility of gas price at time $t$ {in AMM $j$}, $Gas_{js}$ is the gas price of transaction {$s$} in AMM $j$,  $\gamma_t$ are time fixed effects, $\gamma_{Uniswap} $ is the  fixed effect for all {Uniswap AMM exchanges}, $\mathbb{1}_{StablePair}$  is the dummy variable for ``stable pairs"  which consist of two stable coins pegged to one US dollar, and $\epsilon_{jt}, \epsilon_{js}$ are error terms. We cluster our standard errors at the AMM level. The coefficients $\kappa_1$ and $\kappa_2$ quantify the differences in {levels and volatility of gas fees} between ``stable pairs" and ``unstable pairs", respectively. 

Table \ref{regression: gasprice} indicates that the 
 weekly gas price volatility of ``stable pairs" is around 40\% lower than that of ``non-stable pairs".  Additionally, 
the gas price {level} for transactions of ``stable pairs" is around 8\% lower than that of ``non-stable pairs". In summary, Table \ref{regression: gasprice} supports our model predictions from Section \ref{subsec: listof results} {that levels and volatility of gas fees are higher for pairs with larger exchange rate volatility.} {Moreover, the coefficient estimates indicate that these relationships are economically significant.}

\begin{table}[tb!] \centering
\caption{
 Results from regressing a binary variable indicating whether or not the AMM contains a ``stable pair" on gas price and gas price volatility. The data covers 30 AMMs for a 25-week period from Dec 22, 2020 to June 20, 2021.  
 The dependent variable in column (a) is the weekly gas price volatility, and the dependent variable in column (b) is the gas prices of transactions. The independent variable is a dummy equal to one if the AMM contains a pair of stable coins pegged to one US dollar. Time fixed effects and exchange fixed effects are included for all regressions. Standard errors are clustered at the AMM level. Asterisks denote significance levels (***=1\%, **=5\%, *=10\%).} \label{regression: gasprice}
\vspace{1cm}
\begin{tabular*}{0.8\textwidth}{@{\extracolsep{\fill}}  c c c}
\\[-1.8ex]\hline
\hline \\[-1.8ex]
& \multicolumn{2}{c}{\textit{Dependent variables:}} \
\cr \cline{2-3}
\\[-1.8ex] & Gas Price Volatility  & Gas Price   \\  &  (a) &   (b) \\
\hline \\[-1.8ex]
 Intercept & 189.48$^{***}$ & 137.34$^{***}$ \\
  & (16.13) & (1.61) \\
  Stable & -64.34$^{***}$ & -10.51$^{***}$ \\
  & (20.02) & (2.78) \\ 
\hline \\[-1.8ex]
Week fixed effects? & yes & no  \\
Day fixed effects? & no & yes  \\
Exchange fixed effects? & yes & yes \\
 Observations & 750 & 4,161,126 \\
 $R^2$ & 0.24 & 0.04 \\
\hline
\hline \\[-1.8ex]
\textit{Note:} & \multicolumn{2}{r}{$^{*}$p$<$0.1; $^{**}$p$<$0.05; $^{***}$p$<$0.01} \\
\end{tabular*}
\end{table}

\section{Conclusion} \label{sec conclu}
 
{
Our paper analyzes the economic incentives behind blockchain-based financial intermediation.} We show theoretically and empirically that exploitable arbitrage opportunities created by token exchange rate volatility limits the adoption of blockchain-based AMMs by liquidity providers. We also argue that the adoption of AMMs impose negative externalities on other decentralized applications built on the same underlying blockchain. 


Our findings inform the liquidity management strategy of token holders. We argue that they should only deposit into AMMs whose token prices are stable and highly correlated, or whose trading volumes are high. Our results warn against providing liquidity into AMMs which manage Altcoins with no intrinsic value. Importantly, we argue that liquidity providers should not fall prey to the fallacy that token value loss from arbitrage is impermanent and keep depositing if exchange rate reversion is likely. Token exchange rate movements, regardless of the direction, leads to permanent loss, and liquidity providers need to account for the opportunity cost of depositing tokens into AMMs. 

Our results have implications for the operation and design of DeFi exchanges. We argue that the curvature of the pricing curve used by the AMM governs the trade-offs between the severity of the arbitrage problem and the investors' willingness to trade. When these two forces are well balanced, a ``liquidity freeze'' is least likely to occur, the deposit efficiency is the highest, and social welfare is maximized. {Our analysis also sheds light on a common misconception among practitioners, and shows that pooling more than two tokens in the same AMM does not alleviate the arbitrage problem.} 

Our paper can be extended along several directions. 
The first extension is to study the governance of AMMs. Typically, decisions for the operations of DeFi projects are proposed, voted and made by investors who hold the governance tokens. However, different voters, such as developers, liquidity providers, and token investors, may have very different incentives, and {their votes often lead to socially inefficient protocols, such as those demanding very high transaction fees.}  
This calls for the design of a mechanism which distributes governance tokens to agents so to minimize agency costs.
Another desirable extension is to design protocols through which DeFi projects internalize the externalities imposed on the underlying blockchains.
Such a protocol would inform the decision of which DeFi projects should be built on the same blockchain versus multiple blockchains. While executing all DeFi projects on a smaller number of blockchains may increase safety, it would also increase congestion and transaction costs. We leave a systematic study of the trade-off between transaction safety and costs for future research.

 
\newpage
\bibliographystyle{rfs}
\bibliography{defi}

\newpage{}

\setcounter{equation}{0}
\setcounter{mylemma}{0}
\renewcommand{\theequation}{A.\arabic{equation}}
\renewcommand{\themylemma}{A\arabic{lemma}}

\appendix



\section{Technical Results and Proofs}

\begin{proof}[Proof of Lemma \ref{adverseselction order1}]

Without loss of generality, we assume the shock hits B tokens. {The case where the shock leads to an appreciation of A tokens can be handled symmetrically.} Recall that $F$ is  twice continuously differentiable, and $F_x>0, F_y>0$.  By the Implicit Function Theorem, we can then rewrite the constraint of the arbitrageur's optimization problem stated in \eqref{eq arbitrageroptimization} as  
\begin{equation} \label{asfunctionarb1}
      \Delta  q_B^{(t,3)} = g^{(t,3)}(\Delta q_A^{(t,3)}), -\infty < \Delta q_A^{(t,3)}  \leq 0,
\end{equation}
  where $g^{(t,3)}$ is a twice differentiable function. In this way, we write the amount of B tokens the arbitrageur can withdraw from the AMM as a function of the negative of the amount of A tokens {deposited} into the AMM by the arbitrageur, before paying the fee. The first order derivative of the above function is the negative of marginal exchange rate: $$\frac{d (g^{(t,3)})}{d (\Delta q_A^{(t,3)})}= - \frac{ F_x}{F_y}\bigg\rvert_{(x,y) = (y_A^{(t,2)}-\Delta q_A^{(t,3)} , y_B^{(t,2)}-\Delta q_B^{(t,3)}) } \leq 0 .$$ By Assumption \ref{curve}, we have $\frac{d^2 (g^{(t,3)})}{d (\Delta q_A^{(t,3)})^2} = -\frac{-F_y^2F_{xx} + 2F_yF_xF_{xy}-F_x^2F_{yy}}{F_y^3}\bigg\rvert_{(x,y) = (y_A^{(t,2)}-\Delta q_A^{(t,3)} , y_B^{(t,2)}-\Delta q_B^{(t,3)}) } < 0, $ i.e.,  $g^{(t,3)}$ is concave.


Using the relations \eqref{asfunctionarb1}, $p_A^{(t,2)} = p_A^{(t,1)}$ and $p_B^{(t,2)} = p_B^{(t,1)}(1+\beta)$ {into the two-variable optimization problem} (\ref{eq arbitrageroptimization}), we obtain an equivalent single variable optimization problem: 

\begin{equation}\label{eq arbitrageroptimization2}
\max_{-\infty < \Delta q_A^{(t,3)}  \leq 0} \quad   p_A^{(t,1)}(1+f) \Delta q_A^{(t,3)} + p_B^{(t,1)}(1+\beta) g^{(t,3)}(\Delta q_A^{(t,3)})
\end{equation}

The first-order derivative of the objective function in \eqref{eq arbitrageroptimization2} is:

\begin{equation}\label{eqderivative arbitrageroptimization2}
   p_A^{(t,1)}(1+f) - (- p_B^{(t,1)}(1+\beta) \frac{d g^{(t,3)}}{d (\Delta q_A^{(t,3)})})
\end{equation}
The first term is the arbitrageur's marginal cost of exchanging A for B tokens, while the second term is the  marginal benefit. By concavity, the above expression  decreases in $\Delta q_A^{(t,3)}$. Moreover, if $\Delta q_A^{(t,3)} = 0 $, $\frac{d g^{(t,3)}}{d (\Delta q_A^{(t,3)})} = -\frac{p_A^{(t,1)}}{p_B^{(t,1)}}$. This is because  at sub-period 1 of period $t$, the liquidity providers deposit their tokens in the AMM at the fair value exchange rate.  
Hence, the above first-order derivative is positive at $\Delta q_A^{(t,3)} = 0 $ if and only if $1+f>1+\beta$.
This means that the marginal cost exceeds the marginal benefit. {The optimal amount of A tokens exchanged for B tokens} is then 0. If $ \Delta q_A^{(t,3)} = 0$, the value of the objective function is exactly 0, because no arbitrage has occurred. 

If $1+\beta > 1+f$, the optimal exchange amount is attained when the marginal cost equals the marginal benefit of exchanging:

\begin{equation}\label{arbitrageur FOC}
   p_A^{(t,1)}(1+f) = (- p_B^{(t,1)}(1+\beta) \frac{d g^{(t,3)}}{d (\Delta q_A^{(t,3)})}).
\end{equation}

We denote the solution of the above equation by $\Delta q_A^{(t,3)*}$.  By concavity of $g^{(t,3)}$, the first order derivative in \eqref{eqderivative arbitrageroptimization2} is decreasing in the interval $( -\infty, 0]$; it is also negative at 0 by $1+\beta > 1+f$. By part 4 of Assumption \ref{curve}, the first order derivative is positive as $\Delta q_A^{(t,3)}  \rightarrow -\infty$. Thus, by continuity of the first-order derivative, there exists a unique $\Delta q_A^{(t,3)*}$ at which the derivative is 0, and  \eqref{arbitrageur FOC} is satisfied. Therefore, the optimal exchange amount of A tokens,  $\Delta q_A^{(t,3)*}$, exists. Moreover, $\Delta q_A^{(t,3)*}$ is unique because \eqref{arbitrageur FOC} admits at most one solution. 

Recall that $1+\beta > 1 +f$, $ p_A^{(t,1)}(1+f) < (- p_B^{(t,1)}(1+\beta) \frac{d (g^{(t,3)})}{d (\Delta q_A^{(t,3)})})$ for $\Delta q_A^{(t,3)} = 0$. This implies that $\Delta q_A^{(t,3)*} \neq 0$ because the arbitrageur's objective function attains a strictly higher value at $\Delta q_A^{(t,3)*} \neq 0$ than at $\Delta q_A^{(t,3)}=0$.  Because the {objective function has value 0} at $\Delta q_A^{(t,3)}=0$, we obtain {$\pi(y_A^{(t,2)},y_B^{(t,2)},p_B^{(t,2)}, p_A^{(t,2)})>0$} if  $\beta > f$.

We next prove that $\pi(y_A^{(t,2)},y_B^{(t,2)},p_B^{(t,2)}, p_A^{(t,2)})$, $|\Delta q_A^{(t,3)*}|$, and $|\Delta q_B^{(t,3)*}|$ increase in $\beta$ and decrease in $f$. 

In the case $\beta \leq f$, the arbitrageur does not trade, and thus $\pi(y_A^{(t,2)},y_B^{(t,2)},p_B^{(t,2)}, p_A^{(t,2)})= 0,$ $|\Delta q_A^{(t,3)*}|= 0$,$ |\Delta q_B^{(t,3)*}| = 0$, and thus  they are all independent of $f$ and $ \beta$. 

For the case $\beta > f$,  we apply the Envelope Theorem and obtain $\frac{\partial \pi}{\partial f } =  p_A^{(t,1)}* \Delta q_A^{(t,3)*} < 0$, and $\frac{\partial \pi}{\partial \beta } = p_B^{(t,1)}* \Delta q_B^{(t,3)*}> 0$. Therefore, $\pi(y_A^{(t,2)},y_B^{(t,2)},p_B^{(t,2)}, p_A^{(t,2)}) $ increases in $\beta$ and decreases in $f$.  

$q_A^{(t,3)*}$ is defined by the condition stated in \eqref{arbitrageur FOC}. By differentiating both sides of \eqref{arbitrageur FOC} with respect to $f$, we have
\begin{align*}
    p_A^{(t,1)} = - p_B^{(t,1)}(1+\beta) \frac{d^2 g^{(t,3)}}{d (\Delta q_A^{(t,3)})^2} \frac{d \Delta q_A^{(t,3)*}}{d f}.
\end{align*}
 Recall that $g^{(t,3)}$ is concave, which means that $\frac{d^2 g^{(t,3)}}{d (\Delta q_A^{(t,3)})^2} < 0.$ This implies that $\frac{d \Delta q_A^{(t,3)*}}{d f} >0$, i.e., $\Delta q_A^{(t,3)*}$ increases in $f$. 
Following the same procedure as above, we can show that $\Delta q_A^{(t,3)*}$  decreases in $\beta$. Since $\Delta q_A^{(t,3)*}<0$, $|\Delta q_A^{(t,3)*}|$ decreases in $f$, and increases in $\beta$. 

Since $\Delta q_B^{(t,3)*} = g^{(t,3)}(\Delta q_A^{(t,3)*})$ and the derivative of $g^{(t,3)}$ is negative, we have that $\Delta q_B^{(t,3)*}$ decreases in $\Delta q_A^{(t,3)*}<0$ and increases in $|\Delta q_A^{(t,3)*}|$. Thus, $\Delta q_B^{(t,3)*}$ decreases in $f$, and increases in $\beta$.


\end{proof}

\begin{mylemma} \label{proportionality1}
For any period $t= 1,2$ and constant $c > 0$, the following properties are true:
\begin{enumerate}
    \item  $\Delta q_B^{(t,3)*}(cy_A^{(t,2)},cy_B^{(t,2)},p_B^{(t,2)}, p_A^{(t,2)}) = c\Delta q_B^{(t,3)*}(y_A^{(t,2)},y_B^{(t,2)},p_B^{(t,2)}, p_A^{(t,2)})$,
    \item $\Delta q_A^{(t,3)*}(cy_A^{(t,2)},cy_B^{(t,2)},p_B^{(t,2)}, p_A^{(t,2)}) = c\Delta q_A^{(t,3)*}(y_A^{(t,2)},y_B^{(t,2)},p_B^{(t,2)}, p_A^{(t,2)})$,
    \item $\pi(cy_A^{(t,2)},cy_B^{(t,2)},p_B^{(t,2)}, p_A^{(t,2)}) = c\pi(y_A^{(t,2)},y_B^{(t,2)},p_B^{(t,2)}, p_A^{(t,2)})$.

\end{enumerate}
\end{mylemma}

\begin{proof}[Proof of Lemma  \ref{proportionality1}]
Recall that the constraint of the arbitrageur's optimization problem defined in \eqref{eq arbitrageroptimization} is:
$$F(y_A^{(t,2)},y_B^{(t,2)}) = F(y_A^{(t,2)}-\Delta q_A^{(t,3)} , y_B^{(t,2)}-\Delta q_B^{(t,3)}).$$

If the state at subperiod 2 of period $t$ is $(cy_A^{(t,2)},cy_B^{(t,2)},p_B^{(t,2)}, p_A^{(t,2)})$, then the constraint of arbitrageur problem  becomes: 
\begin{align*}
    F(cy_A^{(t,2)},cy_B^{(t,2)}) &= F(cy_A^{(t,2)}-\Delta q_A^{(t,3)} ,c y_B^{(t,2)}-\Delta q_B^{(t,3)}) \\
   \iff c^l F(y_A^{(t,2)},y_B^{(t,2)}) &= c^lF(y_A^{(t,2)}-\frac{\Delta q_A^{(t,3)}}{c} , y_B^{(t,2)}-\frac{\Delta q_B^{(t,3)}}{c})\\
      \iff F(y_A^{(t,2)},y_B^{(t,2)}) &= F(y_A^{(t,2)}-\frac{\Delta q_A^{(t,3)}}{c} , y_B^{(t,2)}-\frac{\Delta q_B^{(t,3)}}{c})
\end{align*}

With the following change of variables, $\Delta q_A^{(t,3)'} = \Delta q_A^{(t,3)}/c, \Delta q_B^{(t,3)'} = \Delta q_B^{(t,3)}/c$, the optimization problem given the state $(cy_A^{(t,2)},cy_B^{(t,2)},p_B^{(t,2)}, p_A^{(t,2)})$ may be rewritten as

\begin{equation}\label{eq arbitrageroptimizationproportion}
\begin{aligned}
\max_{\Delta q_A^{(t,3)'}, \Delta q_B^{(t,3)'}} \quad &  c \; p_A^{(t,2)}(1+f) \Delta q_A^{(t,3)'} + c \; p_B^{(t,2)}\Delta q_B^{(t,3)'}\\
\textrm{s.t.} \quad & F(y_A^{(t,2)},y_B^{(t,2)}) = F(y_A^{(t,2)}-\Delta q_A^{(t,3)'} , y_B^{(t,2)}-\Delta q_B^{(t,3)'}) \\
  &\Delta q_A^{(t,3)'} \leq 0, \Delta q_B^{(t,3)'} \geq 0,    \\
\end{aligned}
\end{equation}
which corresponds to the optimization problem stated in \eqref{eq arbitrageroptimization} except that the objective function is now multiplied by a constant $c$. Therefore, the maximum value for the  optimization problem in \eqref{eq arbitrageroptimizationproportion} is $c$ times the maximum of the  optimization problem stated in \eqref{eq arbitrageroptimization}.

Moreover, the optimization problem \eqref{eq arbitrageroptimizationproportion} admits the same solution as the optimization problem stated in \eqref{eq arbitrageroptimization}, and given by $(\Delta q_A^{(t,3)*}(y_A^{(t,2)},y_B^{(t,2)},p_B^{(t,2)}, p_A^{(t,2)}), \Delta q_B^{(t,3)*}(y_A^{(t,2)},y_B^{(t,2)},p_B^{(t,2)}, p_A^{(t,2)}))$.  After multiplying $\Delta q_A^{(t,3)*}(y_A^{(t,2)},y_B^{(t,2)},p_B^{(t,2)}, p_A^{(t,2)})$ and $\Delta q_B^{(t,3)*}(y_A^{(t,2)},y_B^{(t,2)},p_B^{(t,2)}, p_A^{(t,2)})$ by $c$, we obtain
\begin{align*}
    \Delta q_A^{(t,3)*}(cy_A^{(t,2)},cy_B^{(t,2)},p_B^{(t,2)}, p_A^{(t,2)}) &= c\Delta q_A^{(t,3)*}(y_A^{(t,2)},y_B^{(t,2)},p_B^{(t,2)}, p_A^{(t,2)}),\\ \Delta q_B^{(t,3)*}(cy_A^{(t,2)},cy_B^{(t,2)},p_B^{(t,2)}, p_A^{(t,2)})&= c\Delta q_B^{(t,3)*}(y_A^{(t,2)},y_B^{(t,2)},p_B^{(t,2)}, p_A^{(t,2)}).
\end{align*}

\end{proof} 

\begin{proof}[Proof of Lemma  \ref{traderoptimization}]
Without loss of generality, we assume that the investor {who} arrives is of ``type A''. {The case where the investor {who} arrives is of ``type B'' can be handled using symmetric arguments.} 

We follow the same procedure as in the proof of Lemma \ref{adverseselction order1}. Recall that $F$ is  twice continuously differentiable, and $F_x>0, F_y>0$. By the Implicit Function Theorem, we can then rewrite the constraint of the investor's optimization problem stated in \eqref{eq: investorincentiveequation} as  
\begin{equation} \label{asfunctioninvestor}
      \Delta  Q_A^{(t,2)} = g^{(t,2)}(\Delta Q_B^{(t,2)}), - \infty < \Delta Q_B^{(t,2)}  \leq 0, 
\end{equation}
 where $g^{(t,2)}$ is a twice differentiable function. In this way, we can write the amount of A tokens the investors can withdraw from the AMM as a function of the negative of the amount of B tokens she has to deposit into the AMM, excluding the fee. The first order derivative of the above function is the negative of the marginal exchange rate: 
$$\frac{d g^{(t,2)}}{d (\Delta Q_B^{(t,2)})}= - \frac{ F_y}{F_x}\bigg\rvert_{(x,y) = (y_A^{(t,1)}-\Delta Q_A^{(t,2)} , y_B^{(t,1)}-\Delta Q_B^{(t,2)}) } < 0 .$$
It follows from Assumption \ref{curve} that the function $g^{(t,2)}$ is concave, that is, $\frac{d^2 g^{(t,2)}}{d (\Delta Q_A^{(t,2)})^2} \leq 0.$

Plugging \eqref{asfunctioninvestor} into the two-variable optimization problem stated  in (\ref{eq: investorincentiveequation}),  we obtain the following equivalent, single-variable, optimization problem: 

\begin{equation}\label{eq investoroptimization2}
\max_{- \infty < \Delta Q_B^{(t,2)}  \leq 0} \quad (1+\alpha)p_A^{(t,1)} \Delta g^{(t,2)}\Delta Q_B^{(t,2)} +(1+f) p_B^{(t,1)} \Delta Q_B^{(t,2)}.
\end{equation}
The first-order derivative of the objective function in \eqref{eq investoroptimization2} is

\begin{equation}\label{eqderivative investoroptimization2}
(1+f) p_B^{(t,1)} - \left(  -(1+\alpha)p_A^{(t,1)}  \frac{d g^{(t,2)}}{d (\Delta Q_B^{(t,2)})} \right)
\end{equation}

The first term is the marginal cost of exchanging a B token foran A token, and the second term is the  marginal benefit of investors. By concavity, the above expression  decreases in $\Delta Q_B^{(t,2)}$. If $\Delta Q_B^{(t,2)} = 0 $, $\frac{d g^{(t,2)}}{d (\Delta Q_B^{(t,2)})} = -\frac{p_B^{(t,1)}}{p_A^{(t,1)}}$ because at sub-period 1 the liquidity providers deposit their tokens in the AMM at the fair rate. Hence, the first-order derivative of the objective function stated in \eqref{eqderivative investoroptimization2} is positive at $\Delta Q_B^{(t,2)} = 0 $ if and only if $1+f>1+\alpha$. If $1+f \geq 1+\alpha$,  the marginal cost exceeds the marginal benefit. Hence, the optimal trading size is zero, i.e., the investor does not trade. 

If $1+\alpha > 1+f$, the optimal exchange amount is obtained by equating the marginal cost and the marginal benefit of exchanging:

\begin{equation}\label{investor FOC}
(1+f) p_B^{(t,1)} = \left(  -(1+\alpha)p_A^{(t,1)}  \frac{d g^{(t,2)}}{d (\Delta Q_B^{(t,2)})} \right).
\end{equation}


We denote the solution of the equation above as $\Delta Q_B^{(t,2)*} $. {Observe that} \eqref{investor FOC} {admits at most one solution, hence if a solution $\Delta Q_B^{(t,2)*}$ exists, it is unique.} The existence of a solution $\Delta Q_B^{(t,2)*}$ {follows from the intermediate value theorem:} the derivative \eqref{eqderivative investoroptimization2} is  continuous, negative if $\Delta Q_B^{(t,2)} = 0$, and positive if  $\Delta Q_B^{(t,2)} \rightarrow -\infty.$  Moreover, $\Delta Q_B^{(t,2)*} \neq 0$ because, if $1+\alpha > 1 +f$, $(1+f) p_B^{(t,1)} < (  -(1+\alpha)p_A^{(t,1)}  \frac{d g^{(t,2)}}{d (\Delta Q_B^{(t,2)})} )$ when $\Delta Q_B^{(t,2)} = 0$. Therefore, the investor's objective function attains a higher value at $\Delta Q_B^{(t,2)*}$ than at $0$, and thus the ({``type A''}) investor's maximum surplus {$s_A(y_A^{(t,1)},y_B^{(t,1)},p_B^{(t,1)}, p_A^{(t,1)})$} is strictly positive if  $\alpha > f$.

We next prove that $s_A(y_A^{(t,1)},y_B^{(t,1)},p_B^{(t,1)}, p_A^{(t,1)})$, $|\Delta Q_A^{(t,2)*}|$, and $|\Delta Q_B^{(t,2)*}|$ are increasing in $\alpha$ and decreasing in $f$. For the case $\alpha \leq f$, the investor does not trade, and thus $s_A(y_A^{(t,2)},y_B^{(t,2)},p_B^{(t,2)}, p_A^{(t,2)})= 0,$ $|\Delta Q_A^{(t,2)*}|= 0$,$ |\Delta Q_B^{(t,2)*}|= 0$. As a result,  $\frac{\partial s_A}{\partial f } = \frac{\partial s_A}{\partial \alpha }  = 0,  \frac{\partial |\Delta Q_A^{(t,2)*}|}{\partial f } = \frac{\partial |\Delta Q_A^{(t,2)*}|}{\partial \alpha } = \frac{\partial |\Delta Q_B^{(t,2)*}|}{\partial f } = \frac{\partial |\Delta Q_B^{(t,2)*}|}{\partial \alpha } = 0$.

For the case $\alpha> f$,  we apply the Envelope Theorem and obtain $\frac{\partial s_A}{\partial f } =  p_B^{(t,1)} \Delta Q_B^{(t,2)*} < 0$, and $\frac{\partial \pi}{\partial \alpha } = p_A^{(t,1)}  \Delta Q_A^{(t,2)*}> 0$. Therefore, $ s_A(y_A^{(t,1)},y_B^{(t,1)},p_B^{(t,1)}, p_A^{(t,1)}) $ increases in $\alpha$ and decreases in $f$. 

Recall the quantities $\Delta Q_B^{(t,2)*},\Delta Q_A^{(t,2)*}$. $\Delta Q_B^{(t,2)*}$ defined by the condition in \eqref{investor FOC}. Differentiating both sides of \eqref{investor FOC} with respect to $f$, we obtain
\begin{align*}
     p_B^{(t,1)} =  -(1+\alpha)p_A^{(t,1)}  \frac{d^2 g^{(t,2)}}{d (\Delta Q_B^{(t,2)})^2} \frac{d \Delta Q_B^{(t,2)*} }{ d f}.
\end{align*}
Recall that $g^{(t,2)}$ is concave, thus $\frac{d^2 g^{(t,2)}}{d (\Delta Q_B^{(t,2)})^2} < 0$. It then follows that $ \frac{d \Delta Q_B^{(t,2)*} }{ d f} >0 $ because $p_B^{(t,1)}>0$. Using the same argument, we can  show $ \frac{d \Delta Q_B^{(t,2)*} }{ d \alpha} < 0 $. Since $\Delta Q_B^{(t,2)*}<0$, we have that $|\Delta Q_B^{(t,2)*}|$ decreases in $f$, and increases in $\alpha$. Since $\Delta Q_A^{(t,2)*} = g^{(t,2)}(\Delta Q_B^{(t,2)*})$, and the derivative of $g^{(t,2)}$ is negative, we have $\frac{\partial \Delta Q_A^{(t,2)*}}{\partial f} = \frac{d g^{(t,2)}}{d \Delta Q_B^{(t,2)*}} \frac{\partial \Delta Q_B^{(t,2)*} }{\partial f} <0, $ and $ \frac{\partial \Delta Q_A^{(t,2)*}}{\partial \alpha} = \frac{d g^{(t,2)}}{d \Delta Q_B^{(t,2)*}} \frac{\partial \Delta Q_B^{(t,2)*} }{\partial \alpha} >0 $. Because $\Delta Q_A^{(t,2)*}>0$ , we deduce that $|\Delta Q_A^{(t,2)*}|$ decreases in $f$ and increases in $\alpha$.
\end{proof}

\begin{mylemma} \label{proportionality2}
For any period $t=1,2$ and constant $c > 0$, the following relations hold:
\begin{enumerate}
    \item  $\Delta Q_B^{(t,2)*}(cy_A^{(t,1)},cy_B^{(t,1)},p_B^{(t,1)}, p_A^{(t,1)}) = c\Delta Q_B^{(t,2)*}(y_A^{(t,1)},y_B^{(t,1)},p_B^{(t,1)}, p_A^{(t,1)})$,
    \item  $\Delta Q_A^{(t,2)*}(cy_A^{(t,1)},cy_B^{(t,1)},p_B^{(t,1)}, p_A^{(t,1)}) = c\Delta Q_A^{(t,2)*}(y_A^{(t,1)},y_B^{(t,1)},p_B^{(t,1)}, p_A^{(t,1)})$,
  \item  $s_A(cy_A^{(t,1)},cy_B^{(t,1)},p_B^{(t,1)}, p_A^{(t,1)}) = c* s_A(y_A^{(t,1)},y_B^{(t,1)},p_B^{(t,1)}, p_A^{(t,1)})$,
  \item $s_B(cy_A^{(t,1)},cy_B^{(t,1)},p_B^{(t,1)}, p_A^{(t,1)}) = c* s_B(y_A^{(t,1)},y_B^{(t,1)},p_B^{(t,1)}, p_A^{(t,1)})$.

\end{enumerate}
\end{mylemma}

\begin{proof}[Proof of Lemma  \ref{proportionality2}]
The proof is analogous to that used to prove Lemma  \ref{proportionality1}.
\end{proof} 


\begin{mylemma} \label{lowest value on curve1}
\begin{equation}\label{lowest value on curve}
\begin{aligned}
\min_{a,b} \quad &  p_A*a  + p_B *b \\
\textrm{s.t.} \quad & F(y_A,y_B) = F(y_A+a , y_B+b). \\
\end{aligned}
\end{equation}
For any $p_A, p_B,y_A,y_B > 0$, there exists a unique solution $(a^*, b^*)$ for the optimization problem stated in \eqref{lowest value on curve}, and it satisfies the condition $\frac{ F_x}{F_y}\bigg\rvert_{(x,y) = (y_A+a^* , y_B+b^*)} =\frac{p_A}{p_B}$. 

\end{mylemma}

\begin{proof}[Proof of Lemma  \ref{lowest value on curve1}]
Recall that $F$ is  twice continuously differentiable, and $F_x>0, F_y>0$. By the Implicit Function Theorem, we can rewrite the constraint of the optimization problem stated in \eqref{lowest value on curve} as  
\begin{equation} 
     b= m(a), a\in \mathbb{R},
\end{equation}
 where $m(a)$ is twice differentiable function, whose first order derivative  is $$m'(a)= - \frac{ F_x}{F_y}\bigg\rvert_{(x,y) = (y_A+a , y_B+b) } < 0 .$$ By Assumption \ref{curve},  $m(a)$ is convex, that is, $m''(a) > 0.$ We then plug $b = m(a)$ into  the optimization problem stated in (\ref{lowest value on curve}), and we obtain an equivalent single variable unconstrained optimization problem  

\begin{equation}
\min_{a} \quad  p_A*a  + p_B *m(a), 
\end{equation}
where the objective function is strictly convex. Hence, the optimization problem admits a unique solution $a^*$ which satisfies the first-order condition:$$p_A  + p_B *m'(a^*) = 0.$$ We then have:
$$- m'(a^*)=  \frac{ F_x}{F_y}\bigg\rvert_{(x,y) = (y_A+a^* , y_B+b^*) } =  \frac{p_A}{p_B} .$$

\end{proof} 


\begin{proof}[Proof of Proposition  \ref{propostion unique equilibrium}]
{We follow the proof strategy of Zermelo's theorem. To establish existence, we identify a strategy profile through backward induction}; if multiple equilibria are never encountered in the process of backward induction, then the identified strategy profile must be the unique subgame perfect equilibrium. 


We start from the last period $t=2$. At sub-period 3 of period 2,  the arguments in Section \ref{sec:adverse} show that the optimal  gas fees that liquidity providers attach  to their exit orders are zero. This is the unique optimal action for all liquidity providers, because submitting a positive gas fee would only decrease their total consumption good after exiting.

If there exists an arbitrage opportunity at sub-period 3 of period 2, then by Lemma \ref{adverseselction order1},  the optimal trading order is unique. By Assumption \ref{assarb}, the arbitrageur pays a gas fee equal to the profit from the optimal order. Hence, in the last period the arbitrageur must choose the optimal trade order specified in Lemma \ref{adverseselction order1}; if instead, the arbitrageur chooses a different order, it ends up with a negative profit in the last period, which is sub-optimal. At sub-period 2, by Lemma \ref{traderoptimization}, if an investor arrives, then it chooses the unique optimal trading order. 

At sub-period 1 of period $2$, each liquidity provider maximizes its one-period payoff. Liquidity provider $i$ holds an amount $e_i^{(1)}$ of consumption good that he can use to construct his optimal portfolio $(w_i^{(2)}y_A^{(2,1)}, w_i^{(2)}y_B^{(2,1)},c_i^{(2,1)},x_{A_i}^{(2,1)},x_{B_i}^{(2,1)})$. He assigns zero weight to the consumption good and B tokens, i.e., $x_{B_i}^{(2,1)} = c_i^{(2,1)} = 0  $, because they both yield expected returns lower than A tokens. Hence, liquidity providers  only allocate their consumption good to A tokens or deposit in the AMM. Since the AMM requires tokens to be deposited at the current fair value exchange rate, i.e., $ \frac{ F_x}{F_y}\bigg\rvert_{(x,y) = (y_A^{(2,1)}, y_B^{(2,1)})} =\frac{p_A^{(2,1)}}{p_B^{(2,1)}}= \frac{p_A^{(1,2)}}{p_B^{(1,2)}}$, the ratio of the A and B tokens deposited, $\frac{y_B^{(2,1)}}{y_A^{(2,1)}}$, is already pinned down by the fair value exchange rate at subperiod 2 of period 1, $\frac{p_A^{(1,2)}}{p_B^{(1,2)}}$. Hence, the liquidity provider only chooses the amount of A tokens to deposit. Suppose  liquidity provider $i$ deposits an amount $w_i^{(2)} y_A^{(2,1)} >0$ and $w_i^{(2)}y_B^{(2,1)} > 0 $  of A and B tokens, respectively, and invests the rest of the consumption good on A tokens. The expected payoff of liquidity provider $i$  is then:
\begin{align}\label{one-period payoff liquidity providers}
         &w_i^{(2)} \expe_{(2,1)}  \bigg[p_A^{(2,3)} y_A^{(2,3)} + p_B^{(2,3)} y_B^{(2,3)}  \bigg]  + \expe_{(2,1)}\bigg[ \frac{p_A^{(2,2)}}{p_A^{(2,1)}} \left(e_i^{(1)}- w_i^{(2)}(p_A^{(2,1)} y_A^{(2,1)} + p_B^{(2,1)} y_B^{(2,1)} )\right)\bigg],  
\end{align}
where $\expe_{(t,s)}[ X ]$ denotes the  {expectation of the random variable $X$ conditional on information available at the end of sub-period s of period t.}


In the above expression, the first term is the expected payoff from depositing in the AMM, and the second term is the expected payoff from investing in A tokens. We may rewrite \eqref{one-period payoff liquidity providers} as follows:
\begin{equation}\label{lpoptimization}
     w_i^{(2)} (p_A^{(2,1)} y_A^{(2,1)} + p_B^{(2,1)} y_B^{(2,1)} ) \expe_{(2,1)} \lrb{1+R_{D}^{(2)}}  \nonumber   +  (e_i^{(1)}- w_i^{(2)}(p_A^{(2,1)} y_A^{(2,1)} + p_B^{(2,1)} y_B^{(2,1)} )) \expe_{(2,1)} \lrb{1+R_{A}^{(2)}}   
\end{equation}
where $w_i^{(2)} (p_A^{(2,1)} y_A^{(2,1)} + p_B^{(2,1)} y_B^{(2,1)} )$ is the amount of consumption good used to exchange for tokens deposited in the AMM,  $(e_i^{(1)}- w_i^{(2)}(p_A^{(2,1)} y_A^{(2,1)} + p_B^{(2,1)} y_B^{(2,1)} ))$ is the amount of consumption good invested in A tokens, the one-period expected return from {deposits made in subperiod 1 of period 2} is: 
$$\expe_{(2,1)} \lrb{R_{D}^{(2)}}{:=} \frac{y_A^{(2,1)}}{(p_A^{(2,1)} y_A^{(2,1)} + p_B^{(2,1)} y_B^{(2,1)} )}\expe_{(2,1)}\bigg[p_A^{(2,3)}\frac{y_A^{(2,3)}}{y_A^{(2,1)}}  + p_B^{(2,3)} \frac{y_B^{(2,3)}}{y_A^{(2,1)}} \bigg] -1,$$ and the one-period expected return from {investing in A tokens} is:
$$\expe_{(2,1)} \lrb{R_{A}^{(2)}} {:=} \expe_{(2,1)}\bigg[ \frac{p_A^{(2,2)}}{p_A^{(2,1)}} \bigg] -1.$$


Conditional on the information at {sub-period 1 of period 2}, $\frac{y_B^{(2,1)}}{y_A^{(2,1)}}$ is independent of the deposit $ y_A^{(2,1)}$ and of $w_i^{(2)}$. This is because the AMM requires tokens to be deposited at the current exchange rate, i.e., $ \frac{ F_x}{F_y}\bigg\rvert_{(x,y) = (y_A^{(2,1)}, y_B^{(2,1)})} =\frac{p_A^{(2,1)}}{p_B^{(2,1)}}= \frac{p_A^{(1,2)}}{p_B^{(1,2)}}$, and this pins down the ratio of the A and B tokens deposited.  At sub-period 2, we have $y_A^{(2,2)} = y_A^{(2,1)}$,and thus $ \frac{y_A^{(2,2)}}{y_A^{(2,1)}} =1$ if the investor does not arrive;  $y_A^{(2,2)} = y_A^{(2,1)} - \Delta Q_A^{(2,2)*}, \frac{y_A^{(2,2)}}{y_A^{(2,1)}} = 1 - \frac{\Delta Q_A^{(2,2)*}}{y_A^{(2,1)}}$  if a ``type A '' investor arrives;  $y_A^{(2,2)} = y_A^{(2,1)} - (1+f)\Delta Q_A^{(2,2)*}, \frac{y_A^{(2,2)}}{y_A^{(2,1)}} = 1 - \frac{(1+f)\Delta Q_A^{(2,2)*}}{y_A^{(2,1)}}$  if a ``type B '' investor arrives. As we know from Lemma \ref{proportionality2} that $\Delta Q_A^{(2,2)*}$ is proportional to $y_A^{(2,1)}$, the random variable $\frac{y_A^{(2,2)}}{y_A^{(2,1)}}$ neither depends on $w_i^{(2)}$ nor on $y_A^{(2,1)}$. Using symmetric arguments, we obtain that $\frac{y_B^{(2,2)}}{y_B^{(2,1)}}$ is independent of  $ w_i^{(2)}$ and $y_A^{(2,1)}$. Since we have also shown that $\frac{y_B^{(2,1)}}{y_A^{(2,1)}}$ is independent of $ w_i^{(2)},y_A^{(2,1)}$, we conclude that $\frac{y_B^{(2,2)}}{y_A^{(2,1)}} = \frac{y_B^{(2,1)}}{y_A^{(2,1)}} \frac{y_B^{(2,2)}}{y_B^{(2,1)}}$ is independent of  $ w_i^{(2)}$ and $,y_A^{(2,1)}$.

Using Lemma \ref{proportionality1} and the same arguments above, we deduce that $\frac{y_A^{(2,3)}}{y_A^{(2,2)}}$ and $\frac{y_B^{(2,3)}}{y_A^{(2,2)}}$  are independent of $ w_i^{(2)}$ and $ y_A^{(2,1)}$. As a result,  $\frac{y_A^{(2,3)}}{y_A^{(2,1)}}= \frac{y_A^{(2,3)}}{y_A^{(2,2)}} \frac{y_A^{(2,2)}}{y_A^{(2,1)}}$, and $\frac{y_B^{(2,3)}}{y_B^{(2,1)}}= \frac{y_B^{(2,3)}}{y_B^{(2,2)}} \frac{y_B^{(2,2)}}{y_B^{(2,1)}}$ are random variables, which are independent of $w_i^{(2)}$ and $ y_A^{(2,1)}$. Therefore, the expected return from deposits $\expe_{(2,1)} \lrb{R_{D}^{(2)}}$  does not depend on the amount of tokens deposited by liquidity provider $i$, $ w_i^{(2)}y_A^{(2,1)}$, and  only depends on token prices at the end of sub-period 2 of period 1, $p_A^{(1,2)}$, and $ p_B^{(1,2)} $. Moreover, $\expe_{(2,1)} \lrb{R_{A}^{(2)}} = ((1-\theta)\kappa_1+\theta \kappa_{com}) \beta$ is a constant. Hence, the expected returns $\expe_{(2,1)} \lrb{R_{D}^{(2)}}, \expe_{(2,1)} \lrb{R_{A}^{(2)}} $ are measurable with respect to the random state $\omega^{(1,2)}$. Hence, conditional on $\omega^{(1,2)}$, $\expe_{(2,1)} \lrb{R_{D}^{(2)}}, \expe_{(2,1)} \lrb{R_{A}^{(2)}}$ are known to liquidity providers. Hence, the liquidity providers' portfolio choice at sub-period 1 of period  2 can also be contingent on these values.

If $\expe_{(2,1)} \lrb{R_{D}^{(2)}}> \expe_{(2,1)} \lrb{R_{A}^{(2)}}$, then the unique optimal action of the liquidity providers is {to use} all their consumption good {to purchase} A and B tokens, in the ratio required by the AMM, and deposit them. 
If instead $\expe_{(2,1)} \lrb{R_{D}^{(2)}}< \expe_{(2,1)} \lrb{R_{A}^{(2)}}$, the unique optimal action of liquidity providers is to use all their consumption good to purchase A tokens and hold them without depositing. If $\expe_{(2,1)} \lrb{R_{D}^{(2)}} = \expe_{(2,1)} \lrb{R_{A}^{(2)}}$, then the liquidity providers are indifferent between depositing in the AMMs or holding A tokens only. By our tie-breaking rule given in Assumption \ref{asstiebreak}, the unique optimal action of the arbitrageur is then to exchange all their consumption good to A tokens, and hold them without depositing. 



We now consider period $1$. We first show that the optimal  gas fees that liquidity providers attach to their exit orders are zero. Accounting for liquidity provider $i$'s optimal action in period 2, {the submitted gas fee does not affect the ratio $\frac{e_i^{(2)}}{e_i^{(1)}}$, which is determined by the fair value exchange rate in the state $\omega^{(1,2)}$. That is, $\expe_{(1,2)} \lrb{\frac{e_i^{(2)}}{e_i^{(1)}}\big| g_{(lp,i)}^{(1)}} = \expe_{(1,2)} \lrb{\frac{e_i^{(2)}}{e_i^{(1)}}}$}. Submitting a nonzero gas fee $g_{(lp,i)}^{(1)} >0 $, would only decrease the amount of consumption good available after period 1 to liquidity provider $i$, $e_i^{(1)}(g_{(lp,i)}^{(1)})$, and thus decrease his expected payoff $\expe_{(1,2)} \lrb{e_i^{(2)} | g_{(lp,i)}^{(1)}} =    e_i^{(1)}(g_{(lp,i)}^{(1)})  \expe_{(1,2)} \lrb{\frac{e_i^{(2)}}{e_i^{(1)}}}$.  

By Lemma \ref{adverseselction order1}, if there exists an arbitrage opportunity at  sub-period 3 of period $1$, the optimal trading order is unique. By the optimality of arbitrageur's action in period $2$, its expected profit from the arbitrage order net of the  gas fee paid in period $2$ is zero. Hence, in period $1$, to maximize {its expected cumulative payoff} $\sum_{t=0}^2 \expe_{(1,2)} \lrb{p_B^{(t,2)}\Delta q_B^{(t,3)}+ p_A^{(t,2)}\Delta q_A^{(t,3)} - g_{arb}^{(t,3)}}$, the arbitrageur has to maximize its one-period payoff in period 1. By Assumption \ref{assarb}, the arbitrageur pays a gas fee equal to the profit from the optimal order. Consequently, the arbitrageur must choose the optimal trade order in period 1, otherwise it ends up with a negative profit in such period, which is sub-optimal. 
At sub-period 2 of period 1, by Lemma \ref{traderoptimization}, if an investor arrives, then it has to choose the unique trading order to maximize its surplus.

At sub-period 1, accounting for the optimal actions subsequently taken from all agents, liquidity provider $i$ has to choose his portfolio for period 1, $(w_i^{(1)}y_A^{(1,1)}, w_i^{(1)}y_B^{(1,1)},c_i^{(1,1)},x_{A_i}^{(1,1)},x_{B_i}^{(1,1)})$, {so to maximize} his expected payoff. Each liquidity provider $i$ chooses $c_i^{(1,1)} = 0$, as holding the consumption good is strictly dominated by holding either A or B tokens. Suppose liquidity provider $i$ chooses the portfolio  $(w_i^{(1)}y_A^{(1,1)}, w_i^{(1)}y_B^{(1,1)}, 0 ,x_{A_i}^{(1,1)},x_{B_i}^{(1,1)})$. By the law of total expectation, his expected payoff can then be written as:
\begin{align}\label{subgame1}
       &\expe_{(1,1)} \lrb{e_i^{(2)}}  \nonumber\\= &     e_i^{(0)}\expe_{(1,1)} \lrb{\frac{e_i^{(2)}}{e_i^{(1)}}\frac{e_i^{(1)}}{e_i^{(0)}}} \nonumber \\
        = &e_i^{(0)}  \sum_{\omega^{(1,2)} \in \Omega}  \prob_{(1,1)}(\omega^{(1,2)}) \ex{\frac{e_i^{(2)}}{e_i^{(1)}}\bigg|\omega^{(1,2)}} \frac{e_i^{(1)}}{e_i^{(0)}}(\omega^{(1,2)}) \nonumber \\
        = & e_i^{(0)} \sum_{\omega^{(1,2)} \in \Omega}  \prob_{(1,1)}(\omega^{(1,2)}) \ex{\frac{e_i^{(2)}}{e_i^{(1)}}\bigg|\omega^{(1,2)}} \bigg( \frac{w_i^{(1)} (p_A^{(1,1)} y_A^{(1,1)} + p_B^{(1,1)} y_B^{(1,1)}) }{e_i^{(0)}}*(R_{D}^{(1)}(\omega^{(1,2)})+1) \nonumber \\ +  & \frac{p_A^{(1,1)} x_{A_i}^{(1,1)} }{e_i^{(0)}}(R_{A}^{(1)}(\omega^{(1,2)})+1) + \frac{p_B^{(1,1)} x_{B_i}^{(1,1)} }{e_i^{(0)}}(R_{B}^{(1)}(\omega^{(1,2)})+1)\bigg),
\end{align}
where $\omega^{(1,2)} \in \Omega$ is the state at sub-period 2 of period 1.
$\ex{\frac{e_i^{(2)}}{e_i^{(1)}}\bigg|\omega^{(1,2)}}$
is then pinned down by the price ratio, $ \frac{p_A^{(1,2)}}{p_B^{(1,2)}}$, and thus by the state $\omega^{(1,2)} \in \Omega$. $\frac{w_i^{(1)} (p_A^{(1,1)} y_A^{(1,1)} + p_B^{(1,1)} y_B^{(1,1)}) }{e_i^{(0)}}, $ $\frac{p_A^{(1,1)} x_{A_i}^{(1,1)} }{e_i^{(0)}}, $ and $\frac{p_B^{(1,1)} x_{B_i}^{(1,1)} }{e_i^{(0)}}$ are the portfolio weights for the deposit, A tokens, and B tokens respectively. The one-period return from deposit in period 1 is: 
$$R_{D}^{(1)}(\omega^{(1,2)}) =: \frac{y_A^{(1,1)}}{(p_A^{(1,1)} y_A^{(1,1)} + p_B^{(1,1)} y_B^{(1,1)} )} \left(p_A^{(1,3)}\frac{y_A^{(1,3)}}{y_A^{(1,1)}}  + p_B^{(1,3)} \frac{y_B^{(1,3)}}{y_A^{(1,1)}}  \right) -1 ,$$ 

The one-period return from {holding A tokens and B tokens} in period 1 are: 
$$R_A^{(1)}(\omega^{(1,2)}) = \frac{p_A^{(1,2)}}{p_A^{(1,1)}}  -1, R_B^{(1)}(\omega^{(1,2)}) = \frac{p_B^{(1,2)}}{p_B^{(1,1)}}  -1.$$ The only randomness in period 1 is due to the events occurring at sub-period 2 of period 1. All returns are measurable respect to {the random state} $\omega^{(1,2)}$, and thus  they are all uniquely determined for {each realization of} $\omega^{(1,2)}$.

Accounting for the optimal actions subsequently taken by agents, the expected return of depositing, investing in A tokens, and  investing in B tokens in period 1 are: 
\begin{align}
    \expe_{(1,1)}\lrb{V_D} &=  \sum_{\omega^{(1,2)} \in \Omega} \prob_{(1,1)}(\omega^{(1,2)}) \ex{\frac{e_i^{(2)}}{e_i^{(1)}}\bigg|\omega^{(1,2)}} R_{D}^{(1)}(\omega^{(1,2)}), \label{V_D} \\
    \expe_{(1,1)}\lrb{V_A} &=  \sum_{\omega^{(1,2)} \in \Omega} \prob_{(1,1)}(\omega^{(1,2)}) \ex{\frac{e_i^{(2)}}{e_i^{(1)}}\bigg|\omega^{(1,2)}} R_{A}^{(1)}(\omega^{(1,2)}), \label{V_A} \\ 
    \expe_{(1,1)}\lrb{V_B} &=  \sum_{\omega^{(1,2)} \in \Omega} \prob_{(1,1)}(\omega^{(1,2)}) \ex{\frac{e_i^{(2)}}{e_i^{(1)}}\bigg|\omega^{(1,2)}}R_{B}^{(1)}(\omega^{(1,2)}). \label{V_B}
\end{align}

If $ \expe_{(1,1)}\lrb{V_D} > \max\left \{\expe_{(1,1)}\lrb{V_A}, \expe_{(1,1)}\lrb{V_B}\right \}$, then liquidity provider $i, i \in \mathcal{N}$, exchanges all its consumption good for A and B tokens in the ratio required by the AMM and deposits them. 

If $ \expe_{(1,1)}\lrb{V_D} \leq \max \left \{\expe_{(1,1)}\lrb{V_A}, \expe_{(1,1)}\lrb{V_B}\right \}$, then liquidity provider $i, i \in \mathcal{N}$, does not deposit, and  has to decide between holding A or B tokens.  If $\expe_{(1,1)}\lrb{V_A} > \expe_{(1,1)}\lrb{V_B}$, then he exchanges all its consumption good {for} A tokens and hold them.  If $\expe_{(1,1)}\lrb{V_A} < \expe_{(1,1)}\lrb{V_B}$, then he exchanges all its consumption good {for} B tokens and hold them.  if $\expe_{(1,1)}\lrb{V_A} = \expe_{(1,1)}\lrb{V_B}$, then any portfolio consisting of a mix of A and B tokens is optimal. 


Therefore, we can identify each player's optimal action at every sub-period of the backward induction algorithm. Hence, the strategy profile constructed through this process is a subgame perfect equilibrium of the game. Also, this equilibrium is unique {if} $\expe_{(1,1)}\lrb{V_A} \neq \expe_{(1,1)}\lrb{V_B}$ {or} $ \expe_{(1,1)}\lrb{V_D} > \max\left \{\expe_{(1,1)}\lrb{V_A}, \expe_{(1,1)}\lrb{V_B}\right \}$.

\end{proof}

\begin{proof}[Proof of Proposition  \ref{freezeequilibrium5}]
As shown in the proof of Proposition \ref{propostion unique equilibrium}, a liquidity provider will deposit in period 2 if and only if $\expe_{(2,1)} \lrb{R_{D}^{(2)}}> \expe_{(2,1)}\lrb{R_{A}^{(2)}}$. The expected one-period return of holding A tokens is $\expe_{(2,1)}\lrb{R_{A}^{(2)}} = (1-\theta)\kappa_1\beta+\theta \kappa_{com}\beta$. It remains to calculate the expected one-period return of depositing, given by $\expe_{(2,1)} \lrb{R_{D}^{(2)}}$.



We next analyze the events that may occur at sub-period 2 of period 2. With probability $\frac{\kappa_I}{2}$, a `` type A'' investor arrives to the AMM, and with probability $\frac{\kappa_I}{2}$, a `` type B'' investor arrives. With probability $\theta$, the prices of two tokens co-move, and there is no arbitrage opportunity. With probability $(1-\theta)\kappa_1(1-\kappa_2)$, the price of an A token increases and the price of a B token stays unchanged, which leads to an arbitrage opportunity for the arbitrageur. Similarly, with probability $(1-\theta)\kappa_2(1-\kappa_1)$, the price of a B token increases and the price of an A token stays unchanged, which again leads to an arbitrage opportunity. 

\paragraph{The return from deposits conditional on the occurrence of a common shock.} If the prices of A and B tokens co-move, the exchange rate stays unchanged, so there is no arbitrage opportunity. Thus, the return from deposits is $\beta$ if the common shock occurs, and $0$ if the common shock does not occur. 

\paragraph{The return from deposits conditional on the occurrence of an idiosyncratic shock.} If both A and B tokens are hit by an idiosyncratic shock, their prices co-move, and the return from deposits is $\beta$. If both A and B tokens are not hit by an idiosyncratic shock, the return is $0$. 

If only the A token is hit by an idiosyncratic shock, there exists an arbitrage opportunity. In this case, the return from deposits is
\begin{equation}
R_{arb_A}^{(2)} := \beta\frac{p_A^{(2,1)}y_A^{(2,1)}}{p_A^{(2,1)} y_A^{(2,1)} + p_B^{(2,1)} y_B^{(2,1)}} -\frac{\pi(y_A^{(2,1)},y_B^{(2,1)},p_B^{(1,2)}, (1+\beta)p_A^{(1,2)})}{p_A^{(2,1)} y_A^{(2,1)} + p_B^{(2,1)} y_B^{(2,1)}},
\label{eq:arba}
\end{equation}
where the first term is the return from B token's appreciation, and the second term is the arbitrage ratio, i.e., the token value loss divided by the initial value of deposit. Alternatively, if only B token is hit by an idiosyncratic shock, then there exists an arbitrage opportunity. In this case, the return of deposit is then:
\begin{equation}
R_{arb_B}^{(2)} :=\beta\frac{p_B^{(2,1)}y_B^{(2,1)}}{p_A^{(2,1)} y_A^{(2,1)} + p_B^{(2,1)} y_B^{(2,1)}} -\frac{\pi(y_A^{(2,1)},y_B^{(2,1)},(1+\beta)p_B^{(1,2)}, p_A^{(1,2)})}{p_A^{(2,1)} y_A^{(2,1)} + p_B^{(2,1)} y_B^{(2,1)}}.
\label{eq:arbb}
\end{equation}

We know from Lemma \ref{adverseselction order1} that $\frac{\partial \pi(y_A^{(2,1)},y_B^{(2,1)},(1+\beta)p_B^{(1,2)}, p_A^{(1,2)})}{\partial \beta } = p_B^{(2,1)} \Delta q_B^{(2,3)*}> 0, \frac{\partial \Delta q_B^{(2,3)*}}{\partial \beta } > 0$.
This implies that $\frac{\pi(y_A^{(2,1)},y_B^{(2,1)},(1+\beta)p_B^{(1,2)}, p_A^{(1,2)})}{p_A^{(2,1)} y_A^{(2,1)} + p_B^{(2,1)} y_B^{(2,1)}}$ increases in $\beta$. Moreover, for any $\beta > M > f$, we have
\begin{align*}
    \frac{\partial \pi(y_A^{(2,1)},y_B^{(2,1)},(1+\beta)p_B^{(1,2)}, p_A^{(1,2)})}{\partial \beta } &= p_B^{(2,1)} \Delta q_B^{(2,3)*}(y_A^{(2,1)},y_B^{(2,1)},(1+\beta)p_B^{(1,2)}, p_A^{(1,2)}) \\ &\geq p_B^{(2,1)} \Delta q_B^{(2,3)*}(y_A^{(2,1)},y_B^{(2,1)},(1+M)p_B^{(1,2)}, p_A^{(1,2)}) > 0,  
\end{align*} 
where we have used that $\frac{\Delta q_B^{(2,3)*}}{\partial \beta } > 0$.  Applying the above inequality, we then have:
\begin{align*}
    \pi(y_A^{(2,1)},y_B^{(2,1)},(1+\beta)p_B^{(1,2)}, p_A^{(1,2)}) &= \int_f^{\beta}\frac{\partial \pi(y_A^{(2,1)},y_B^{(2,1)},(1+\beta)p_B^{(1,2)}, p_A^{(1,2)})}{\partial \beta }  \\  &\geq \int_M^{\beta}\frac{\partial \pi(y_A^{(2,1)},y_B^{(2,1)},(1+\beta)p_B^{(1,2)}, p_A^{(1,2)})}{\partial \beta } \\ &\geq (\beta -M) \Delta q_B^{(2,3)*}(y_A^{(2,1)},y_B^{(2,1)},(1+M)p_B^{(1,2)}, p_A^{(1,2)}).
\end{align*}
As a result, $\frac{\pi(y_A^{(2,1)},y_B^{(2,1)},(1+\beta)p_B^{(1,2)}, p_A^{(1,2)})}{p_A^{(2,1)} y_A^{(2,1)} + p_B^{(2,1)} y_B^{(2,1)}}$ converges to $\infty$ when $\beta \rightarrow \infty.$ Following the same procedure, we can also show that  $\frac{\pi(y_A^{(2,1)},y_B^{(2,1)},p_B^{(1,2)}, (1+\beta)p_A^{(1,2)})}{p_A^{(2,1)} y_A^{(2,1)} + p_B^{(2,1)} y_B^{(2,1)}}$ increases in $\beta$ and  converges to $\infty$ when $\beta \rightarrow \infty.$

\paragraph{The return from deposits conditional on the investor's arrival.}We only consider the case where a ``type A'' investor arrives to the AMM. The case where  a ``type B'' investor arrives can be obtained following the same procedure.  By Lemma  \ref{traderoptimization}, the investor chooses its optimal trading sizes,  $\Delta Q_A^{(2,2)*}, \Delta Q_B^{(2,2)*}$. If the trading sizes are nonzero, then after the investor trades, the spot rate at which A tokens are exchanged for B tokens is higher than the fair value exchange rate: $$ \frac{ F_x}{F_y}\bigg\rvert_{(x,y) = (y_A^{(2,2)},y_B^{(2,2)})}  > \frac{(1+\alpha)p_A^{(2,1)}}{(1+f)p_B^{(2,1)}}  > \frac{p_A^{(2,1)}}{p_B^{(2,1)}} = \frac{p_A^{(2,2)}}{p_B^{(2,2)}}.$$ 

If $\frac{1}{1+f} \frac{ F_x}{F_y}\bigg\rvert_{(x,y) = (y_A^{(2,2)},y_B^{(2,2)})}  < \frac{p_A^{(2,1)}}{p_B^{(2,1)}}$, then there is no arbitrage opportunity after the investor trades. This is because for the arbitrageur, the marginal benefit of trading is lower than the trading cost. The return from the deposit in this case is
\begin{align} \label{return of fees1}
    \frac{p_A^{(2,3)}{y_A^{(2,3)}} + p_B^{(2,3)} {y_B^{(2,3)}}}{p_A^{(2,1)}{y_A^{(2,1)}} + p_B^{(2,1)} {y_B^{(2,1)}}} -1 &= \frac{p_A^{(2,1)}({y_A^{(2,1)}}-\Delta Q_A^{(2,2)*})  + p_B^{(2,1)} ({y_B^{(2,1)}}-(1+f)\Delta Q_B^{(2,2)*})}{p_A^{(2,1)}{y_A^{(2,1)}} + p_B^{(2,1)} {y_B^{(2,1)}}} - 1 \nonumber \\ &= -\frac{p_A^{(2,1)}\Delta Q_A^{(2,2)*}  + p_B^{(2,1)} (1+f)\Delta Q_B^{(2,2)*}}{p_A^{(2,1)}{y_A^{(2,1)}} + p_B^{(2,1)} {y_B^{(2,1)}}} \nonumber\\ 
     &= \frac{ \int_{\Delta Q_B^{(2,2)*}}^0 p_A^{(2,1)}(g^{(2,2)}(x))'  + p_B^{(2,1) } (1+f)dx}{p_A^{(2,1)}{y_A^{(2,1)}} + p_B^{(2,1)} {y_B^{(2,1)}}}>0.
\end{align}

We next show that the above expression increases in $\alpha$. The integrand $ p_A^{(2,1)}(g^{(2,2)}(x)')  + p_B^{(2,1) } ((1+f)) > 0$ for any $x <0 $ as the  marginal exchange rate $-g^{(2,2)}(x)' < \frac{p_B^{(2,1) } }{p_A^{(2,1)}}$. This implies that the expression in \eqref{return of fees1} increases in $|\Delta Q_B^{(2,2)*}|$. The integrand is invariant to $\alpha$, and  $|\Delta Q_B^{(2,2)*}|$ increases in $\alpha$ by Lemma \ref{traderoptimization}. Therefore, the return from deposits given in \eqref{return of fees1} increases in $\alpha$.

If $\frac{1}{1+f} \frac{ F_x}{F_y}\bigg\rvert_{(x,y) = (y_A^{(2,2)},y_B^{(2,2)})}  > \frac{p_A^{(2,1)}}{p_B^{(2,1)}}$, then  there exists an arbitrage opportunity because the marginal benefit of trading is larger than the marginal cost. The arbitrageur chooses the trading sizes $\Delta q_A^{(2,3)*}, \Delta q_B^{(2,3)*}$ such that the marginal benefit breaks even with the marginal cost:
\begin{equation}\label{breakeven condition arbitrageur}
    \frac{1}{1+f} \frac{ F_x}{F_y}\bigg\rvert_{(x,y) = (y_A^{(2,2)}-\Delta q_A^{(2,3)*},y_B^{(2,2)}- \Delta q_B^{(2,3)*})}  = \frac{p_A^{(2,1)}}{p_B^{(2,1)}},
\end{equation}
where $\Delta q_A^{(2,3)*}, \Delta q_B^{(2,3)*}$ satisfies the constraint, $F(y_A^{(2,2)}-\Delta q_A^{(2,3)*}, y_B^{(2,2)}- \Delta q_B^{(2,3)*}) = F(y_A^{(2,2)}, {y_B^{(2,2)}})$, and \eqref{breakeven condition arbitrageur} pins down the ratio of A to B tokens, $ \frac{y_A^{(2,2)}-\Delta q_A^{(2,3)*}}{y_B^{(2,2)}- \Delta q_B^{(2,3)*}}$. 


We next analyze the sensitivity of the function  $F(y_A^{(2,2)}-\Delta q_A^{(2,3)*}, y_B^{(2,2)}- \Delta q_B^{(2,3)*})$  to $\alpha$:
\begin{align*}
  \frac{d F(y_A^{(2,2)}-\Delta q_A^{(2,3)*}, y_B^{(2,2)}- \Delta q_B^{(2,3)*})}{d \alpha}  &= \frac{d F(y_A^{(2,2)}, {y_B^{(2,2)}})}{d \alpha}\\ &=\frac{d F(y_A^{(2,1)}-\Delta Q_A^{(2,2)*}, {y_B^{(2,1)}}-(1+f)\Delta Q_B^{(2,2)*})}{d \alpha}\\ &=- F_x\frac{d \Delta Q_A^{(2,2)*}}{d \alpha} - (1+f) F_y  \frac{d \Delta Q_B^{(2,2)*}}{d \alpha} \\  &= - f F_y  \frac{d \Delta Q_B^{(2,2)*}}{d \alpha} > 0,
\end{align*}
where the last inequality follows from Lemma \ref{traderoptimization}, and we have also used the condition $- F_x\frac{d \Delta Q_A^{(2,2)*}}{d \alpha} - F_y  \frac{d \Delta Q_B^{(2,2)*}}{d \alpha} = 0 $ which is achieved by differentiating the equation $F(y_A^{(2,1)}-\Delta Q_A^{(2,2)*}, {y_B^{(2,1)}}-\Delta Q_B^{(2,2)*}) = F(y_A^{(2,1)}, {y_B^{(2,1)}})$ on both sides with respect to $\alpha$.

Hence, we have shown that $F(y_A^{(2,2)}-\Delta q_A^{(2,3)*}, y_B^{(2,2)}- \Delta q_B^{(2,3)*})$ is increasing in $\alpha$. Thus, by the third property of $F$ stated in Assumption \ref{curve} and the fact that the ratio of A and B tokens, $ \frac{y_A^{(2,2)}-\Delta q_A^{(2,3)*}}{y_B^{(2,2)}- \Delta q_B^{(2,3)*}}$ is already determined, we know that {both} $y_A^{(2,2)}-\Delta q_A^{(2,3)*}$ {and} $y_B^{(2,2)}- \Delta q_B^{(2,3)*}$ {are increasing in} $\alpha$. 

We then show that $-q_A^{(2,3)*}$ increases in $\alpha$. The quantity $-q_A^{(2,3)*}$ is pinned down by \eqref{breakeven condition arbitrageur} and the constraint $F(y_A^{(2,2)}-\Delta q_A^{(2,3)*}, y_B^{(2,2)}- \Delta q_B^{(2,3)*}) = F(y_A^{(2,2)}, {y_B^{(2,2)}})$. Differentiating \eqref{breakeven condition arbitrageur}  with respect to $\alpha$, we have:
\begin{align*}
    &\frac{1}{1+f} F_{xx} \left(\frac{d y_A^{(2,2)}}{  d \alpha} +     \frac{d (-\Delta q_A^{(2,3)*)}}{d  \alpha}  \right) +\frac{1}{1+f} F_{xy} \left(\frac{d y_B^{(2,2)}}{  d \alpha} +     \frac{d (-\Delta q_B^{(2,3)*)}}{d  \alpha} \right) \\ &-  \frac{p_A^{(2,1)}}{p_B^{(2,1)}} \left( F_{yy} \left(\frac{d y_B^{(2,2)}}{  d \alpha} +  \frac{d (-\Delta q_B^{(2,3)*)}}{d  \alpha}  \right) -  F_{xy} \left(\frac{d y_A^{(2,2)}}{  d \alpha} +     \frac{d (-\Delta q_A^{(2,3)*)}}{d  \alpha}  \right)\right) = 0 .
\end{align*}
Recall that $y_A^{(2,2)} = (y_A^{(2,1)}-\Delta Q_A^{(2,2)*})$, $y_B^{(2,2)} = {y_B^{(2,1)}}-(1+f)\Delta Q_B^{(2,2)*} $. We can then rewrite the equation above as
\begin{align*}
    & \left(\frac{1}{1+f} F_{xx} - \frac{p_A^{(2,1)}}{p_B^{(2,1)}} F_{xy} \right)\frac{d (y_A^{(2,1)}-\Delta Q_A^{(2,2)*})}{  d \alpha} +    \left(\frac{1}{1+f} F_{xy} - \frac{p_A^{(2,1)}}{p_B^{(2,1)}} F_{yy} \right)\frac{d ( {y_B^{(2,1)}}-(1+f)\Delta Q_B^{(2,2)*})}{  d \alpha} \\ &+  \left(\frac{1}{1+f} F_{xx} - \frac{p_A^{(2,1)}}{p_B^{(2,1)}} F_{xy} \right)\frac{d (-\Delta q_A^{(2,3)*})}{  d \alpha} +  \left(\frac{1}{1+f} F_{xy} - \frac{p_A^{(2,1)}}{p_B^{(2,1)}} F_{yy} \right) \frac{d (-\Delta q_B^{(2,3)*})}{  d (-\Delta q_A^{(2,3)*})} \frac{d (-\Delta q_A^{(2,3)*})}{  d \alpha}  = 0. 
\end{align*}
By Assumption \ref{curve}, $F_{xx}<0, F_{xy}>0, F_{yy}<0$; by Lemma \ref{traderoptimization}, we have $\frac{d (y_A^{(2,1)}-\Delta Q_A^{(2,2)*})}{  d \alpha} > 0,$ and $ \frac{d ( {y_B^{(2,1)}}-(1+f)\Delta Q_B^{(2,2)*})}{  d \alpha} < 0$. By the constraint $F(y_A^{(2,2)}-\Delta q_A^{(2,3)*}, y_B^{(2,2)}- \Delta q_B^{(2,3)*}) = F(y_A^{(2,2)}, {y_B^{(2,2)}})$, we have $\frac{d (-\Delta q_B^{(2,3)*})}{  d (-\Delta q_A^{(2,3)*})} < 0$. Combining those conditions, we conclude that $\frac{d (-\Delta q_A^{(2,3)*})}{  d \alpha} > 0$.



For the case where arbitrage occurs after the investor trades, the return from deposits is:
\begin{align} \label{return of fees2}
    &\frac{p_A^{(2,3)}{y_A^{(2,3)}} + p_B^{(2,3)} {y_B^{(2,3)}}}{p_A^{(2,1)}{y_A^{(2,1)}} + p_B^{(2,1)} {y_B^{(2,1)}}} -1 \nonumber \\ = &\frac{p_A^{(2,1)}({y_A^{(2,1)}}-\Delta Q_A^{(2,2)*}- (1+f)\Delta q_A^{(2,3)*})  + p_B^{(2,1)} ({y_B^{(2,1)}}-(1+f)\Delta Q_B^{(2,2)*}- \Delta q_B^{(2,3)*})}{p_A^{(2,1)}{y_A^{(2,1)}} + p_B^{(2,1)} {y_B^{(2,1)}}} - 1 \nonumber \\
    = &\frac{p_A^{(2,1)}({y_A^{(2,1)}}-\Delta Q_A^{(2,2)*}- \Delta q_A^{(2,3)*})  + p_B^{(2,1)} ({y_B^{(2,1)}}-(1+f)\Delta Q_B^{(2,2)*}- \Delta q_B^{(2,3)*})}{p_A^{(2,1)}{y_A^{(2,1)}} + p_B^{(2,1)} {y_B^{(2,1)}}} \nonumber \\ +&\frac{p_A^{(2,1)}(-f \Delta q_A^{(2,3)*}) }{p_A^{(2,1)}{y_A^{(2,1)}} + p_B^{(2,1)} {y_B^{(2,1)}}} - 1,
\end{align}
where the first equality follows from the conditions $y_A^{(2,3)} = {y_A^{(2,1)}}-\Delta Q_A^{(2,2)*}- (1+f)\Delta q_A^{(2,3)*} $ and $y_B^{(2,3)} = {y_B^{(2,1)}}-\Delta Q_B^{(2,2)*}- (1+f)\Delta q_B^{(2,3)*}$. Those two conditions reflects the fact that the deposit in the AMM is only altered by the investor's trade and the arbitrageur's trade.  Recall we have shown that both the first and the second term of \eqref{return of fees2} is increasing in $\alpha$, and thus the return from deposits when the investor arrives is increasing in $\alpha$. 

We next show that the first term in \eqref{return of fees2}, given by $\frac{p_A^{(2,1)}({y_A^{(2,2)}}- \Delta q_A^{(2,3)*})  + p_B^{(2,1)} ({y_B^{(2,2)}}- \Delta q_B^{(2,3)*})}{p_A^{(2,1)}{y_A^{(2,1)}} + p_B^{(2,1)} {y_B^{(2,1)}}} = \frac{p_A^{(2,1)}({y_A^{(2,1)}}-\Delta Q_A^{(2,2)*}- \Delta q_A^{(2,3)*})  + p_B^{(2,1)} ({y_B^{(2,1)}}-(1+f)\Delta Q_B^{(2,2)*}- \Delta q_B^{(2,3)*})}{p_A^{(2,1)}{y_A^{(2,1)}} + p_B^{(2,1)} {y_B^{(2,1)}}} $, is greater than 1. Since the AMM requires the liquidity providers to deposit at the fair value exchange rate, the condition  $ \frac{ F_x}{F_y}\bigg\rvert_{(x,y) = (y_A^{(2,1)},y_B^{(2,1)})} = \frac{p_A^{(2,1)}}{p_B^{(2,1)}}$ must hold. We also have $ \frac{ F_x}{F_y}\bigg\rvert_{(x,y) = (y_A^{(2,2)}-\Delta q_A^{(2,3)*},y_B^{(2,2)}- \Delta q_B^{(2,3)*})} \neq \frac{p_A^{(2,1)}}{p_B^{(2,1)}}$ by  \eqref{breakeven condition arbitrageur}. Moreover, we  have the following inequality: 
\begin{align*}
    F(y_A^{(2,2)}-\Delta q_A^{(2,3)*},y_B^{(2,2)}- \Delta q_B^{(2,3)*}) = F(y_A^{(2,2)},y_B^{(2,2)}) &= F(y_A^{(2,1)}-\Delta Q_A^{(2,2)*}, {y_B^{(2,1)}}-(1+f)\Delta Q_B^{(2,2)*})\\ & > F(y_A^{(2,1)}-\Delta Q_A^{(2,2)*}, {y_B^{(2,1)}}-\Delta Q_B^{(2,2)*})\\ & = F(y_A^{(2,1)}, {y_B^{(2,1)}}),
\end{align*}
where the inequality follows from $F_y>0$ and $\Delta Q_B^{(2,2)*} < 0,$. The equalities hold because the value of the pricing function, without accounting for the fee, must stay unchanged for each transaction. The above result suggests that the value of pricing function increases after the investors and traders trade. 

By Assumption \ref{curve}, we can write $ F(y_A^{(2,2)}-\Delta q_A^{(2,3)*},y_B^{(2,2)}- \Delta q_B^{(2,3)*})$ as $ c^l F(\frac{y_A^{(2,2)}-\Delta q_A^{(2,3)*}}{c},\frac{y_B^{(2,2)}- \Delta q_B^{(2,3)*}}{c})$ where $c>1,l>0, F(\frac{y_A^{(2,2)}-\Delta q_A^{(2,3)*}}{c},\frac{y_B^{(2,2)}- \Delta q_B^{(2,3)*}}{c}) = F(y_A^{(2,1)}, {y_B^{(2,1)}})$. It follows from Lemma \ref{lowest value on curve1} that
\begin{align*}
    p_A^{(2,1)}{y_A^{(2,1)}} + p_B^{(2,1)} {y_B^{(2,1)}} &\leq  p_A^{(2,1)}{\left(\frac{y_A^{(2,2)}-\Delta q_A^{(2,3)*}}{c}\right)} + p_B^{(2,1)} {\left(\frac{y_B^{(2,2)}- \Delta q_B^{(2,3)*}}{c}\right)} \\
    &< p_A^{(2,1)}{({y_A^{(2,2)}-\Delta q_A^{(2,3)*}})} + p_B^{(2,1)} {({y_B^{(2,2)}- \Delta q_B^{(2,3)*}})},  
\end{align*}
which means that the first term in \eqref{return of fees2}  is greater than 1. Moreover, since $\Delta q_A^{(2,3)*} <0,$ we know that the second term, $\frac{p_A^{(2,1)}(-f \Delta q_A^{(2,3)*}) }{p_A^{(2,1)}{y_A^{(2,1)}} + p_B^{(2,1)} {y_B^{(2,1)}}}$, is positive.  It then follows that the return from deposits stated in \eqref{return of fees2} is positive.

We denote the return from deposit when ``type A '' investors and `` type B '' investors arrive as $R_{inv_A}^{(2)}$ and $R_{inv_B}^{(2)}$ respectively. Using similar arguments as above, we can conclude that they are both increasing in $\alpha$. By Lemma \ref{proportionality2}, they are independent of the amount of deposits and only depend of the token price ratio at the start of the period, $\frac{p_A^{(2,1)}}{p_B^{(2,1)}}= \frac{p_A^{(1,2)}}{p_B^{(1,2)}}$.

\paragraph{Liquidity freeze threshold. } We  compare the expectation of the one-period return from depositing at period 2, $\expe_{(2,1)} \lrb{R_{D}^{(2)}}$, with the expectation of the one-period return from holding A tokens at period 2, $\expe_{(2,1)}\lrb{R_{A}^{(2)}}$.  By the law of total expectation, we have the following expressions: 
\begin{align}\label{total probability}
    &\expe_{(2,1)} \lrb{R_{D}^{(2)}}- \expe_{(2,1)}\lrb{R_{A}^{(2)}} \nonumber\\
    =& \frac{\kappa_I}{2}R_{inv_A}^{(2)} + \frac{\kappa_I}{2}R_{inv_B}^{(2)}+ (1-\theta)\left( \kappa_1(1-\kappa_2)R_{arb_A}^{(2)} +\kappa_2(1-\kappa_1)R_{arb_B}^{(2)} + \kappa_1\kappa_2 \beta\right)  \nonumber\\ +& \theta\kappa_{com} \beta -((1-\theta)\kappa_1+\theta \kappa_{com}) \beta \nonumber\\
    =& \frac{\kappa_I}{2}R_{inv_A}^{(2)} + \frac{\kappa_I}{2}R_{inv_B}^{(2)} - (1-\theta)\bigg( \kappa_1(1-\kappa_2) \frac{\pi(y_A^{(2,1)},y_B^{(2,1)},p_B^{(1,2)}, (1+\beta)p_A^{(1,2)})}{p_A^{(2,1)} y_A^{(2,1)} + p_B^{(2,1)}y_B^{(2,1)}} \nonumber \\ +&\kappa_2(1-\kappa_1)\frac{\pi(y_A^{(2,1)},y_B^{(2,1)},(1+\beta)p_B^{(1,2)}, p_A^{(1,2)})}{p_A^{(2,1)} y_A^{(2,1)} + p_B^{(2,1)}y_B^{(2,1)}} \bigg)-  (1-\theta)(\kappa_1-\kappa_2) \beta \frac{p_B^{(2,1)}y_B^{(2,1)}}{p_A^{(2,1)} y_A^{(2,1)} + p_B^{(2,1)} y_B^{(2,1)}},
\end{align}
where, {in the last expression}, we have used the expressions of $R_{arb_A}^{(2)}$ and $R_{arb_B}^{(2)}$ given, respectively in equations~\eqref{eq:arba} and~\eqref{eq:arbb}. The first two terms in the last expression of Eq.~\eqref{total probability} are the expected returns from the fee revenue, the third term is the expected token value loss due to arbitrage, and the last term is the opportunity cost of holding B tokens. The first two terms are non-negative, and the remaining terms are non-positive. All of them are independent of the amount of deposits and pinned down by the token price ratio at the start of the period, $\frac{p_A^{(2,1)}}{p_B^{(2,1)}}= \frac{p_A^{(1,2)}}{p_B^{(1,2)}}$. If \eqref{total probability} is positive, then the liquidity providers deposit their tokens. 

If $\frac{\kappa_I}{2}R_{inv_A}^{(2)} + \frac{\kappa_I}{2}R_{inv_B}^{(2)} = 0$, then \eqref{total probability} is non-positive, so there exists liquidity freeze for any $\beta \geq 0$.

Next, consider the case $\frac{\kappa_I}{2}R_{inv_A}^{(2)} + \frac{\kappa_I}{2}R_{inv_B}^{(2)} > 0$. If $\theta = 1$, that is, the token prices always co-move, then there does not exist a liquidity freeze for any $\beta \in [0, \infty)$. This is because \eqref{total probability} is positive. If  $\theta < 1$, then \eqref{total probability} is positive when $\beta \rightarrow 0 $, and negative when $\beta \rightarrow \infty $. Since the first two terms are independent of $\beta$, while the last three terms decrease in $\beta$, the last expression in \eqref{total probability} is monotonically decreasing in $\beta$. It is also continuous in $\beta$ because all terms are differentiable respect to $\beta$. Therefore, there exists a threshold 
$\beta_{frz}$ such that \eqref{total probability} is nonpositive if and only if $\beta > \beta_{frz}$. Since \eqref{total probability} is pinned down by the ratio $\frac{p_A^{(1,2)}}{p_B^{(1,2)}}$, so is the threshold $\beta_{frz}$. 

Therefore, there exists a critical threshold $\beta_{frz}\left(\frac{p_A^{(1,2)}}{p_B^{(1,2)}}\right)$, where $\beta_{frz}: \mathbb{R} \rightarrow \mathbb{R} \cup \{+\infty\}$ is a function of the fair token exchange rate at the start of period 2, such that a ``liquidity freeze'' occurs in period 2 if and only if $\beta \geq \beta_{frz}\left(\frac{p_A^{(1,2)}}{p_B^{(1,2)}}\right)$.

There exist three possible realizations of $\frac{p_A^{(1,2)}}{p_B^{(1,2)}}$ in period 2. These are $\frac{p_A^{(0)}}{p_B^{(0)}}, \frac{(1+\beta)p_A^{(0)}}{p_B^{(0)}}, \frac{p_A^{(0)}}{(1+\beta)p_B^{(0)}}$, {where $p_A^{(0)}, p_B^{(0)}$ are the token price at the initial state $\omega^{(0)}$}. A ``liquidity freeze'' occurs surely in period 2 if and only if $$\beta \geq \overline{\beta_{frz}} := \max \left \{\beta_{frz}\left({\frac{p_A^{(0)}}{p_B^{(0)}}}\right),\beta_{frz}\left(\frac{(1+\beta)p_A^{(0)}}{p_B^{(0)}}\right),\beta_{frz}\left( \frac{p_A^{(0)}}{(1+\beta)p_B^{(0)}}\right) \right \}.$$

We then show that ``Liquidity freeze'' occurs in period 1 if $\beta \geq \overline{\beta_{frz}}.$ If $\beta \geq \overline{\beta_{frz}}$, then ``Liquidity freeze'' occurs surely in period 2, and $\ex{\frac{e_i^{(2)}}{e_i^{(1)}}\bigg|\omega^{(1,2)}} = \ex{1+R_{A}^{(2)}}= 1+ ((1-\theta)\kappa_1+\theta \kappa_{com}) \beta$ for any $\omega^{(1,2)} \in \Omega$. Plugging these expressions into \eqref{V_D} and \eqref{V_A},  we obtain 
$$ \expe_{(1,1)}\lrb{V_D} = \expe_{(1,1)} \lrb{R_{D}^{(1)}} \left(1+\ex{R_{A}^{(2)}}\right)= \expe_{(2,1)} \lrb{R_{D}^{(2)}\bigg| \frac{p_A^{(1,2)}}{p_B^{(1,2)}} = 
\frac{p_A^{(0)}}{p_B^{(0)}}} \left(1+ \ex{R_{A}^{(2)}} \right),$$  where $\expe_{(1,1)} \lrb{R_{D}^{(1)}} = \expe_{(2,1)} \lrb{R_{D}^{(2)}\bigg| \frac{p_A^{(1,2)}}{p_B^{(1,2)}}= 
\frac{p_A^{(0)}}{p_B^{(0)}}}$ because if the initial fair value exchange rate, $\frac{p_A^{(0)}}{p_B^{(0)}}$ is the same as the fair value exchange rate at sub-period 2 of period 1, $\frac{p_A^{(1,2)}}{p_B^{(1,2)}}$, then the expected one-period return of depositing is the same for both periods. 
$$ \expe_{(1,1)}\lrb{V_A} = \ex{R_{A}^{(1)}}(1+\ex{R_{A}^{(2)}}) =  \ex{R_{A}^{(2)}}(1+\ex{R_{A}^{(2)}}). $$

Since $\beta \geq \overline{\beta_{frz}} \geq \beta_{frz}\left(\frac{p_A^{(0)}}{p_B^{(0)}}\right)$, we have $\ex{R_{A}^{(2)}} \geq   \expe_{(2,1)} \lrb{R_{D}^{(2)}\bigg| \frac{p_A^{(1,2)}}{p_B^{(1,2)}} = 
\frac{p_A^{(0)}}{p_B^{(0)}}}$. This means that $ \expe_{(1,1)}\lrb{V_D} \leq  \expe_{(1,1)}\lrb{V_A} $, i.e.,  
 ``Liquidity freeze'' also occurs in period 1. 
 Therefore, ``Liquidity freeze'' occurs surely in both period 1 and 2 if and only if $\beta \geq \overline{\beta_{frz}}.$
 
 \paragraph{Comparative Statics.} We show that for any arbitrary $\frac{p_A^{(1,2)}}{p_B^{(1,2)}}$,  $\beta_{frz}\left(\frac{p_A^{(1,2)}}{p_B^{(1,2)}}\right)$ increases in $\alpha, \kappa_I, \theta, \kappa_2 $, and decreases in $ \kappa_1.$ 
 
 For the case $\frac{\kappa_I}{2}R_{inv_A}^{(2)} + \frac{\kappa_I}{2}R_{inv_B}^{(2)} = 0$, $\beta_{frz}\left(\frac{p_A^{(1,2)}}{p_B^{(1,2)}}\right) = 0$, which is invariant of all the parameters locally. For the case $\frac{\kappa_I}{2}R_{inv_A}^{(2)} + \frac{\kappa_I}{2}R_{inv_B}^{(2)} > 0$ and $\theta = 1$, $\beta_{frz}\left(\frac{p_A^{(1,2)}}{p_B^{(1,2)}}\right) = +\infty $, which is also invariant of all other  parameters except $\theta$ locally. 
 
 We then consider the case $\frac{\kappa_I}{2}R_{inv_A}^{(2)} + \frac{\kappa_I}{2}R_{inv_B}^{(2)} > 0$   and $\theta < 1$. $\beta_{frz}\left(\frac{p_A^{(1,2)}}{p_B^{(1,2)}}\right)$ is defined by the equation: $\expe_{(2,1)} \lrb{R_{D}^{(2)}- R_{A}^{(2)}} = 0$. Using the relation~\eqref{total probability}, we  take the partial derivative of $\expe_{(2,1)} \lrb{R_{D}^{(2)}- R_{A}^{(2)}} $  with respect to $\alpha, \kappa_I, \theta$, and obtain
 \begin{align}
     \frac{\partial \expe_{(2,1)} \lrb{R_{D}^{(2)}- R_{A}^{(2)}}}{\partial\alpha} &= \frac{\kappa_I}{2}\left(\frac{\partial  R_{inv_A}^{(2)}}{\partial\alpha} +\frac{\partial R_{inv_B}^{(2)}}{\partial\alpha} \right) >  0,  \\
     \frac{\partial \expe_{(2,1)} \lrb{R_{D}^{(2)}- R_{A}^{(2)}}}{\partial\kappa_I} &= \frac{1}{2}R_{inv_A}^{(2)} + \frac{1}{2}R_{inv_B}^{(2)} > 0, \\
     \frac{\partial \expe_{(2,1)} \lrb{R_{D}^{(2)}- R_{A}^{(2)}}}{\partial \theta} &=  - \bigg( \kappa_1(1-\kappa_2) \frac{\pi(y_A^{(2,1)},y_B^{(2,1)},p_B^{(1,2)}, (1+\beta)p_A^{(1,2)})}{p_A^{(2,1)} y_A^{(2,1)} + p_B^{(2,1)}y_B^{(2,1)}} \nonumber \\ & +\kappa_2(1-\kappa_1)\frac{\pi(y_A^{(2,1)},y_B^{(2,1)},(1+\beta)p_B^{(1,2)}, p_A^{(1,2)})}{p_A^{(2,1)} y_A^{(2,1)} p_B^{(2,1)}y_B^{(2,1)}} + (\kappa_1-\kappa_2) \beta \frac{p_B^{(2,1)}y_B^{(2,1)}}{p_A^{(2,1)} y_A^{(2,1)} + p_B^{(2,1)} y_B^{(2,1)}} \bigg) \nonumber \\ &> 0. 
 \end{align}
Recall that $\frac{\partial \pi(y_A^{(2,1)},y_B^{(2,1)},p_B^{(1,2)}, (1+\beta)p_A^{(1,2)})}{\partial \beta}>0, \frac{\partial \pi(y_A^{(2,1)},y_B^{(2,1)},p_B^{(1,2)}, (1+\beta)p_A^{(1,2)})}{\partial \beta}>0$ from Lemma \ref{adverse selection does happen3}, so we have
 \begin{align*}
     \frac{\partial \expe_{(2,1)} \lrb{R_{D}^{(2)}- R_{A}^{(2)}}}{\partial \beta} &= - \frac{(1-\theta)}{p_A^{(2,1)} y_A^{(2,1)} + p_B^{(2,1)} y_B^{(2,1)}}\bigg( \kappa_1(1-\kappa_2) \frac{\partial \pi(y_A^{(2,1)},y_B^{(2,1)},p_B^{(1,2)}, (1+\beta)p_A^{(1,2)})}{\partial \beta}\\ +&\kappa_2(1-\kappa_1)\frac{\partial \pi(y_A^{(2,1)},y_B^{(2,1)},(1+\beta)p_B^{(1,2)}, p_A^{(1,2)})}{\partial \beta} +(\kappa_1-\kappa_2)  ({p_B^{(2,1)}y_B^{(2,1)}})\bigg)<0.
 \end{align*}
 Thus, applying the Implicit Function Theorem to the curve $\expe_{(2,1)} \lrb{R_{D}^{(2)}- R_{A}^{(2)}} = 0$, we deduce $\frac{\partial \beta_{frz}\left(\frac{p_A^{(1,2)}}{p_B^{(1,2)}}\right)}{\partial \alpha} > 0, \frac{\partial \beta_{frz}\left(\frac{p_A^{(1,2)}}{p_B^{(1,2)}}\right)}{\partial \kappa_I} > 0, \frac{\partial \beta_{frz}\left(\frac{p_A^{(1,2)}}{p_B^{(1,2)}}\right)}{\partial \theta} > 0.$
 
 We then consider the partial derivative of $\expe_{(2,1)} \lrb{R_{D}^{(2)}- R_{A}^{(2)}}$ with respect to $\kappa_1$ and $\kappa_2$.
 \begin{align}
      \frac{\partial \expe_{(2,1)} \lrb{R_{D}^{(2)}- R_{A}^{(2)}}}{\partial \kappa_1} &= - (1-\theta)\bigg( (1-\kappa_2) \frac{\pi(y_A^{(2,1)},y_B^{(2,1)},p_B^{(1,2)}, (1+\beta)p_A^{(1,2)})}{p_A^{(2,1)} y_A^{(2,1)} + p_B^{(2,1)}y_B^{(2,1)}} \nonumber \\ -&\kappa_2\frac{\pi(y_A^{(2,1)},y_B^{(2,1)},(1+\beta)p_B^{(1,2)}, p_A^{(1,2)})}{p_A^{(2,1)} y_A^{(2,1)} + p_B^{(2,1)}y_B^{(2,1)}} + \beta \frac{p_B^{(2,1)}y_B^{(2,1)}}{p_A^{(2,1)} y_A^{(2,1)} + p_B^{(2,1)} y_B^{(2,1)}} \bigg) < 0,
 \end{align}
 where we have used the inequality $0 \leq \pi(y_A^{(2,1)},y_B^{(2,1)},(1+\beta)p_B^{(1,2)}, p_A^{(1,2)}) \leq (\beta - f) p_B^{(2,1)} y_B^{(2,1)}. $ This inequality can be seen as follows:
      \begin{align*}
     \pi(y_A^{(2,1)},y_B^{(2,1)},(1+\beta)p_B^{(1,2)}, p_A^{(1,2)}) &=  p_A^{(2,1)} (1+f) \Delta q_A^{(2,3)*} + p_B^{(2,1) } (1+\beta) \Delta q_B^{(2,3)*} \\
     &\leq  -p_A^{(2,1)} (1+f) \frac{p_B^{(2,1) }}{ p_A^{(2,1)} } \Delta q_B^{(2,3)*} + p_B^{(2,1) } (1+\beta) \Delta q_B^{(2,3)*}  \\
     &=(\beta - f) p_B^{(2,1)} \Delta q_B^{(2,3)*} \\
     &<  (\beta - f) p_B^{(2,1)}   y_B^{(2,1)}.
     \end{align*}
     In the above expressions, the first inequality holds since  the marginal exchange rate from an A token to a B token is smaller than the spot exchange rate $\frac{p_A^{(1,2)}}{p_B^{(1,2)}}$, so the marginal cost for each extra B token acquired is larger than $(1+f) {p_B^{(2,1) }}$ for the arbitrageur. The second inequality holds because the amount of B tokens the arbitrageur can withdraw from the AMM is smaller than the total amount of B tokens deposited in the AMM. 


 It follows from a similar calculation that $\frac{\partial \expe_{(2,1)} \lrb{R_{D}^{(2)}- R_{A}^{(2)}}}{\partial \kappa_2}> 0$. 
By the Implicit Function Theorem, we have $\frac{\partial \beta_{frz}\left(\frac{p_A^{(1,2)}}{p_B^{(1,2)}}\right)}{\partial \kappa_1} < 0, \frac{\partial \beta_{frz}\left(\frac{p_A^{(1,2)}}{p_B^{(1,2)}}\right)}{\partial \kappa_2} > 0.$

 {As}
 $\overline{\beta_{frz}} =\max \left \{\beta_{frz}\left({\frac{p_A^{(1,2)}}{p_B^{(1,2)}}}\right),\beta_{frz}\left(\frac{(1+\beta)p_A^{(0)}}{p_B^{(0)}}\right),\beta_{frz}\left( \frac{p_A^{(0)}}{(1+\beta)p_B^{(0)}}\right) \right \}$, {we deduce that} $\overline{\beta_{frz}} $ also increases in $\alpha, \kappa_I, \theta , \kappa_2$, and decreases in $ \kappa_1$.
 
\end{proof}
\begin{mylemma} \label{lemma impermenent loss}
The ``impermanent loss'' increases in the difference between initial fair value exchange rate and the fair value exchange rate at the end of investment horizon, i.e., $\frac{\partial  IL\left(\frac{p_A^{(0)}}{p_B^{(0)}},\frac{p_A^{(0)}}{(1+\beta)p_B^{(0)}} \right) }{\partial \beta} > 0$, and $\frac{\partial  IL\left(\frac{p_A^{(0)}}{p_B^{(0)}},\frac{(1+\beta)p_A^{(0)}}{p_B^{(0)}} \right) }{\partial \beta} > 0$.

\end{mylemma}

\begin{proof}[Proof of Lemma  \ref{lemma impermenent loss}]
We only prove $\frac{\partial  IL\left(\frac{p_A^{(0)}}{p_B^{(0)}},\frac{p_A^{(0)}}{(1+\beta)p_B^{(0)}} \right) }{\partial \beta} > 0$. The inequality $\frac{\partial  IL\left(\frac{p_A^{(0)}}{p_B^{(0)}},\frac{(1+\beta)p_A^{(0)}}{p_B^{(0)}} \right) }{\partial \beta} > 0$ can be proven by following the same procedure. First, recall that $IL\left(\frac{p_A^{(0)}}{p_B^{(0)}},\frac{p_A^{(0)}}{(1+\beta)p_B^{(0)}} \right) $ is defined as:
\begin{equation}\label{impermanent loss2}
 IL\left(\frac{p_A^{(0)}}{p_B^{(0)}}, \frac{p_A^{(0)}}{(1+\beta)p_B^{(0)}} \right) := 1- \frac{p_A^{(0)}x_2 + (1+\beta)p_B^{(0)}y_2}{p_A^{(0)}x_1 + (1+\beta)p_B^{(0)}y_1},
  \end{equation}
where $x_1,y_1,x_2,y_2>0$ are specified by the following constraints: $$F(x_1,y_1)=F(x_2,y_2), \frac{F_x(x_1,y_1)}{F_y(x_1,y_1)} = \frac{p_A^{(0)}}{p_B^{(0)}}, \frac{F_x(x_2,y_2)}{F_y(x_2,y_2)} = \frac{p_A^{(0)}}{(1+\beta)p_B^{(0)}}.$$

Note that  $x_1,y_1$ do not depend on $\beta$. We differentiate \eqref{impermanent loss2} with respect to $\beta,$ and obtain
\begin{align*}
    &- \frac{ \frac{\partial (p_A^{(0)}x_2 + (1+\beta)p_B^{(0)}y_2)}{\partial \beta}(p_A^{(0)}x_1 + (1+\beta)p_B^{(0)}y_1) - (p_B^{(0)}y_1)(p_A^{(0)}x_2 + (1+\beta)p_B^{(0)}y_2)}{ (({p_A^{(0)}x_1 + (1+\beta)p_B^{(0)}y_1})^2}\\ &= - \frac{ p_B^{(0)}y_2(p_A^{(0)}x_1 + (1+\beta)p_B^{(0)}y_1) - p_B^{(0)}y_1(p_A^{(0)}x_2 + (1+\beta)p_B^{(0)}y_2)}{ (({p_A^{(0)}x_1 + (1+\beta)p_B^{(0)}y_1})^2} \\ &= - \frac{ p_B^{(0)}p_A^{(0)} (x_1y_2 -x_2y_1)}{ (({p_A^{(0)}x_1 + (1+\beta)p_B^{(0)}y_1})^2}
\end{align*}
where we have used $\frac{\partial (p_A^{(0)}x_2 + (1+\beta)p_B^{(0)}y_2)}{\partial \beta} = p_B^{(0)}y_2$. Observe that $x_2$ and $y_2$ changes with $\beta$, but the equality just mentioned holds by Lemma \ref{lowest value on curve1}. As $\frac{F_x(x_2,y_2)}{F_y(x_2,y_2)} = \frac{p_A^{(0)}}{(1+\beta)p_B^{(0)}}$ is satisfied, $(x_2,y_2)$ is the solution to the optimization problem:
\begin{equation}
\begin{aligned}
\min_{x_2,y_2} \quad & p_A^{(0)}x_2 + (1+\beta)p_B^{(0)}y_2 \\
\textrm{s.t.} \quad & F(x_1,y_1)=F(x_2,y_2). \\
\end{aligned}
\end{equation}
Therefore, we can apply the envelope theorem and obtain $\frac{\partial (p_A^{(0)}x_2 + (1+\beta)p_B^{(0)}y_2)}{\partial \beta} = p_B^{(0)}y_2$. Moreover, as shown in the proof of Lemma \ref{lowest value on curve1}, we can rewrite $F(x_1,y_1)=F(x_2,y_2)$ as 

\begin{equation} 
     y_2 - y_1 = m(x_2 - x_1), x_2 -x_1\in \mathbb{R},
\end{equation}
 where $m$ is a twice differentiable function, whose first order derivative  is $m'(x_2 - x_1)= - \frac{ F_x(x_2,y_2)}{F_y(x_2,y_2)}  < 0$.  By Assumption \ref{curve},  $m$ is convex, that is, $m'' > 0.$ 
When $x_2 - x_1 = 0$, we have $m'(0)= - \frac{ F_x(x_1,y_1)}{F_y(x_1,y_1)}.$ Since $m'' > 0$ and $-m'(0) = \frac{F_x(x_1,y_1)}{F_y(x_1,y_1)} = \frac{p_A^{(0)}}{p_B^{(0)}} > \frac{p_A^{(0)}}{(1+\beta)p_B^{(0)}} = \frac{F_x(x_2,y_2)}{F_y(x_2,y_2)} = -m'(x_2 - x_1)$, we have $x_2 - x_1>0$. Moreover, as $m'<0$, we have $ y_2 - y_1 = m(x_2 - x_1) < m(0) = 0$, so $y_2 - y_1<0$. Hence, we obtain  $x_1y_2 -x_2y_1<0 $.  Therefore, we obtain $$\frac{\partial  IL\left(\frac{p_A^{(0)}}{p_B^{(0)}},\frac{p_A^{(0)}}{(1+\beta)p_B^{(0)}} \right) }{\partial \beta} = - \frac{ p_B^{(0)}p_A^{(0)} (x_1y_2 -x_2y_1)}{ (({p_A^{(0)}x_1 + (1+\beta)p_B^{(0)}y_1})^2}> 0.$$


\end{proof} 

\begin{proof}[Proof of Corollary  \ref{cor1}]
The probability of token exchange reversion is exactly the probability that an A token is hit by an idiosyncratic shock while a B token is not hit by an idiosyncratic shock:
$$(1-\theta)\kappa_1(1-\kappa_2),$$
which clearly decreases in $\kappa_2$.




The probability that the fair value exchange rate remains $\frac{p_A^{(0)}}{(1+\beta)p_B^{(0)}}$ is $(1-\theta)(\kappa_1\kappa_2 +(1-\kappa_2)(1-\kappa_1)) +\theta$. The probability that the fair value exchange rate becomes $\frac{p_A^{(0)}}{(1+\beta)^2p_B^{(0)}}$ after period 2 is $(1-\theta)\kappa_2(1-\kappa_1)$.   By the law of total expectation, the expected ``impermanent loss'' is 
$$IL\left(\frac{p_A^{(0)}}{p_B^{(0)}}, \frac{p_A^{(0)}}{(1+\beta)p_B^{(0)}} \right) \left((1-\theta)(\kappa_1\kappa_2 +(1-\kappa_2)(1-\kappa_1)) +\theta\right) + IL\left(\frac{p_A^{(0)}}{p_B^{(0)}},\frac{p_A^{(0)}}{(1+\beta)^2p_B^{(0)}} \right) (1-\theta)\kappa_2(1-\kappa_1), $$
whose partial derivative with respect to $\kappa_2$ is then
$$IL\left(\frac{p_A^{(0)}}{p_B^{(0)}}, \frac{p_A^{(0)}}{(1+\beta)p_B^{(0)}} \right) (1-\theta)\kappa_1   +\left( IL\left(\frac{p_A^{(0)}}{p_B^{(0)}},\frac{p_A^{(0)}}{(1+\beta)^2p_B^{(0)}}\right)-IL\left(\frac{p_A^{(0)}}{p_B^{(0)}}, \frac{p_A^{(0)}}{(1+\beta)p_B^{(0)}} \right) \right) (1-\theta)(1-\kappa_1)>0, $$
where we have used the inequality $ IL\left(\frac{p_A^{(0)}}{p_B^{(0)}},\frac{p_A^{(0)}}{(1+\beta)^2p_B^{(0)}}\right) > IL\left(\frac{p_A^{(0)}}{p_B^{(0)}}, \frac{p_A^{(0)}}{(1+\beta)p_B^{(0)}} \right)$, which holds because of Lemma \ref{lemma impermenent loss} and the condition $(1+\beta)^2 > (1+\beta)$. Hence, the expected ``impermanent loss'' is increasing in the parameter $\kappa_2$. 

The expected marginal fee revenue from deposits, $\frac{\kappa_I}{2}R_{inv_A}^{(2)} + \frac{\kappa_I}{2}R_{inv_B}^{(2)} $, is independent of $\kappa_2$ because ${\kappa_I},R_{inv_A}^{(2)}, R_{inv_B}^{(2)}  $ do not depend on the parameter $\kappa_2$.

As shown in the proof of Proposition \ref{freezeequilibrium5}, liquidity providers choose to deposit in period 2 if and only $\beta < \beta_{frz}^{(2)}= \beta_{frz}\left( \frac{p_A^{(0)}}{(1+\beta)p_B^{(0)}}\right)$. We have shown in the proof of Proposition \ref{freezeequilibrium5} that $\frac{\partial \beta_{frz}^{(2)}} {\partial \kappa_2} > 0$, hence the threshold  $\beta_{frz}^{(2)}$ increases in $\kappa_2$. 

\end{proof}

\begin{proof}[Proof of Proposition  \ref{feesgoup6}]

By Assumption~\ref{assarb}, {in every period $t=1,2$,} the gas fee attached to the arbitrage order is equal to the largest profit possible from the arbitrage, {given by} $ \pi(y_A^{(t,2)},y_B^{(t,2)},p_B^{(t,2)}, p_A^{(t,2)})$. 
By Lemma \ref{proportionality2}, $$ \pi(y_A^{(t,2)},y_B^{(t,2)},p_B^{(t,2)}, p_A^{(t,2)})  = y_A^{(t,1)} \pi\left(\frac{y_A^{(t,2)}}{y_A^{(t,1)}},\frac{y_B^{(t,2)}}{y_A^{(t,1)}},p_B^{(t,2)}, p_A^{(t,2)} \right) ,$$
where {at sub-period 1 of period $t$},  $\pi\left(\frac{y_A^{(t,2)}}{y_A^{(t,1)}},\frac{y_B^{(t,2)}}{y_A^{(t,1)}},p_B^{(t,2)}, p_A^{(t,2)}\right) > 0 $ is a random variable that does not depend on the amount of  A tokens $y_A^{(t,1)}$ deposited in the AMM. The required ratio of deposits $\frac{ y_A^{(t,1)}}{y_B^{(t,1)}}$ is also uniquely pinned down by the token price ratio before sub-period 1 of period $t$, which means that $y_B^{(t,1)}$ is a constant multiple of $y_A^{(t,1)}$. Hence, $\pi\left(\frac{y_A^{(t,2)}}{y_A^{(t,1)}},\frac{y_B^{(t,2)}}{y_A^{(t,1)}},p_B^{(t,2)}, p_A^{(t,2)}\right)$ also does not depend on the amount of B tokens deposited, $y_B^{(t,1)}$. We have that
\begin{align*}
    \expe_{(t,1)} [g_{arb}^{(t,3)}] &= \expe_{(t,1)} [\pi(y_A^{(t,2)},y_B^{(t,2)},p_B^{(t,2)}, p_A^{(t,2)})] \\
    &=  \expe_{(t,1)} \left[y_A^{(t,1)} \pi\left( \frac{y_A^{(t,2)}}{y_A^{(t,1)}},\frac{y_B^{(t,2)}}{y_A^{(t,1)}},p_B^{(t,2)}, p_A^{(t,2)}\right) \right]\\
    &= y_A^{(t,1)} \expe_{(t,1)} \left[\pi\left( \frac{y_A^{(t,2)}}{y_A^{(t,1)}},\frac{y_B^{(t,2)}}{y_A^{(t,1)}},p_B^{(t,2)}, p_A^{(t,2)})  \right)\right],
\end{align*}
which increases in $y_A^{(t,1)}$ because $\pi\left( \frac{y_A^{(t,2)}}{y_A^{(t,1)}},\frac{y_B^{(t,2)}}{y_A^{(t,1)}},p_B^{(t,2)}, p_A^{(t,2)}\right) $ is a non-negative random variable, and thus $ \expe_{(t,1)} \left[\pi\left( \frac{y_A^{(t,2)}}{y_A^{(t,1)}},\frac{y_B^{(t,2)}}{y_A^{(t,1)}},p_B^{(t,2)}, p_A^{(t,2)}\right)\right] \geq 0$. The quantity $\expe_{(t,1)} [g_{arb}^{(t,3)}]$ is also increasing in $y_B^{(t,1)}$ because $y_B^{(t,1)}$ is just $y_A^{(t,1)}$ multiplied by a constant.


 Using a similar argument, we find that $ Var_{(t,1)} [g_{arb}^{(t,3)}] = (y_A^{(t,1)})^2 Var_{(t,1)} \left[\pi\left( \frac{y_A^{(t,2)}}{y_A^{(t,1)}},\frac{y_B^{(t,2)}}{y_A^{(t,1)}},p_B^{(t,2)}, p_A^{(t,2)} \right)\right]$, which implies that the variance of gas fee increases both in $y_A^{(t,1)}$ and in $y_B^{(t,1)}$.

\end{proof}


\begin{proof}[Proof of Lemma  \ref{curvature lemma1}]
Recall that liquidity providers must deposit their tokens at the spot price 
$ \frac{ F_x}{F_y}\bigg\rvert_{(x,y) = (y_A^{(t,1)}, y_B^{(t,1)})} =\frac{p_A^{(t,1)}}{p_B^{(t,1)}}$. This condition guarantees that a liquidity provider deposits A tokens and B tokens with same value, i.e., $\frac{p_A^{(t,1)}}{p_B^{(t,1)}} = \frac{y_B^{(t,1)}}{y_A^{(t,1)}}$ for $t=1,2$. 

We consider the case when a shock only hits the B token {in  period $t,t\in \{1,2\}$}. The case where a shocks hits the A token can be proven via symmetric arguments. We plug the specification of the pricing curve, $F^{(t)}_k (x,y) = (1-k) \; A \; F_0^{(t)} (
x,y) + k \; F_1 (x,y) = (1-k) \; A \; ( p_A^{(t,1)} x +p_B^{(t,1)} y) + k \; xy$, into the arbitrageur's optimization problem \eqref{eq arbitrageroptimization}. After straightforward algebraic manipulations, we obtain the following arbitrageur's optimization problem in period $t$:
 \begin{equation} \label{arbitrageur optimization curvature}
\begin{aligned}
\max_{\Delta q_A^{(t,3)}, \Delta q_B^{(t,3)}} \quad &  p_A^{(t,2)}(1+f) \Delta q_A^{(t,3)} + p_B^{(t,2)}\Delta q_B^{(t,3)}\\
\textrm{s.t.} \quad & q_B^{(t,3)} = \frac{q_A^{(t,3)}y_B^{(t,2)}}{kq_A^{(t,3)}-y_A^{(t,2)}},
\\
  &\Delta q_A^{(t,3)} \leq 0, y_B^{(t,2)} \geq \Delta q_B^{(t,3)} \geq 0.   \\
\end{aligned}
\end{equation}
{Plugging $ q_B^{(t,3)} = \frac{q_A^{(t,3)}y_B^{(t,2)}}{kq_A^{(t,3)}-y_A^{(t,2)}}$ into $y_B^{(t,2)} \geq \Delta q_B^{(t,3)}\geq 0$, we obtain $\frac{y_A^{(t,2)}}{-1+k}\leq \Delta q_A^{(t,3)} \leq 0$.  We then use $\frac{q_A^{(t,3)}y_B^{(t,2)}}{kq_A^{(t,3)}-y_A^{(t,2)}}$ to replace $q_B^{(t,3)}$ } in \eqref{arbitrageur optimization curvature},  which leads to the following equivalent  single variable and unconstrained optimization problem:
 \begin{equation}\label{optimization arbitrage curvature}
\begin{aligned}
\max_{\frac{y_A^{(t,2)}}{-1+k}\leq \Delta q_A^{(t,3)} \leq 0} \quad &  p_A^{(t,1)}(1+f) \Delta q_A^{(t,3)} + (1+\beta) p_B^{(t,1)} \frac{\Delta q_A^{(t,3)}y_B^{(t,2)}}{k \Delta q_A^{(t,3)}-y_A^{(t,2)}}
\end{aligned}
\end{equation}

As in the more general case of the proof of Lemma \ref{adverseselction order1}, the optimal trading amount $\Delta q_A^{(t,3)} =0$ if $\beta \leq f$. If $\beta > f$, then the optimal trading amount is achieved either when the marginal benefit is equal to the marginal cost, i.e.
$$p_A^{(t,1)}(1+f) -  (1+\beta) p_B^{(t,1)}\frac{y_A^{(t,2)} y_B^{(t,2)}}{(k \Delta q_A^{(t,3)*}-y_A^{(t,2)})^2} = 0,$$ which leads to
\begin{equation}\label{order2}
  \Delta  q_A^{(t,3)*} = \frac{y_A^{(t,2)}}{k}\left(1-\sqrt{\frac{1+\beta}{1+f}}\right)
\end{equation}
or when all the B tokens in the AMM have been withdrawn, i.e.,
$$ p_A^{(t,1)}(1+f) -  (1+\beta) p_B^{(t,1)}\frac{y_A^{(t,2)} y_B^{(t,2)}}{(k \Delta q_A^{(t,3)*}-y_A^{(t,2)})^2} < 0,  \frac{\Delta q_A^{(t,3)*}y_B^{(t,2)}}{k \Delta q_A^{(t,3)*}-y_A^{(t,2)}} = y_B^{(t,2)},$$ which leads to the following solution
\begin{equation}\label{order1}
   \Delta q_A^{(t,3)*} =\frac{y_A^{(t,2)}}{-1+k}.
\end{equation}
The latter case occurs if $1+\beta > \frac{1+f}{(1-k)^2}$, i.e., if the curvature {of the pricing function} is sufficiently small. 

We then plug the optimal trading {amount} into \eqref{optimization arbitrage curvature}. If $\beta \leq f$,  we have $\pi(y_A^{(t,2)},y_B^{(t,2)},p_B^{(t,2)}, p_A^{(t,1)}) = 0$. If $\frac{1+f}{(1-k)^2} -1 >\beta > f,$  plugging \eqref{order2} into the objective function given by \eqref{optimization arbitrage curvature}, we obtain
 \begin{equation} \label{arbitrage loss curevature 1}
\begin{aligned} 
\pi(y_A^{(t,2)},y_B^{(t,2)},p_B^{(t,2)}, p_A^{(t,2)}) &= \frac{(   \sqrt{\frac{1+\beta}{1+f}}-1)( \sqrt{{(1+\beta)}{(1+f)}}p_B^{(t,1)}y_B^{(t,2)}- p_A^{(t,1)}(1+f)  y_A^{(t,2)})}{k} \\
&= \frac{( \sqrt{1+\beta}- \sqrt{1+f})^2 p_A^{(t,1)} y_A^{(t,1)}}{k},
\end{aligned}
\end{equation}
where to obtain the second equality, we use the identity $\frac{p_A^{(t,1)}}{p_B^{(t,1)}} = \frac{y_B^{(t,1)}}{y_A^{(t,1)}}$, and the fact that $y_A^{(t,2)} = y_A^{(t,1)} , y_B^{(t,2)} = y_B^{(t,1)} $ if only a token price shock arrives at sub-period 2 of period $t$. Applying the above identities again, we have that if an arbitrage opportunity occurs in period $t$, {then the realized arbitrage loss ratio is}:
\begin{equation} \label{realized arb ratio 1}
    {\frac{\pi(y_A^{(t,2)},y_B^{(t,2)},p_B^{(t,2)}, p_A^{(t,2)})}{p_A^{(t,1)}y_A^{(t,1)}+p_B^{(t,1)}y_B^{(t,1)}}} = \frac{( \sqrt{1+\beta}- \sqrt{1+f})^2 }{2k}.
\end{equation}
The probability that an arbitrage occurs in period $t$ is $ (1-\theta)(\kappa_1 (1- \kappa_2) + \kappa_2 (1- \kappa_1)) $, so we have
$$\expe_{(t,1)}\bigg[\frac{\pi(y_A^{(t,2)},y_B^{(t,2)},p_B^{(t,2)}, p_A^{(t,2)})}{p_A^{(t,1)}y_A^{(t,1)}+p_B^{(t,1)}y_B^{(t,1)}}\bigg] = (1-\theta)(\kappa_1 (1- \kappa_2) + \kappa_2 (1- \kappa_1)) \frac{( \sqrt{1+\beta}- \sqrt{1+f})^2 }{2k}$$
The above ratio is obviously decreasing in $k$. If $\beta \geq \frac{1+f}{(1-k)^2} -1 ,$ we plug \eqref{order1} into the objective function given by \eqref{optimization arbitrage curvature}. Following the same procedure as above, we obtain 
\begin{align}\label{arbitrage loss curevature 3}
     \pi(y_A^{(t,2)},y_B^{(t,2)},p_B^{(t,2)}, p_A^{(t,2)}) &= {(1+\beta)}p_B^{(t,1)}y_B^{(t,2)} - (1+f) p_A^{(t,1)} \frac{y_A^{(t,2)}}{1-k} \nonumber\\
     &=  {(1+\beta)}p_B^{(t,1)}y_B^{(t,1)} - (1+f) p_A^{(t,1)} \frac{y_A^{(t,1)}}{1-k}, 
\end{align}
where to obtain the last equality, we use the identity $\frac{p_A^{(t,1)}}{p_B^{(t,1)}} = \frac{y_B^{(t,1)}}{y_A^{(t,1)}}$, and the fact that $y_A^{(t,2)} = y_A^{(t,1)} , y_B^{(t,2)} = y_B^{(t,1)} $ if only a token price shock arrives at sub-period 2 of period $t$. The probability that an arbitrage occurs in period $t$ is $ (1-\theta)(\kappa_1 (1- \kappa_2) + \kappa_2 (1- \kappa_1)) $, so we have
 \begin{equation} \label{arbitrage loss curevature 2}
\begin{aligned}
\expe_{(t,1)}\bigg[\frac{\pi(y_A^{(t,2)},y_B^{(t,2)},p_B^{(t,2)}, p_A^{(t,2)})}{p_A^{(t,1)}y_A^{(t,1)}+p_B^{(t,1)}y_B^{(t,1)}}\bigg] = &= (1-\theta)(\kappa_1 (1- \kappa_2) + \kappa_2 (1- \kappa_1))  \frac{{(1+\beta)}p_B^{(t,1)}y_B^{(t,2)} - (1+f) p_A^{(t,1)} \frac{y_A^{(t,2)}}{1-k}}{p_A^{(t,1)}y_A^{(t,1)}+p_B^{(t,1)}y_B^{(t,1)}} \\ &= (1-\theta)(\kappa_1 (1- \kappa_2) + \kappa_2 (1- \kappa_1))  \frac{(1+\beta) - \frac{1+f}{1-k}}{2},
\end{aligned}
\end{equation}
where we use the identity $\frac{p_A^{(t,1)}}{p_B^{(t,1)}} = \frac{y_B^{(t,1)}}{y_A^{(t,1)}}$. Clearly, the above expectation is decreasing in $k$. 

We then calculate the investors' maximum surplus for ``type A'' investors and ``type B'' investors, respectively given by {$s_A (y_A^{(t,1)},y_B^{(t,1)},p_B^{(t,1)}, p_A^{(t,1)}),$ and $ s_B (y_A^{(t,1)},y_B^{(t,1)},p_B^{(t,1)}, p_A^{(t,1)}).$} We consider the case where a ``type B'' investor arrives. The case where a ``type A'' investor arrives can be easily handled with symmetric arguments. We proceed as above, and plug the expression of the pricing function $F^{(t)}_k (x,y)$ into the investor's optimization problem \eqref{traderoptimization}. This leads to the following equivalent single-variable optimization problem:
 
  \begin{equation}\label{optimization investor curvature}
\begin{aligned}
\max_{\frac{y_A^{(t,1)}}{-1+k}\leq \Delta Q_A^{(t,2)} \leq 0} \quad &  p_A^{(t,1)}(1+f) \Delta Q_A^{(t,2)}  + (1+\alpha) p_B^{(t,1)} \frac{\Delta Q_A^{(t,2)} y_B^{(t,1)}}{k\Delta Q_A^{(t,2)} -y_A^{(t,1)}}
\end{aligned}
\end{equation}
 
If $\alpha \leq f$, then the optimal trading amount is $\Delta Q_A^{(t,2)*} =0$. If $\frac{1+f}{(1-k)^2} -1 > \alpha > f$, then the optimal trading amount is achieved  when the marginal benefit is equal to marginal cost, i.e.,
\begin{equation} \label{eq optimal trading amount}
   \Delta Q_A^{(t,2)*}   = \frac{y_A^{(t,1)}}{k}\left(1-\sqrt{\frac{1+\alpha}{1+f}}\right).
\end{equation}
If $1+\alpha > \frac{1+f}{(1-k)^2}$, then the optimal trading amount is achieved {when all B tokens in the AMM have been withdrawn}, i.e.,
\begin{equation}\label{eq optimal trading amount2}
   \Delta Q_A^{(t,2)*} =\frac{y_A^{(t,1)}}{-1+k}.
\end{equation}

If $\alpha \leq f,$ we have that $s_B(y_A^{(t,1)},y_B^{(t,1)},p_B^{(t,1)}, p_A^{(t,1)})= 0$. If $\frac{1+f}{(1-k)^2} -1 > \alpha > f,$ then {the (``type B'') investor's maximum surplus is given by}
$$s_B (y_A^{(t,1)},y_B^{(t,1)},p_B^{(t,1)}, p_A^{(t,1)}) = \frac{( \sqrt{1+\alpha}- \sqrt{1+f})^2}{k}p_A^{(t,1)}y_A^{(t,1)}.$$ Using the identity $\frac{p_A^{(t,1)}}{p_B^{(t,1)}} = \frac{y_B^{(t,1)}}{y_A^{(t,1)}}$, we obtain
 \begin{equation}
\begin{aligned}
\frac{s_B (y_A^{(t,1)},y_B^{(t,1)},p_B^{(t,1)}, p_A^{(t,1)})}{p_A^{(t,1)}y_A^{(t,1)}+p_B^{(t,1)}y_B^{(t,1)}}
&=  \frac{( \sqrt{1+\alpha}- \sqrt{1+f})^2}{2k}.
\end{aligned}
\end{equation}
The above expression obviously decreases in $k$. If $ \alpha \geq \frac{1+f}{(1-k)^2} -1 ,$  then we have {the  investor's maximum surplus:}
$$s_B (y_A^{(t,1)},y_B^{(t,1)},p_B^{(t,1)}, p_A^{(t,1)}) = \left((1+\alpha) - \frac{1+f}{1-k}\right) p_A^{(t,1)}y_A^{(t,1)},$$ and 
 \begin{equation}\label{surplus ratio 2}
\begin{aligned}
\frac{s_B (y_A^{(t,1)},y_B^{(t,1)},p_B^{(t,1)}, p_A^{(t,1)})}{p_A^{(t,1)}y_A^{(t,1)}+p_B^{(t,1)}y_B^{(t,1)}}&= \frac{1}{2} \left((1+\alpha) - \frac{1+f}{1-k}\right),
\end{aligned}
\end{equation}
which is also decreasing in $k$.

{Following the same procedure above, we can show that for a ``type A'' investor arriving in period $t$, $\frac{s_A (y_A^{(t,1)},y_B^{(t,1)},p_B^{(t,1)}, p_A^{(t,1)})}{p_A^{(t,1)}y_A^{(t,1)}+p_B^{(t,1)}y_B^{(t,1)}} = \frac{s_B (y_A^{(t,1)},y_B^{(t,1)},p_B^{(t,1)}, p_A^{(t,1)})}{p_A^{(t,1)}y_A^{(t,1)}+p_B^{(t,1)}y_B^{(t,1)}}$, and both of these quantities are decreasing in $k$. Therefore,  $\frac{\sum_{i=A,B}s_i(y_A^{(t,1)},y_B^{(t,1)},p_B^{(t,1)}, p_A^{(t,1)})}{p_A^{(t,1)}y_A^{(t,1)}+p_B^{(t,1)}y_B^{(t,1)}} $ is decreasing in $k$. }

\end{proof}

\begin{proof}[Proof of Proposition  \ref{curvature propostion}]

We follow the proof strategy of Proposition  \ref{freezeequilibrium5}. Let $k_1 := 1 - \sqrt{\frac{1+f}{1+\alpha}}$, which {yields the condition} $ \alpha = \frac{1+f}{(1-k_1)^2} -1 $. Let $k_2 := 1 - \sqrt{\frac{1+f}{1+\beta}}$, which {yields} the condition $\beta = \frac{1+f}{(1-k_2)^2} -1 $. It follows from the assumption $\alpha \geq \beta $ that $k_1 \geq  k_2$.
   
We distinguish between two cases: (1) there does not exist an arbitrage opportunity after the investor arrives and trades in sub-period 2 of period $t, t=1,2$ , (2) there exists an arbitrage  opportunity after the investor arrives and trades in sub-period 2 of period $t, t=1,2$. We only consider case (1), as case (2) follows from similar arguments and using a slightly different expression for the return. 

Observe that case (1) occurs if $\sqrt{1-\frac{f}{\sqrt{\alpha+1} \sqrt{f+1}}}-\frac{ \sqrt{f+1}}{\sqrt{\alpha+1}} \leq 0$.  
As in the more general case considered in Proposition  \ref{curvature propostion}, the liquidity provider will deposit in  period 2 if and only if $\expe_{(2,1)} \lrb{R_{D}^{(2)}}> \expe_{(2,1)}\lrb{R_{A}^{(1)}}$. The expected one-period return from holding A tokens is $\expe_{(2,1)}\lrb{R_{A}^{(1)}} = (\kappa_1(1-\theta)+\kappa_{com}\theta)\beta$. It remains to calculate the expected one-period return from depositing $\expe_{(2,1)} \lrb{R_{D}^{(2)}}$.

We first consider the case where a ``type B'' investor arrives to the AMM, and calculate the return from depositing. Recall from \eqref{return of fees1} that 
\begin{align} \label{investor return curvature general}
   R_{inv_B}^{(2)} &=    \frac{(1+f) p_A^{(2,1)}(-\Delta Q_A^{(2,2)*})  + p_B^{(2,1)} (-\Delta Q_B^{(2,2)*})}{p_A^{(2,1)}{y_A^{(2,1)}} + p_B^{(2,1)} {y_B^{(2,1)}}}
\end{align}
If $k>k_1$, plugging $\Delta Q_A^{(2,2)*} = \frac{y_A^{(2,1)}}{k}\left(1-\sqrt{\frac{1+\alpha}{1+f}}\right)$ from \eqref{eq optimal trading amount}  into \eqref{investor return curvature general} and using the relation $Q_B^{(2,2)*} = \frac{\Delta Q_A^{(2,2)*} y_B^{(2,2)}}{k\Delta Q_A^{(2,2)*} -y_A^{(2,2)}}$,   we obtain: 
\begin{align} \label{invesotr return curvature1}
   R_{inv_B}^{(2)} &=    \frac{(1+f) p_A^{(2,1)}\left(\frac{y_A^{(2,1)}}{k}\left(\sqrt{\frac{1+\alpha}{1+f}}-1\right)\right)  - p_B^{(2,1)} \frac{y_B^{(2,1)}}{k \sqrt{\frac{1+\alpha}{1+f}}}\left(\sqrt{\frac{1+\alpha}{1+f}}-1\right)}{p_A^{(2,1)}{y_A^{(2,1)}} + p_B^{(2,1)} {y_B^{(2,1)}}}
   \nonumber\\
   &=  \frac{(1+f - \sqrt{\frac{1+f}{1+\alpha}}) (\sqrt{\frac{1+\alpha}{1+f}}-1) }{2k}
\end{align}
If $k< k_1$, plugging $\Delta Q_A^{(2,2)*} =\frac{y_A^{(2,1)}}{-1+k}$ from \eqref{eq optimal trading amount2} into \eqref{investor return curvature general} and using the relationship $Q_B^{(2,2)*} = \frac{\Delta Q_A^{(2,2)*} y_B^{(2,2)}}{k\Delta Q_A^{(2,2)*} -y_A^{(2,2)}}$, we have: 
\begin{align} \label{invesotr return curvature2}
   R_{inv_B}^{(2)} &=    \frac{(1+f) p_A^{(2,1)}\frac{y_A^{(2,1)}}{1-k}  - p_B^{(2,1)} y_B^{(2,1)}}{p_A^{(2,1)}{y_A^{(2,1)}} + p_B^{(2,1)} {y_B^{(2,1)}}}
   \nonumber\\
   &=  \frac{1}{2}\left({\frac{1+f}{1-k}-1 }\right)
\end{align}

Following the same procedure above, we obtain  $R_{inv_A}^{(2)} = R_{inv_B}^{(2)}$.

We then consider the event where only the B token is hit by an idiosyncratic shock. Then there exists an arbitrage opportunity. Recall that in this case, the return from depositing is then:
$$  R_{arb_B}^{(2)} =:\beta\frac{p_B^{(2,1)}y_B^{(2,1)}}{p_A^{(2,1)} y_A^{(2,1)} + p_B^{(2,1)} y_B^{(2,1)}} -\frac{\pi(y_A^{(2,1)},y_B^{(2,1)},(1+\beta)p_B^{(1,2)}, p_A^{(1,2)})}{p_A^{(2,1)} y_A^{(2,1)} + p_B^{(2,1)} y_B^{(2,1)}}$$
    
If $k>k_2$, using \eqref{arbitrage loss curevature 1}, the identity $\frac{p_A^{(2,1)}}{p_B^{(2,1)}} = \frac{y_B^{(2,1)}}{y_A^{(2,1)}}$, and the fact that $y_A^{(2,2)} = y_A^{(2,1)} , y_B^{(2,2)} = y_B^{(2,1)} $ when no investor arrives at sub-period 2 of period 2,  we obtain: 
\begin{align} \label{return deposit arb curvature 1}
   R_{arb_B}^{(2)} &=   \frac{\beta}{2} - \frac{( \sqrt{1+\beta}- \sqrt{1+f})^2 }{2k}.
\end{align}
Similarly, if $k< k_2$, using \eqref{arbitrage loss curevature 3} and the above identities, we have: 
\begin{align} \label{return deposit arb curvature 2}
   R_{arb_B}^{(2)} &=    \frac{\beta}{2} - \frac{(1+\beta - \frac{1+f}{1-k}) }{2}.
\end{align}

Following the same procedure above, we obtain $ R_{arb_B}^{(2)} =  R_{arb_A}^{(2)}$.

We now  compare the expectation of the one-period return from depositing at period 2, $\expe_{(2,1)} \lrb{R_{D}^{(2)}}$, with the expectation of the one-period return from holding an A tokens at period 2, $\expe_{(2,1)}\lrb{R_{A}^{(2)}}$. Recall from \eqref{total probability} that we have:
\begin{align}\label{total probability2}
    &\expe_{(2,1)} \lrb{R_{D}^{(2)}}- \expe_{(2,1)}\lrb{R_{A}^{(2)}} \nonumber\\
     =& \frac{\kappa_I}{2}R_{inv_A}^{(2)} + \frac{\kappa_I}{2}R_{inv_B}^{(2)}+ (1-\theta)\left( \kappa_1(1-\kappa_2)R_{arb_A}^{(2)} +\kappa_2(1-\kappa_1)R_{arb_B}^{(2)} + \kappa_1\kappa_2 \beta\right)  \nonumber\\ +& \theta\kappa_{com} \beta -((1-\theta)\kappa_1+\theta \kappa_{com}) \beta 
\end{align}

If $k \geq k_1 \geq k_2$, plugging \eqref{invesotr return curvature1} and \eqref{return deposit arb curvature 1} into \eqref{total probability2}, and using the identities  $ R_{arb_B}^{(2)} =  R_{arb_A}^{(2)}$ and $R_{inv_A}^{(2)} = R_{inv_B}^{(2)}$,  we obtain: 
\begin{align}\label{payoff case 1}
    &\expe_{(2,1)} \lrb{R_{D}^{(2)}}- \expe_{(2,1)}\lrb{R_{A}^{(2)}}      = \frac{1}{k} \tau_1 -   (1-\theta)(\kappa_1-\kappa_2) \beta,
\end{align}
where $$\tau_1  = \bigg({\kappa_I} \frac{(1+f - \sqrt{\frac{1+f}{1+\alpha}}) (\sqrt{\frac{1+\alpha}{1+f}}-1) }{2} -  (1-\theta)\frac{ (\sqrt{1+\beta}- \sqrt{1+f})^2}{2}(\kappa_1(1-\kappa_2)+\kappa_2(1-\kappa_1)) \bigg).  $$

If $\tau_1 < 0$, then all the terms in \eqref{payoff case 1} are negative, and we have $\expe_{(2,1)} \lrb{R_{D}^{(2)}}- \expe_{(2,1)}\lrb{R_{A}^{(2)}} < 0,$ that is, the expected one-period return from depositing is smaller than the expected one-period return from holding A token. As a result, there is a liquidity freeze. The liquidity providers' expected return at period 2 is then $\expe_{(2,1)}\lrb{R_{A}^{(2)}} = (\kappa_1(1-\theta)+\kappa_{com}\theta)\beta $, which is constant with respect to $k$. 

If $\tau_1 > 0$, then $\expe_{(2,1)} \lrb{R_{D}^{(2)}} - \expe_{(2,1)}\lrb{R_{A}^{(2)}} $ decreases in $k$. Since $\expe_{(2,1)}\lrb{R_{A}^{(2)}}$ is constant in $k$, we also have that $\expe_{(2,1)} \lrb{R_{D}^{(2)}} $ decreases in $k$. Thus, the liquidity providers' expected return at period 2, $\max\left \{\expe_{(2,1)} \lrb{R_{D}^{(2)}}, \expe_{(2,1)}\lrb{R_{A}^{(2)}} \right \}$ also decreases in $k$.

If $ k_1 \geq k \geq k_2 $, plugging \eqref{invesotr return curvature2} and \eqref{return deposit arb curvature 1} into \eqref{total probability2}, and using the identities  $ R_{arb_B}^{(2)} =  R_{arb_A}^{(2)}$ and $R_{inv_A}^{(2)} = R_{inv_B}^{(2)}$,   we obtain 
\begin{align}\label{payoff case 2}
    \expe_{(2,1)} \lrb{R_{D}^{(2)}}- \expe_{(2,1)}\lrb{R_{A}^{(2)}}      =  \tau_2 - (1-\theta)(\kappa_1-\kappa_2) \beta,
\end{align}
where $$\tau_2  = \frac{\kappa_I}{2} \left({\frac{1+f}{1-k}-1 }\right) -  \frac{1}{k} (1-\theta)\frac{ \sqrt{1+\beta}- \sqrt{1+f})^2}{2}(\kappa_1(1-\kappa_2)+\kappa_2(1-\kappa_1)),$$ which  increases in $k$ since all the terms in $\tau_2$ increases in $k$, and $(1-\theta)(\kappa_1-\kappa_2) \beta$ is constant with respect to $k$.

If $ k_2 \geq k \geq 0 $, plugging \eqref{invesotr return curvature1} and \eqref{return deposit arb curvature 2} into \eqref{total probability2}, and using the identities  $ R_{arb_B}^{(2)} =  R_{arb_A}^{(2)}$ and $R_{inv_A}^{(2)} = R_{inv_B}^{(2)}$,  we have: 
\begin{align}\label{payoff case 3}
    \expe_{(2,1)} \lrb{R_{D}^{(2)}}- \expe_{(2,1)}\lrb{R_{A}^{(2)}}      =  \tau_3 - (1-\theta)(\kappa_1-\kappa_2) \beta,
\end{align}
where $$\tau_3  = \bigg(\frac{\kappa_I}{2} \left({\frac{1+f}{1-k}-1 }\right) -  (1-\theta) \frac{(1+\beta - \frac{1+f}{1-k}) }{2}(\kappa_1(1-\kappa_2)+\kappa_2(1-\kappa_1)) \bigg),  $$ which  increases in $k$ since all the terms in $\tau_3$ increases in $k$, and $(1-\theta)(\kappa_1-\kappa_2) \beta$ is constant in $k$.

Therefore, $\expe_{(2,1)} \lrb{R_{D}^{(2)}}- \expe_{(2,1)}\lrb{R_{A}^{(2)}} $ increases in $k$ for $k< k_1$. Since $\expe_{(2,1)}\lrb{R_{A}^{(2)}}$ is constant in $k$, we have the expected return from depositing, $\expe_{(2,1)} \lrb{R_{D}^{(2)}}$ increases in $k$. Moreover, we also have  the liquidity providers' expected return at period 2, $\max\left \{\expe_{(2,1)} \lrb{R_{D}^{(2)}}, \expe_{(2,1)}\lrb{R_{A}^{(2)}} \right\}$ increases in $k$ for $k<k_1$.

The liquidity providers' expected return at period 2, $\max\left \{\expe_{(2,1)} \lrb{R_{D}^{(2)}}, \expe_{(2,1)}\lrb{R_{A}^{(2)}} \right \}$ is maximized at $k^* = k_1,$ since it increases in $k$ on the interval $[0,k_1]$ and decreases in $k$ on the interval $[k_1,1]$. Since  $\expe_{(2,1)} \lrb{R_{D}^{(2)}}- \expe_{(2,1)}\lrb{R_{A}^{(2)}} $  is also maximized at $k^* = k_1,$   if a liquidity freeze occurs at period 2 for $k = k^*$, i.e, $\expe_{(2,1)} \lrb{R_{D}^{(2)}}- \expe_{(2,1)}\lrb{R_{A}^{(2)}}<0$, then $\expe_{(2,1)} \lrb{R_{D}^{(2)}}- \expe_{(2,1)}\lrb{R_{A}^{(2)}}<0$ for other $k \in [0,1]$, which means that a liquidity freeze also occurs for other $k \in [0,1]$.   Moreover, $\max\left \{\expe_{(2,1)} \lrb{R_{D}^{(2)}}, \expe_{(2,1)}\lrb{R_{A}^{(2)}} \right\}$ only depends on fixed parameters, which means that it does not depend on the state $\omega^{(1,3)}$. Plugging $\ex{\frac{e_i^{(2)}}{e_i^{(1)}}\bigg|\omega^{(1,2)}} =\max\left \{\expe_{(2,1)} \lrb{R_{D}^{(2)}}, \expe_{(2,1)}\lrb{R_{A}^{(2)}} \right\}+1$  into \eqref{V_D}, \eqref{V_A}, and \eqref{V_B}, we have
\begin{align}
    \expe_{(1,1)}\lrb{V_D} &=  \expe_{(1,1)} \lrb{R_{D}^{(1)}} \left(\max\left\{\expe_{(2,1)} \lrb{R_{D}^{(2)}}, \expe_{(2,1)}\lrb{R_{A}^{(2)}} \right\}+1\right), \\
    \expe_{(1,1)}\lrb{V_A} &= \expe_{(1,1)} \lrb{R_{A}^{(1)}} \left(\max\left\{\expe_{(2,1)} \lrb{R_{D}^{(2)}}, \expe_{(2,1)}\lrb{R_{A}^{(2)}} \right\}+1\right),  \\ 
    \expe_{(1,1)}\lrb{V_B} &= \expe_{(1,1)} \lrb{R_{B}^{(1)}} \left(\max\left\{\expe_{(2,1)} \lrb{R_{D}^{(2)}}, \expe_{(2,1)}\lrb{R_{A}^{(2)}} \right\}+1\right).
\end{align}
The above expressions indicate that each liquidity provider only needs to maximize its one-period return in period 1. Hence, liquidity providers choose their portfolios in period 1 exactly the same as they choose their portfolios in period 2, which also maximize the one-period return. Hence, the aggregate payoff of liquidity provider $i$ is:
$$\left(\max\left\{\expe_{(2,1)} \lrb{R_{D}^{(2)}}, \expe_{(2,1)}\lrb{R_{A}^{(2)}} \right\}+1\right)^2 e_i^{(0)},$$
which is maximized at $k^*$.

Because the optimal actions are the same in both periods, a liquidity freeze occurs at period 1 if and only if it occurs at period 2. This implies that in period 1, a liquidity freeze is also least likely when$k=k^*$. As a result, for any period $t=1,2$,  if a liquidity freeze occurs for $k^*$, it also occurs for any other $k\in[0,1]$.
\end{proof}

\begin{proof}[Proof of Proposition  \ref{proposition welfare}]

If there exists a liquidity freeze, then there is no deposit made at the AMM, and consequently the capital efficiency is 0. 

We next consider the case where deposits are nonzero at equilibrium. If the investor does not arrive in period $t$, then $|\Delta Q_A^{(t,2)*}| = |\Delta Q_B^{(t,2)*}|=0 $, and the realized capital efficiency is zero. The probability of an investor arriving is constant with respect to $k$. In order to show that the expected capital efficiency at period $t$ is maximized at $k^*$, it suffices to show that upon the arrival of an investor, the realized capital efficiency is maximized at $k^*$. We again consider the case where a ``type B " investor arrives, and omit the case of a ``type A'' investor arriving, because it follows from similar arguments. Recall that the capital efficiency is defined as:
 \begin{equation}\label{capital efficiency 1}
      {\frac{p_A^{(t,1)} |\Delta Q_A^{(t,2)*}| +p_B^{(t,1)} |\Delta Q_B^{(t,2)*}|}{p_A^{(t,1)}y_A^{(t,1)}+p_B^{(t,1)}y_B^{(t,1)}}} =    {\frac{p_A^{(t,1)} |\Delta Q_A^{(t,2)*}| +p_B^{(t,1)} |\frac{\Delta Q_A^{(t,2)*} y_B^{(t,1)}}{k\Delta Q_A^{(t,2)*} -y_A^{(t,1)}}|}{p_A^{(t,1)}y_A^{(t,1)}+p_B^{(t,1)}y_B^{(t,1)}}},
 \end{equation}
where we have used the relation $\Delta Q_B^{(t,2)*} = \frac{\Delta Q_A^{(t,2)*} y_B^{(t,1)}}{k\Delta Q_A^{(t,2)*} -y_A^{(t,1)}}$.

If $k \geq k^*, $  plugging $\Delta Q_A^{(t,2)*} = \frac{y_A^{(t,1)}}{k}\left(1-\sqrt{\frac{1+\alpha}{1+f}}\right)$ from \eqref{eq optimal trading amount} into \eqref{capital efficiency 1}, we obtain that the realized capital efficiency when  a ``type B" investor arrives is
\begin{align*}
   \frac{ p_A^{(t,1)}\frac{y_A^{(t,1)}}{k}\left(\sqrt{\frac{1+\alpha}{1+f}}-1\right)  + p_B^{(t,1)} \frac{y_B^{(t,1)}}{k \sqrt{\frac{1+\alpha}{1+f}}}\left(\sqrt{\frac{1+\alpha}{1+f}}-1\right)}{p_A^{(t,1)}{y_A^{(t,1)}} + p_B^{(t,1)} {y_B^{(t,1)}}},
\end{align*}
 which  decreases in $k$.

 If $k < k^*, $ plugging $\Delta Q_A^{(t,2)*} =\frac{y_A^{(t,1)}}{-1+k}$ from \eqref{eq optimal trading amount2} into \eqref{capital efficiency 1},  we obtain
\begin{align*}
 \frac{ p_A^{(t,1)}\frac{y_A^{(t,1)}}{1-k}  + p_B^{(t,1)} y_B^{(t,1)}}{p_A^{(t,1)}{y_A^{(t,1)}} + p_B^{(t,1)} {y_B^{(t,1)}}},
\end{align*}
 which increases in $k$. Therefore, the capital efficiency is maximized at $k=k^*.$

We then show that the expected social welfare is maximized at $k=k^*.$ Recall that the social welfare $W$ is defined as the sum of all agents' expected payoffs:
$$W= \sum_{i =1}^n \ex{e_i^{(2)}}+ \sum_{t=0}^2 \ex{p_B^{(t,2)}\Delta q_B^{(t,3)}+ p_A^{(t,2)}\Delta q_A^{(t,3)} - g_{arb}^{(t,3)}} +  \sum_{t=0}^2 \ex{{\frac{\kappa_I \sum_{i=A,B}s_i(y_A^{(t,1)},y_B^{(t,1)},p_B^{(t,1)}, p_A^{(t,1)})}{2}}}, $$
where the first term is the cumulative expected payoff of liquidity providers, the second term is the expected payoff of the arbitrageur, and the third term is the expected payoff of  investors. 

For $k>k^*$, we have shown  in the proof of Lemma \ref{curvature lemma1} that ``type A'' and ``type B'' investors' maximum surplus is decreasing in $k$. We have also proven that the liquidity providers' aggregate payoff is decreasing in $k$ in Proposition \ref{curvature propostion}. Moreover, we also shown that the equilibrium payoff of  the arbitrageur is 0. Therefore, the sum of all agents' expected payoffs decreases in $k$, for $k>k^*.$

As shown in the proof of Proposition \ref{curvature propostion},  the one-period return from depositing is the same in both periods. Hence, we denote it by $\ex{R_{D}}$ and omit the time superscript.  Similarly, we denote the one-period return from holding  A tokens by $\ex{R_{A}}$.

We now consider the case where $k_2 \leq k\leq k^*$. Recall from the proof of Proposition \ref{curvature propostion} that the liquidity providers deposit in both periods if and only if $\ex{R_{D} - R_{A}}>0$. If $\ex{R_{D} - R_{A}} > 0$ then liquidity providers do not deposit, and only hold A tokens in their portfolio. In such case,  a liquidity freeze occurs in both periods, and both the arbitrageur and investors have zero payoff. Then, the expected social welfare  is $$\sum_{i=1}^n e_i^{(0)}(1+ \ex{R_{A}})^2 = \sum_1^n e_i^{(0)}(1+(\kappa_1(1-\theta)+\kappa_{com}\theta)\beta)^2. $$ If  $\ex{R_{D}- R_{A}} > 0$ then the liquidity providers deposit in both periods, and their expected return in both periods are $\ex{R_{D}}$. The expected social welfare in this case may be written as:
\begin{dmath} \label{social welfare2}
\sum_{i=1}^n e_i^{(0)} (1+\ex{R_{D}})^2 +  \ex{\sum_{i=1}^n e_i^{(0)}\frac{\kappa_I \sum_{i=A,B}s_i(y_A^{(1,1)},y_B^{(1,1)},p_B^{(1,1)}, p_A^{(1,1)})}{2({p_A^{(1,1)}y_A^{(1,1)}+p_B^{(1,1)}y_B^{(1,1)}})}} + \ex{  \sum_{i=1}^n e_i^{(1)}\frac{\kappa_I \sum_{i=A,B}s_i(y_A^{(2,1)},y_B^{(2,1)},p_B^{(2,1)}, p_A^{(2,1)})}{2({p_A^{(2,1)}y_A^{(2,1)}+p_B^{(2,1)}y_B^{(2,1)}})}}
\end{dmath}
where the first term is the the cumulative expected payoff of  liquidity providers, the second and the third terms are the expected  payoff of investors in period 1 and 2 respectively. 

From \eqref{surplus ratio 2} and the condition $\frac{s_A (y_A^{(t,1)},y_B^{(t,1)},p_B^{(t,1)}, p_A^{(t,1)})}{p_A^{(t,1)}y_A^{(t,1)}+p_B^{(t,1)}y_B^{(t,1)}} = \frac{s_B (y_A^{(t,1)},y_B^{(t,1)},p_B^{(t,1)}, p_A^{(t,1)})}{p_A^{(t,1)}y_A^{(t,1)}+p_B^{(t,1)}y_B^{(t,1)}}$, we know that ${\frac{\kappa_I \sum_{i=A,B}s_i(y_A^{(t,1)},y_B^{(t,1)},p_B^{(t,1)}, p_A^{(t,1)})}{2({p_A^{(t,1)}y_A^{(t,1)}+p_B^{(t,1)}y_B^{(t,1)}})}} = \frac{\kappa_I}{2} \left((1+\alpha) - \frac{1+f}{1-k}\right)$ is {the same} for both periods $t=1,2$. Since  liquidity providers deposit in period 1, we also have  $ \ex{  \sum_{i=1}^n e_i^{(1)}} = \sum_{i=1}^n e_i^{(0)} (1+\ex{R_{D}}) $. Using the conditions above, we can rewrite \eqref{social welfare2} as 
$$ \sum_{i=1}^n e_i^{(0)} ((1+\ex{R_{D}}) \tau_4 + \tau_4 +1),$$ where
\begin{align*}
    \tau_4 &= \ex{R_{D}} + {\frac{\kappa_I \sum_{i=A,B}s_i(y_A^{(t,1)},y_B^{(t,1)},p_B^{(t,1)}, p_A^{(t,1)})}{2({p_A^{(t,1)}y_A^{(t,1)}+p_B^{(t,1)}y_B^{(t,1)}})}}
\end{align*} is the expected return of liquidity providers plus  the investors'  surplus ratio multiplied by the investors' arrival probability. Observe that $\tau_4 \geq \ex{R_{D}}$ because $s_i(y_A^{(t,1)},y_B^{(t,1)},p_B^{(t,1)}, p_A^{(t,1)}) >0$ by Lemma \ref{traderoptimization}. We then use \eqref{surplus ratio 2}, \eqref{payoff case 2}, along with the conditions $ \ex{R_{A}} =(\kappa_1(1-\theta)+\kappa_{com}\theta)\beta)$ and $\frac{s_A (y_A^{(t,1)},y_B^{(t,1)},p_B^{(t,1)}, p_A^{(t,1)})}{p_A^{(t,1)}y_A^{(t,1)}+p_B^{(t,1)}y_B^{(t,1)}} = \frac{s_B (y_A^{(t,1)},y_B^{(t,1)},p_B^{(t,1)}, p_A^{(t,1)})}{p_A^{(t,1)}y_A^{(t,1)}+p_B^{(t,1)}y_B^{(t,1)}}$, and rewrite $\tau_4$ as
$$\tau_4 =\frac{\kappa_I}{2} \alpha- \frac{1}{k} (1-\theta)\frac{ (\sqrt{1+\beta}- \sqrt{1+f})^2}{2}(\kappa_1(1-\kappa_2)+\kappa_2(1-\kappa_1)) + (\kappa_1(1-\theta)+\kappa_{com}\theta)\beta - (1-\theta)(\kappa_1-\kappa_2) \beta.$$
Clearly, the expression above is increasing in $k$.

Combining the two cases discussed above, we can write the expected social welfare as
\begin{align} \label{welfare case1}
     W &=  \mathbb{1}_{\ex{R_{D}- R_{A}} \leq 0} \sum_{i=1}^n e_i^{(0)} (1+\ex{R_{A}})^2+ \mathbb{1}_{\ex{R_{D}- R_{A}} > 0} \sum_{i=1}^n e_i^{(0)} ((1+\ex{R_{D}}) \tau_4 + \tau_4 +1) \nonumber
     \\&= (1+(\ex{R_{A}})^2) + \mathbb{1}_{\ex{R_{D}- R_{A}} > 0} \sum_{i=1}^n e_i^{(0)} ((2+\ex{R_{D}}) \tau_4 +1-(1+(\ex{R_{A}})^2)). 
\end{align}
   Moreover, using the inequality $\tau_4 \geq \ex{R_{D}}$, we deduce
  \begin{align*}
      &(2+\ex{R_{D}}) \tau_4 +1-(1+(\ex{R_{A}})^2) \\
      \geq & (1+\ex{R_{D}}) ^2-(1+(\ex{R_{A}})^2)\\ \geq& 0,
  \end{align*}

  Because $\tau_4$ is increases in $k$, so is $\sum_{i=1}^n e_i^{(0)} ((2+\ex{R_{D}}) \tau_4 +1-(1+(\ex{R_{A}})^2))$. We also have that $\mathbb{1}_{\ex{R_{D}- R_{A}} > 0} \geq 0$ is increasing in $k$, as we have shown $\ex{R_{D}- R_{A}}$ increases in $k$ for $k_2 \leq k\leq k^*$ in the proof of Proposition \ref{curvature propostion}. Therefore, the expected social welfare $W$ given in (\ref{welfare case1}) is increasing in $k$ for $k_2 \leq k\leq k^*$.

If $ k\leq k_2$, we can prove that $W$ is increasing in $k$ using a similar procedure. Therefore, the expected social welfare is maximized at $k^*$.
\end{proof}


\begin{proof}[Proof of Proposition  \ref{more than one token proposition}]
If $\beta \leq f$, then no arbitrage occurs, and the arbitrage loss ratio for both AMMs is zero.

Next, we consider the case $\beta >f$. Consider the AMM which manages A and B tokens only. We are considering an AMM utilizing a constant product function, that is, a special case of $F_k^{(t)}$ where we set $k=1$. Plugging $k=1$ into \eqref{arbitrage loss curevature 1}, we have that the realized token value loss when an arbitrage opportunity occurs at period $t, t\in\{1,2\}$, is given by 
$$\pi_{AB}(y_A^{(t,2)},y_B^{(t,2)},p_B^{(t,2)}, p_A^{(t,2)}) = ( \sqrt{1+\beta}- \sqrt{1+f})^2 p_A^{(t,1)} y_A^{(t,1)}.$$ The probability that an arbitrage opportunity occurs is $2(1-\theta)(1-\kappa)\kappa$. At period $t$, the expected arbitrage ratio of the AMM that pools two tokens is
$$\expe_{(t,1)}\bigg[\frac{\pi_{AB}(y_A^{(t,2)},y_B^{(t,2)},p_B^{(t,2)}, p_A^{(t,2)})}{p_A^{(t,1)}y_A^{(t,1)}+p_B^{(t,1)}y_B^{(t,1)}}\bigg]= (1-\theta)(1-\kappa)\kappa ( \sqrt{1+\beta}- \sqrt{1+f})^2,$$
{where we have used that liquidity providers must deposit at the spot rate, and thus the ratio of deposits is uniquely pinned down by the relations $p_A^{(t,1)}y_A^{(t,1)} = p_B^{(t,1)}y_B^{(t,1)}$}.

We then calculate the arbitrage ratio of the AMM that pools three tokens. An arbitrage occurs if the token price shock hits only one or two tokens. The probability that only one token is hit by an idiosyncratic shock is $3(1-\theta)(1-\kappa)^2\kappa$, and the probability that  two tokens are hit by idiosyncratic shock{s} is $3(1-\theta)(1-\kappa)\kappa^2.$ Suppose that only token B is hit by a price shock, then the arbitrageur solves the following optimization problem:
\begin{equation}\label{eq arbitrageroptimization 3tokens}
\begin{aligned}
\max_{\Delta q_A^{(t,3)}, \Delta q_B^{(t,3)}, \Delta q_C^{(t,3)}} \quad &  p_A^{(t,1)}(1+f) \Delta q_A^{(t,3)} + (1+\beta)p_B^{(t,1)}\Delta q_B^{(t,3)} + (1+f)p_C^{(t,1)}\Delta q_C^{(t,3)}\\
\textrm{s.t.} \quad & y_A^{(t,2)} y_B^{(t,2)} y_C^{(t,2)} = (y_A^{(t,2)}-\Delta q_A^{(t,3)}) ( y_B^{(t,2)}-\Delta q_B^{(t,3)}) ( y_C^{(t,2)}-\Delta q_C^{(t,3)}) \\
  &\Delta q_A^{(t,3)} \leq 0, \Delta q_B^{(t,3)} \geq 0,  \Delta q_C^{(t,3)} \leq 0.   \\
\end{aligned}
\end{equation}
The arbitrage profit is maximized if the following first-order conditions are satisfied:
$$\frac{(1+\beta)p_B^{(t,1)} }{(1+f)p_A^{(t,1)}} = \frac{(y_A^{(t,2)}-\Delta q_A^{(t,3)})}{( y_B^{(t,2)}-\Delta q_B^{(t,3)})}, \frac{(1+\beta)p_B^{(t,1)} }{(1+f)p_C^{(t,1)}} = \frac{(y_C^{(t,2)}-\Delta q_C^{(t,3)})}{( y_B^{(t,2)}-\Delta q_B^{(t,3)})}.$$ The relations above hold if the marginal benefit of {exchanging tokens} is equal to the marginal cost of the arbitrageur. 

Using the condition above, the identities 
$p_A^{(t,1)}y_A^{(t,1)} = p_B^{(t,1)}y_B^{(t,1)} = p_C^{(t,1)}y_C^{(t,1)} , y_A^{(t,1)} = y_A^{(t,2)}, y_B^{(t,1)} = y_B^{(t,2)},  y_C^{(t,1)} = y_C^{(t,2)}$ and the constraint $y_A^{(t,2)} y_B^{(t,2)} y_C^{(t,2)} = (y_A^{(t,2)}-\Delta q_A^{(t,3)}) ( y_B^{(t,2)}-\Delta q_B^{(t,3)}) ( y_C^{(t,2)}-\Delta q_C^{(t,3)})$, we obtain
\begin{equation}
\begin{aligned}\label{case of three tokens results}
p_A^{(t,1)}(y_A^{(t,2)}-\Delta q_A^{(t,3)}) &= p_C^{(t,1)}(y_C^{(t,2)}-\Delta q_C^{(t,3)}) = \left(\frac{1+\beta}{1+f}\right)^{\frac{1}{3}}p_A^{(t,1)}y_A^{(t,1)},\\
p_B^{(t,1)}(y_B^{(t,2)}-\Delta q_B^{(t,3)}) &=  \left(\frac{1+\beta}{1+f}\right)^{-\frac{2}{3}}p_A^{(t,1)}y_A^{(t,1)}
\end{aligned}
\end{equation}

Plugging \eqref{case of three tokens results} into the objective function given in \eqref{eq arbitrageroptimization 3tokens}, we then obtain the realized arbitrage loss when only token B is hit by a shock:
\begin{align*}
    \pi_{ABC}(y_A^{(t,2)},y_B^{(t,2)},y_C^{(t,2)},p_C^{(t,2)}, p_B^{(t,2)}, p_A^{(t,2)}) &= p_A^{(t,1)}y_A^{(t,1)}(3+\beta+2f - 3(1+f)^{\frac{2}{3}} (1+\beta)^{\frac{1}{3}}) \\
    &\geq p_A^{(t,1)}y_A^{(t,1)}(2+\beta+f - 2(1+f)^{\frac{1}{2}} (1+\beta)^{\frac{1}{2}})\\
       &= p_A^{(t,1)}y_A^{(t,1)}( \sqrt{1+\beta}- \sqrt{1+f})^2,
\end{align*}
where we have used the geometric inequality $1+f + 2(1+f)^{\frac{1}{2}} (1+\beta)^{\frac{1}{2}}) \geq 3(1+f)^{\frac{2}{3}} (1+\beta)^{\frac{1}{3}}$.  

Following an analogous procedure, we can verify that the following inequality holds if two tokens are hit by price shocks:
\begin{equation} \label{comparison}
     \pi_{ABC}(y_A^{(t,2)},y_B^{(t,2)},y_C^{(t,2)},p_C^{(t,2)}, p_B^{(t,2)}, p_A^{(t,2)}) \geq p_A^{(t,1)}y_A^{(t,1)}( \sqrt{1+\beta}- \sqrt{1+f})^2.
\end{equation}

Recall that the probability that only one token is hit by a shock is $3(1-\theta)(1-\kappa)^2\kappa$, and the probability  that two tokens are hit by idiosyncratic shocks is  $3(1-\theta)(1-\kappa)\kappa^2.$ Applying the inequality in \eqref{comparison} and the identity $p_A^{(t,1)}y_A^{(t,1)} = p_B^{(t,1)}y_B^{(t,1)} = p_C^{(t,1)}y_C^{(t,1)}$, we have:
\begin{align*}
      &\expe_{(t,1)}\bigg[\frac{\pi_{ABC}(y_A^{(t,2)},y_B^{(t,2)},y_C^{(t,2)},p_C^{(t,2)}, p_B^{(t,2)}, p_A^{(t,2)})}{p_A^{(t,1)}y_A^{(t,1)}+p_B^{(t,1)}y_B^{(t,1)}+p_C^{(t,1)}y_C^{(t,1)}}\bigg] \\ &\geq (1-\theta)((1-\kappa)^2\kappa +  (1-\kappa)\kappa^2)( \sqrt{1+\beta}- \sqrt{1+f})^2   \\ &= (1-\theta)(1-\kappa)\kappa ( \sqrt{1+\beta}- \sqrt{1+f})^2 \\ &= \expe_{(t,1)}\bigg[\frac{\pi_{AB}(y_A^{(t,2)},y_B^{(t,2)},p_B^{(t,2)}, p_A^{(t,2)})}{p_A^{(t,1)}y_A^{(t,1)}+p_B^{(t,1)}y_B^{(t,1)}}\bigg]
\end{align*}

\end{proof}

\end{document}